\documentclass[aps,prd,floatfix,nofootinbib,superscriptaddress,amsmath,amssymb,twocolumn,notitlepage]{revtex4-1}
\usepackage{graphicx}
\usepackage{flushend}
\usepackage{hyperref}
\usepackage{natbib}
\usepackage{float}

\usepackage[english]{babel}
\usepackage{amsmath,amssymb,mathrsfs}
\usepackage{bigints}
\usepackage{subfig}
\usepackage{appendix}
\usepackage{bm}
\usepackage[utf8]{inputenc}
\usepackage{grffile} 
\usepackage[font=small,labelfont=bf,justification=raggedright,format=plain,singlelinecheck=off]{caption}
\usepackage{mathtools}
\usepackage{tabularx}
\usepackage{enumitem}
\usepackage{comment}

\newcommand{\be}{\begin{equation}}
\newcommand{\ee}{\end{equation}}
\newcommand{\ba}{\begin{eqnarray}}
\newcommand{\ea}{\end{eqnarray}}

\newcommand{\beq}{\begin{equation}}
\newcommand{\eeq}{\end{equation}}  
\newcommand{\bea}{\begin{eqnarray}}
\newcommand{\eea}{\end{eqnarray}}
\newcommand{\beqa}{\begin{eqnarray}}
\newcommand{\eeqa}{\end{eqnarray}}
\newcommand{\bseq}{\begin{subequations}}
\newcommand{\eseq}{\end{subequations}}

\def\snn{{\sqrt{ s_{\rm NN}}}}
\def\e{{\rm e}}
\def\di{{\rm d}}

\def\0{{\boldsymbol 0}}

\def\detmet{\vert g \vert^{\frac{1}{2}}}
\def\GeV{\mathrm{GeV}}
\def\fm{\mathrm{fm}}
\def\MeV{\mathrm{MeV}}
\def\fm{\mathrm{fm}}
\def\d{\mathrm{d}}

\begin{document}

\preprint{}

\title{Magnetic fields in heavy ion collisions: flow and charge transport } 


\author{Gabriele Inghirami}
\affiliation{Department of Physics,  P.O. Box 35, 40014, University of Jyv\"{a}skyl\"{a}, Finland}
\affiliation{Helsinki Institute of Physics, P.O. Box 64, 00014 University of Helsinki, Finland}

\author{Mark Mace}
\affiliation{Department of Physics,  P.O. Box 35, 40014, University of Jyv\"{a}skyl\"{a}, Finland}
\affiliation{Helsinki Institute of Physics, P.O. Box 64, 00014 University of Helsinki, Finland}

\author{Yuji Hirono}
\affiliation{Asia Pacific Center for Theoretical Physics, Pohang, Gyeongbuk 37673, Korea
}
\affiliation{Department of Physics, POSTECH, Pohang, Gyeongbuk 37673, Korea
}

\author{Luca {Del~Zanna}}
\affiliation{Dipartimento di Fisica e Astronomia, Universit\`a di Firenze, Via G. Sansone 1, I-50019 Sesto F.no (Firenze), Italy}
\affiliation{INAF - Osservatorio Astrofisico di Arcetri, L.go E. Fermi 5, I-50125 Firenze, Italy}
\affiliation{INFN - Sezione di Firenze, Via G. Sansone 1, I-50019 Sesto F.no (Firenze), Italy}

\author{Dmitri E.~Kharzeev}
\affiliation{Department of Physics and Astronomy, Stony Brook University, Stony Brook, New York 11794-3800, USA}
\affiliation{Department of Physics, Brookhaven National Laboratory, Upton, New York 11973-5000, USA}
\affiliation{RIKEN-BNL Research Center, Brookhaven National Laboratory, Upton, New York 11973-5000, USA}

\author{Marcus Bleicher}
\affiliation{Frankfurt Institute for Advanced Studies, Ruth-Moufang-Stra{\ss}e 1, 60438 Frankfurt am Main, Germany}
\affiliation{Institute for Theoretical Physics, Goethe-Universit{\"a}t, Max-von-Laue-Stra{\ss}e 1, 60438 Frankfurt am Main, Germany}
\affiliation{GSI Helmholtzzentrum f\"ur Schwerionenforschung GmbH,Planckstra{\ss}e 1, 64291 Darmstadt, Germany}
\affiliation{John von Neumann Institute for Computing, Forschungszentrum J{\"u}lich, 52425 J{\"u}lich, Germany}

\date{\today}

\begin{abstract}
At the earliest times after a heavy-ion collision, the magnetic field
created by the spectator nucleons will generate an extremely strong,
albeit rapidly decreasing in time, magnetic field.
The impact of this magnetic field may have detectable consequences, and
is believed to drive anomalous transport effects like the Chiral
Magnetic Effect (CME).
We detail an exploratory study on the effects of a dynamical magnetic
field on the hydrodynamic medium created in the collisions of two
ultrarelativistic heavy-ions, using the framework of numerical ideal
MagnetoHydroDynamics (MHD) with the ECHO-QGP code.
In this study, we consider a magnetic field captured in a
conducting medium, where the conductivity can receive contributions
from the electromagnetic conductivity $\sigma$ and the chiral magnetic
conductivity $\sigma_{\chi}$.
We first study the elliptic flow of pions, which we show is relatively
unchanged by the introduction of a magnetic field. 
However, by
increasing the magnitude of the magnetic field, we find evidence for an
enhancement of the elliptic flow in peripheral collisions. This effect is stronger at RHIC than the LHC, and it is
evident already at intermediate collision centralities. Next, we
explore the impact of the chiral magnetic conductivity on electric
charges produced at the edges of the fireball. This initial
$\sigma_\chi$ can be understood as a long-wavelength effective
description of chiral fermion production. We then demonstrate that this
chiral charge, when transported by the MHD medium, produces a charge dipole perpendicular to the reaction plane which extends a few units in rapidity. 
Assuming charge conservation at the freeze-out surface, we show that the produced
charge imbalance can have measurable effects on some experimental observables, like $v_1$ or $\langle \sin \phi \rangle$. This  demonstrates the ability of a MHD fluid to transport the signature of the initial chiral magnetic fields to late times. We also comment on the limitations of the ideal MHD approximation and detail how further development of a dissipative-resistive model can provide a more realistic description of the QGP.

\end{abstract}


\maketitle

\section*{Introduction}
\label{sec:intro}

The collision of two ultrarelativistic nuclei deposits enough energy such that the constituents of the nucleons become liberated, forming a strongly interacting plasma, the Quark Gluon Plasma (QGP); phenomenological studies suggest that this is the most perfect fluid created in nature~\cite{PhysRevLett.99.172301}. In the earliest moments after the collision, the system is subjected to what is expected to be the strongest magnetic field created in nature~\cite{Kharzeev:2007jp,Skokov:2009qp,Zhong:2014cda}, of the order of $eB\sim (m_\pi)^2\sim 10^{14}~\text{T}$. 
This magnetic field is generated by the spectator protons in the collision, 
and may be captured if the medium produced has finite electric conductivity~\cite{Gursoy:2014aka}. 
In order to quantify the effects of the magnetic field, one must study the mutual interactions of this magnetic field and the matter produced in these collisions in a consistent framework.
For the QGP, ab initio studies in terms of the fundamental theory, Quantum Chromodynamics (QCD), remain elusive.

Instead, a long-wavelength effective description of the QGP phase of the system in terms of relativistic hydrodynamics has proven very successful~\cite{Jaiswal:2016hex}. Coupling this mutual interaction of relativistic hydrodynamics and a magnetic field results in the effective description known as MagnetoHydroDynamics (MHD). 
MHD is an established and very successful tool to describe the complex dynamics of the electromagnetic fields in astrophysics: e.g. the magnetic fields around and within compact objects~\cite{Porth:2019wxk,Olmi:2016avl,Pili17,Olmi:2019xco}, protostellar accretion disks~\cite{1991ApJ...376..214B}, and jets~\cite{Rubini14} or in the solar wind~\cite{GombosiExtendedMHDmodeling2018,Priest14}. In this work we extend the use of the MHD framework also to the realm of heavy-ion collisions. In this context, the magnetic fields exhibit very short characteristic timescales, of order $\sim$ 1 fm or even less, a circumstance that might put into question the applicability of MHD. Actually, especially in our case, in which, for simplicity, we embrace the ideal MHD limit, that is the limit of infinite electrical conductivity (compared to the time scale of interest), subtle inconsistencies might appear~\cite{Florkowski:2018ubm}.  The basic assumption of infinite electrical conductivity translates into imposing a null electric field in the fluid comoving frame at any time, implying a zero relaxation time and, therefore, hidden causality problems. This issue calls for the future addition of resistive effects into the code~\cite{Palenzuela21042009}, including a proper numerical treatment~\cite{pareschi_russo} of the terms associated to the electric field with characteristic evolution time much shorter than the rest of the system. However, we recall that ideal hydrodynamics~\cite{Kolb:2000sd} has been successfully exploited to explain most of the features of the observed anisotropic collective flows~\cite{Adler:2001nb,Adams:2004bi} before being superseded by viscous hydrodynamics~\cite{Heinz:2005bw,Romatschke:2007mq} and before investigating its foundations and the extent of its applicability~\cite{Denicol:2012cn,Niemi:2014wta,Molnar:2016vvu,Floerchinger:2017cii,Strickland:2018ayk}. Similarly, we find legitimate to use the ideal MHD framework for some preliminary analysis, leaving an improvement of the model to follow up projects and keeping a close eye on the progress toward a deeper understanding of its foundations~\cite{Denicol:2018rbw,Denicol:2019iyh}.
The use of fully fledged MHD in heavy-ion collisions would open a new possibility to explore the electromagnetic properties of the QGP, like the electrical conductivity~\cite{Steinert:2013fza,Amato:2013naa,Aarts:2014nba,Greif:2014oia,Hattori:2016cnt}, $\sigma$, 
through the comparison with experimental data. 
However, despite the limits of our model, in this paper we will elucidate a number of novel implications of interacting magnetic fields and a perfect fluid on the observables of heavy-ion collisions. 

Another exciting prospect is the ability to develop a greater understanding of anomalous transport phenomena in heavy-ion collisions, like the Chiral Magnetic Effect (CME)~\cite{Kharzeev:2007jp,Fukushima:2008xe}, the interest of the present study, and the Chiral Magnetic Wave~\cite{Kharzeev:2010gd}. 
In heavy-ion collisions, topological configurations of the gluon fields generate chiral fermions through the chiral anomaly of QCD~\cite{tHooft:1976rip}.
In the presence of a strong external magnetic field, local chirality imbalance will then generate an electric current; this is the CME. 
While there are a number of potential sources of these topological configurations throughout the pre-equilibrium~\cite{Kharzeev:2001ev,Lappi:2006fp,Mace:2016svc,Lappi:2017skr} and QGP stage of the evolution~\cite{Moore:2010jd},
due to the very short lifetime of the magnetic field, it is likely that the best window for the generation of a CME signal is during the pre-equilibrium stage. 
To this end, recent non-perturbative studies suggest that topological sphaleron configurations may be produced abundantly enough at the earliest time after a heavy-ion to create a viable CME signal~\cite{Mace:2016svc,Mace:2016shq}.
In line with the long-wavelength description of the QGP in terms of a fluid, we can consider that the chiral charge produced during the pre-equilibrium stage is effectively encoded in $\sigma_\chi$. Its effects are then incorporated into the MHD description, whereby we can study the charge separation in the fluid and the effects on the produced particles. In particular, we will demonstrate that a clear electric charge dipole, which extends a few units in rapidity, is observed. 

Previous simplified studies have investigated the effects of the electromagnetic fields on the hydrodynamic description of heavy ion collisions~\cite{Deng:2012pc,Pang:2016yuh,Roy:2017yvg}, albeit neglecting the back-reaction of the fluid on the fields. Only recently has the back-reaction from the conducting current been consistently taken into account within the MagnetoHydroDynamics (MHD) framework~\cite{Inghirami:2016iru,Das:2017qfi}, albeit in the ideal limit of infinite electrical conductivity.
We here assume the same framework, based  of the numerically-validated ECHO-QGP code~\cite{Inghirami:2016iru}. 
While the description of the initial magnetic field is still an active topic of investigation (c.f.~\cite{Holliday:2016lbx,Peroutka:2017esw}), we consider initial magnetic fields which are the result of an electrically and/or chiral-magnetically conducting medium, with electric conductivity $\sigma$ and CME conductivity $\sigma_\chi$. 
As the magnetic field is expected to decrease rapidly as a function of time, it is interesting to understand what effect a strong magnetic field may have on the produced particles and observables like elliptic flow. 
While only the order of magnitude of the initial field is known, signatures of this strong field may aid in determining the relevant field strength. 
To this end, in this paper, we study ideal (3+1)D MHD simulations using geometrical Glauber initial conditions~\cite{Miller:2007ri} and we explore how varying basic parameters like the impact parameter, the conductivity of the medium in the pre-equilibrium phase, the freeze-out temperature and the magnitude of the initial magnetic field affects the elliptic flow of pions at mid-rapidity. The primary interest of this exploratory work is to understand the implications that dynamical magnetic fields may have on QGP observables; we do not wish to obfuscate this by including now-standard phenomenological features, such as viscosity or hadronic cascades. We therefore refrain from making theory-to-data comparison.

The outline of this paper is as follows: first, in Section~\ref{sec:ic}, we recall the framework that we are using and we briefly present how we compute the initial conditions for the magnetic field. In Section~\ref{sec:ell_flow}, we present and discuss the results of our simulations. In particular, we perform exploratory studies of the magnetic field influence on the elliptic flow of pions. Then in Section~\ref{sec:charge_cme}, we consider the electric charge imbalance created by an initially helical magnetic field, and discuss the implications for searches for the CME.  In Section~\ref{sec:issues}, we discuss known issues with our approach, and how these may be addressed. Finally, in Section~\ref{sec:concl}, we discuss the limits of our current approach and discuss future directions which will enable direct comparison with experimental data through appropriate observables.
%


\section{Setup of the numerical simulations}
\label{sec:ic}
We perform (3+1) dimensional numerical simulations of magnetic fields coupled to ideal relativistic hydrodynamics in Milne/Bjorken coordinates. The formalism implemented into the code as described in Ref.~\cite{Inghirami:2016iru}, is summarized in App.~\ref{app:ideal_mhd}. The initial energy density distribution is computed according to a geometrical Glauber model~\cite{DelZanna:2013eua,Becattini:2015ska}, whose details can be found in App.~\ref{app:initial_edens}.
\subsection{Initial magnetic field}
We compute the initial electromagnetic field produced by a point charge moving at constant velocity in a medium with electric conductivity  $\sigma$ and chiral magnetic conductivity $\sigma_{\chi}$ as in Ref.~\cite{Li:2016tel}:
\begin{eqnarray}
	B_{\phi}\left(t,\mathbf{x}\right) & = & \frac{Q}{4\pi}\cdot\frac{v\gamma x_{T}}{\Delta^{3/2}}\left(1+\frac{\sigma v\gamma}{2}\sqrt{\Delta}\right)e^{A}, \label{Eq:inBp} \\
	B_{r}\left(t,\mathbf{x}\right) & = & -\sigma_{\chi}\frac{Q}{8\pi} \frac{v\gamma^{2}x_{T}}{\Delta^{3/2}} \left[\gamma(vt-z)+A\sqrt{\Delta}\right]e^{A}, \label{Eq:inBr} \\
	B_{z}\left(t,\mathbf{x}\right) & = & \sigma_{\chi}\frac{Q}{8\pi} \frac{v\gamma}{\Delta^{3/2}} \Big[\gamma^{2}(vt-z)^{2}\left(1+\frac{\sigma v\gamma}{2}\sqrt{\Delta}\right) \nonumber \\
	&+& \Delta\left(1-\frac{\sigma v\gamma}{2}\sqrt{\Delta}\right)\Big] e^{A} \label{Eq:inBz}
\end{eqnarray}
where $\Delta\equiv\gamma^{2}(vt-z)^{2}+x_{T}^{2}$, $A\equiv(\sigma v\gamma/2)[\gamma(vt-z)-\sqrt{\Delta}]$. We would like to stress that assuming that the conductivity is just a scalar quantity, constant both in space and time, clearly represents an oversimplification~\cite{Aarts:2014nba,Greif:2014oia,Hattori:2016cnt} of the properties of a system out of equilibrium and undergoing an extremely fast dynamical evolution. Moreover, this procedure relies on a semi-classic scheme that lacks some quantum features~\cite{Peroutka:2017bzu}. However, we recall that the scope of the present works is not to provide a complete description of the QGP formation and evolution, but just to perform an exploratory evaluation of how the magnetic fields might modify the collective flows of charged pions and induce an electric charge separation with respect to the reaction plane. Therefore, despite the strong approximations at the basis of their derivation, we consider Eqs.~(\ref{Eq:inBp}-\ref{Eq:inBz}) sufficiently good for our purposes. As we mention also in Section \ref{eflow}, we consider the possibility that by using this initialization procedure we might underestimate the magnitudes of the magnetic fields, so, to partially compensate this issue, we perform several additional simulations with the values of the magnetic field components amplified up to four times.\\
Then, we approximate the two nuclei as uniformly charged spheres that are Lorentz-contracted to disks and, by numerical integration over each of them, as explained in detail in Ref.~\cite{Tuchin:2013apa}, we compute the spatial distribution of the initial magnetic field. We assume that the collision does not affect the motion and the distribution of the electric charges contained in the nuclei. 
If we also assume that the fluid has initial null velocity in Milne/Bjorken coordinates, the transformation of the magnetic field components $B^i$ from Cartesian to Milne/Bjorken coordinates components $\tilde{B}^i$ is given by:
\be
\tilde{B}^x=B^x/\cosh(\eta),\,  \tilde{B}^y=B^y/\cosh(\eta), \,  \tilde{B}^{\eta}=B^z/\tau.
\label{Eq:B_from_Mink_to_Bjorken}
\ee
The derivation of Eqs.~(\ref{Eq:B_from_Mink_to_Bjorken}) is provided in App.~\ref{app:transf_law}. \\
Of course, in the context of a future more refined model, all these approximations introduced in the pre-equilibrium dynamics should be reconsidered~\cite{Greif:2017irh}, possibly allowing for a non null initial velocity field influenced by the electromagnetic fields. Nevertheless, the development of an adequate framework for the initial conditions, combining both QCD~\cite{Niemi:2015qia} and QED, possibly with random topological charge densities~\cite{Tuchin:2020pbg} and assuming non constant and non uniform conductivity, requires a considerable effort, which goes far beyond the goals of this study and that must be tackled within a separate dedicated project.\\

\begin{figure*}[h!]
	\begin{minipage}[b]{0.48\textwidth}
		\includegraphics[width=1\textwidth]{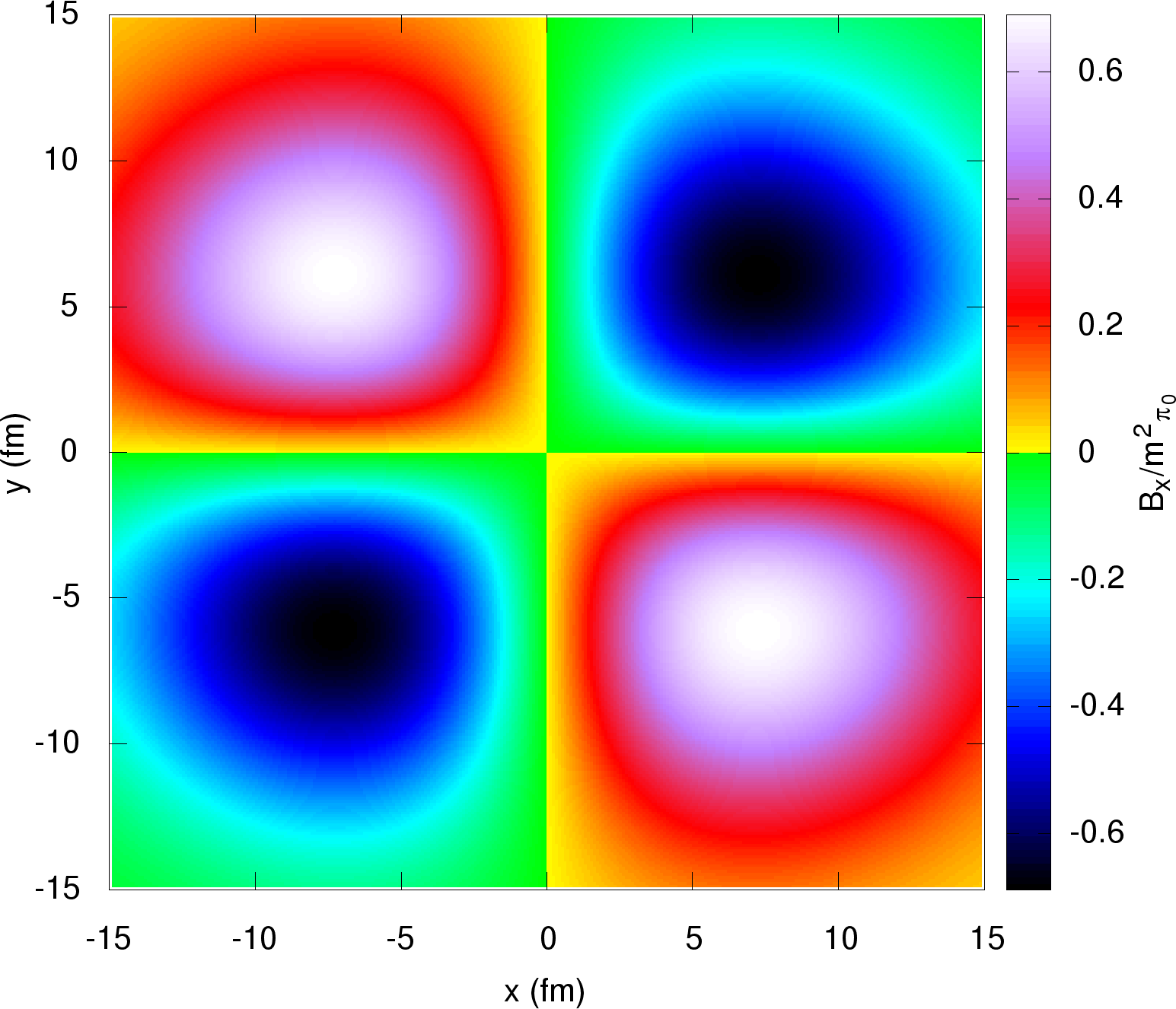}
	\end{minipage}
	\hspace{3mm}
	\begin{minipage}[b]{0.48\textwidth}
		\includegraphics[width=1\textwidth]{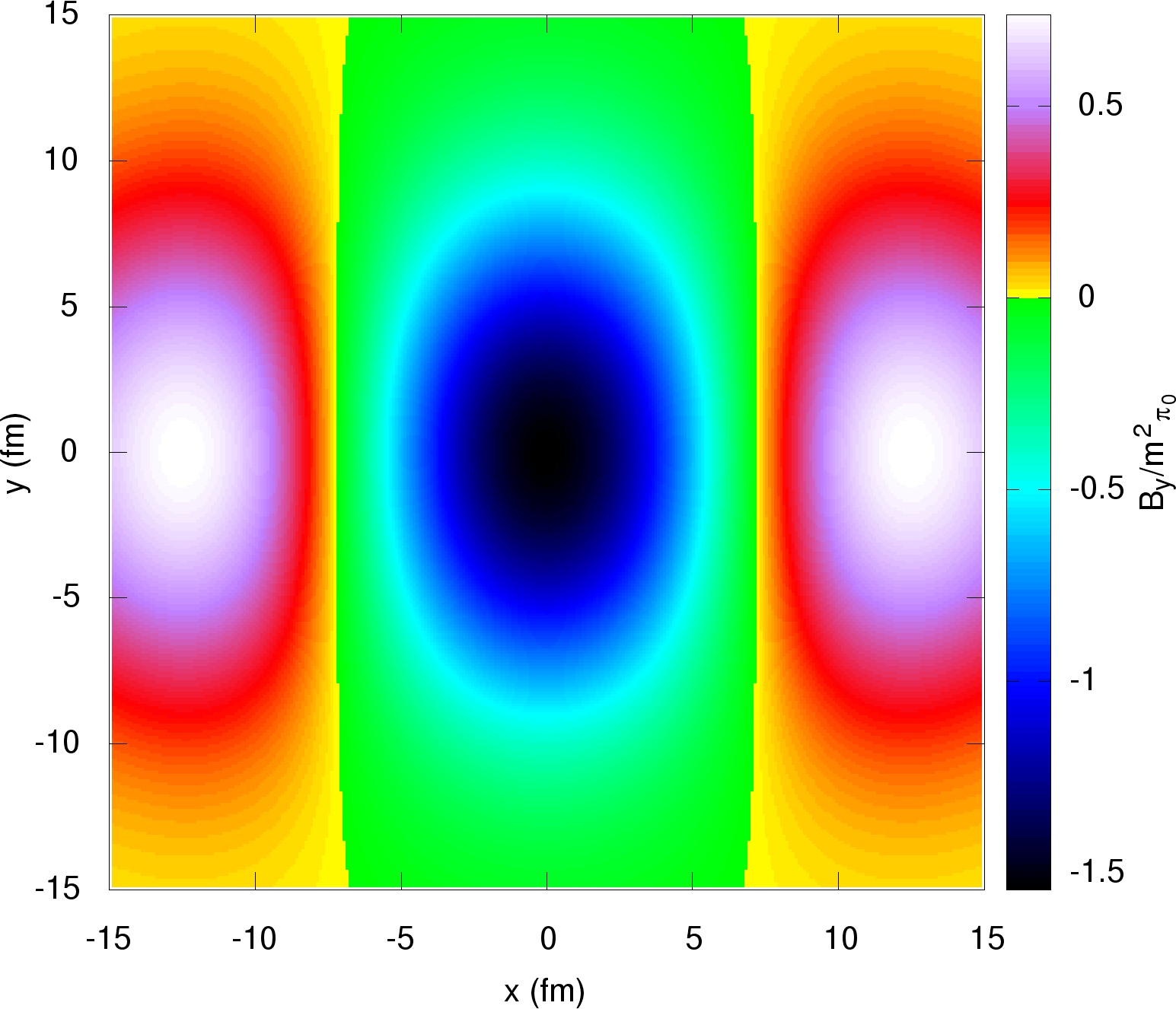}
	\end{minipage}
	\begin{minipage}[b]{0.48\textwidth}
		\includegraphics[width=1\textwidth]{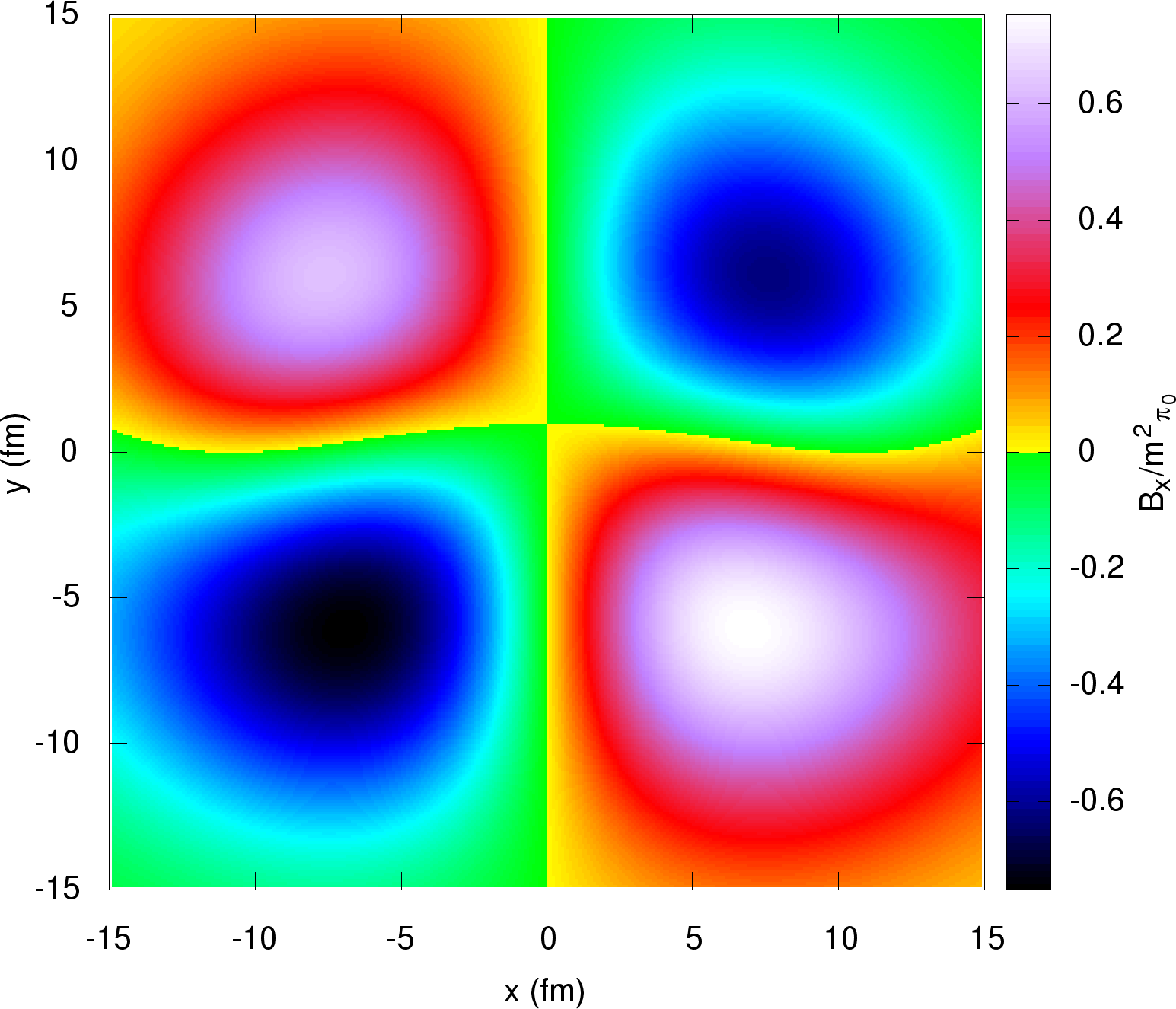}
	\end{minipage}
	\hspace{3mm}
	\begin{minipage}[b]{0.48\textwidth}
		\includegraphics[width=1\textwidth]{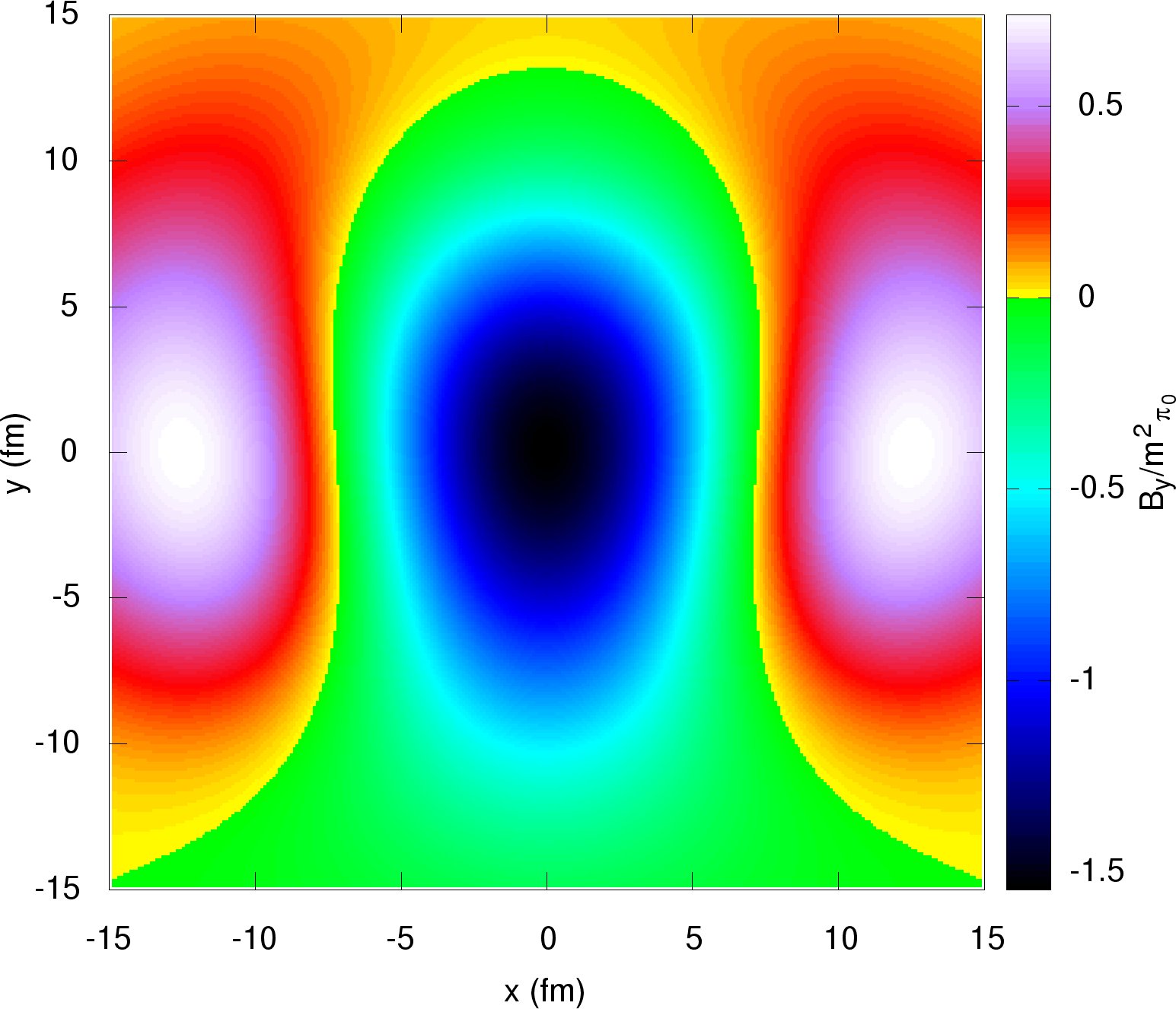}
	\end{minipage}
	\caption{Initial magnetic field components (left figure: $x$, right figure: $y$) at $\tau_0=0.4\,fm/c$ and $\eta=0$. Inclusion of a contribution of chiral origin: top row no, bottom row yes. The figures refer to Au+Au collisions at $\snn=200\,\textrm{GeV}$ with an impact parameter $b=12\,\textrm{fm}$, assuming a medium with constant electrical conductivity $\sigma=5.8\,\textrm{MeV}$ and a chiral magnetic conductivity $\sigma_{\chi}=1.5\,\textrm{MeV}$.}
	\label{initial_B_RHIC_b12}
\end{figure*}

We note that the addition of a magnetic field of chiral origin explicitly breaks the rotational symmetry of $B_{x,y}$ in the transverse plane, as the comparison between the top (no chiral $\vec{B}$) and the bottom (with chiral $\vec{B}$) rows of Fig.~(\ref{initial_B_RHIC_b12}) shows. However in the regions where the differences between the plots are more evident, the magnitude of the magnetic field is small and it is reasonable to expect only minor effects on bulk observables. 

\subsection{Parameter set}
The aim of this study is to develop an understanding of the impact of a dynamical magnetic field to the standard hydrodynamics description of heavy-ion collisions. For the present study, we do not fine tune the parameters to exactly reproduce experimental data and instead employ standard values taken from the existing literature, in particular from Ref.~\cite{Pang:2016yuh}. Unless specified otherwise, we adopt the same values for the conductivities from Ref.~\cite{Li:2016tel}. At the beginning of the numerical simulations, we assume that the fluid contains the same magnetic field computed as explained in the previous section, but the electrical conductivity of the medium becomes infinite, thus allowing us to apply the ideal MHD formalism. Unfortunately, right now our code is not able to handle a finite conductivity and, rather than unnecessarily extend this approximation also to the initial conditions, we preferred to maintain a more realistic scenario for them, albeit at the cost of introducing an inconsistency in the time evolution of the properties of the medium. By virtue of the ideal MHD formulation, we neglect the contribution of the initial electric fields and we assume that the fluid has vanishing initial velocity. The parameters used in the simulations are summarized in Table~\ref{parameters_table}.
\begin{table}
\begin{tabular}{| l | l | l |}
    
	\hline
	Parameter&RHIC&LHC\\
	\hline
	Ion&$^{197}$Au&$^{208}$Pb\\
	$\snn$ ($\textrm{GeV}$) & $200$& $2760$\\
	$\tau_0$ ($\textrm{fm/c}$) & $0.4$ & $0.2$\\
	$T_{f.o.}$ ($\textrm{MeV}$)& $154$ & $154$\\
	$\epsilon_0$  ($\textrm{GeV}/\textrm{fm}^3$)& $55$ & $413.9$\\
	$\sigma_{in.}$ ($\textrm{mb}$) & $42$ & $64$\\
	$\alpha_{bin.coll.}$ & $0.05$ & $0.05$\\
	$\eta_{flat}$ & $5.9$ & $7.0$\\
	$\sigma_{\eta}$ & $0.4$ & $0.6$\\
	$E_{min}$ ($\textrm{MeV}/\textrm{fm}^3$) & $50$ & $100$\\
	\hline	
\end{tabular}
\caption{List of the main parameters used in the numerical simulations.}
\label{parameters_table}
\end{table}
Our standard grid resolution is $0.2\,\textrm{fm}$ along $x$ and $y$ transverse directions and $0.2$ along the longitudinal $\eta$ direction. The grid resolution is doubled ($0.1$) when we evaluate also the electric charge density.

\section{Computation of azimuthal anisotropy observables}
\label{sec:ell_flow}

So-called `flow' observables, characterized by the Fourier harmonics of the particle spectra~\cite{Borghini:2001vi}, have become key measurements for characterizing the QGP. It is expected that these quantities characterize both the expansion as well as azimuthal anisotropies developed through the fluid evolution. Therefore, it is a reasonable starting point to understand the influence of dynamical magnetic fields interacting with the fluid by studying these flow harmonics.
Our main observables are the elliptic flow $v_2(p_T)$ at mid-rapidity and the directed flow $v_1(y)$ of charged pions. We compute the thermal spectra with the Cooper-Frye prescription~\cite{Cooper:1974mv} as described in Ref.~\cite{DelZanna:2013eua}. 
We extend this formulation by preserving local charge conservation across the freezeout hypersurface in the local rest frame, resulting in a charge dependent spectrum; this will be discussed in greater detail later and in App.~\ref{app:chargedepfo}.

\subsection{Elliptic flow as a function of centrality}{\label{eflow}}
Let us start with the exploration of the centrality dependence of the effect of the magnetic fields. To this aim, we simulate Au+Au collisions at $\snn=200~\mathrm{GeV}$ for the impact parameter $b=2,5,8,10,12\,\textrm{fm}$. Given the uncertainties in the properties and in the dynamics of the medium in the pre-equilibrium phase, it is possible that the magnitude of the initial magnetic fields may be underestimated by Eqs.~(\ref{Eq:inBp}-\ref{Eq:inBz}). To investigate this uncertainty in the strength of the initial magnetic field, we perform systematic studies by increasing the initial field $B_0$ (obtained from Eqs.~(\ref{Eq:inBp}-\ref{Eq:inBz})) by factors of 2 through 4.

\begin{figure*}[ht!]
\begin{center}
	\begin{minipage}[b]{0.37\textwidth}
		\includegraphics[width=1\textwidth]{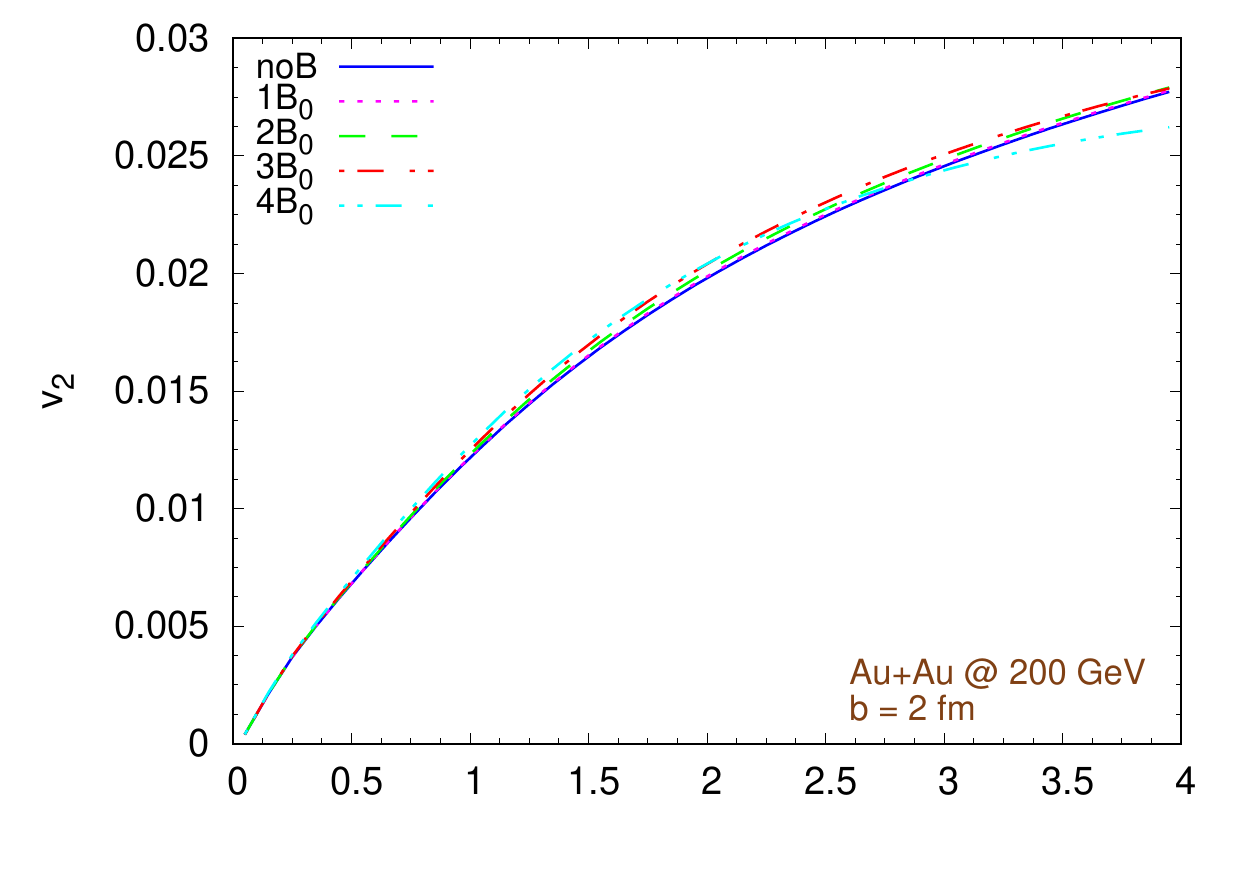}
	\end{minipage}
	\hspace{6mm}
	\begin{minipage}[b]{0.37\textwidth}
		\includegraphics[width=1\textwidth]{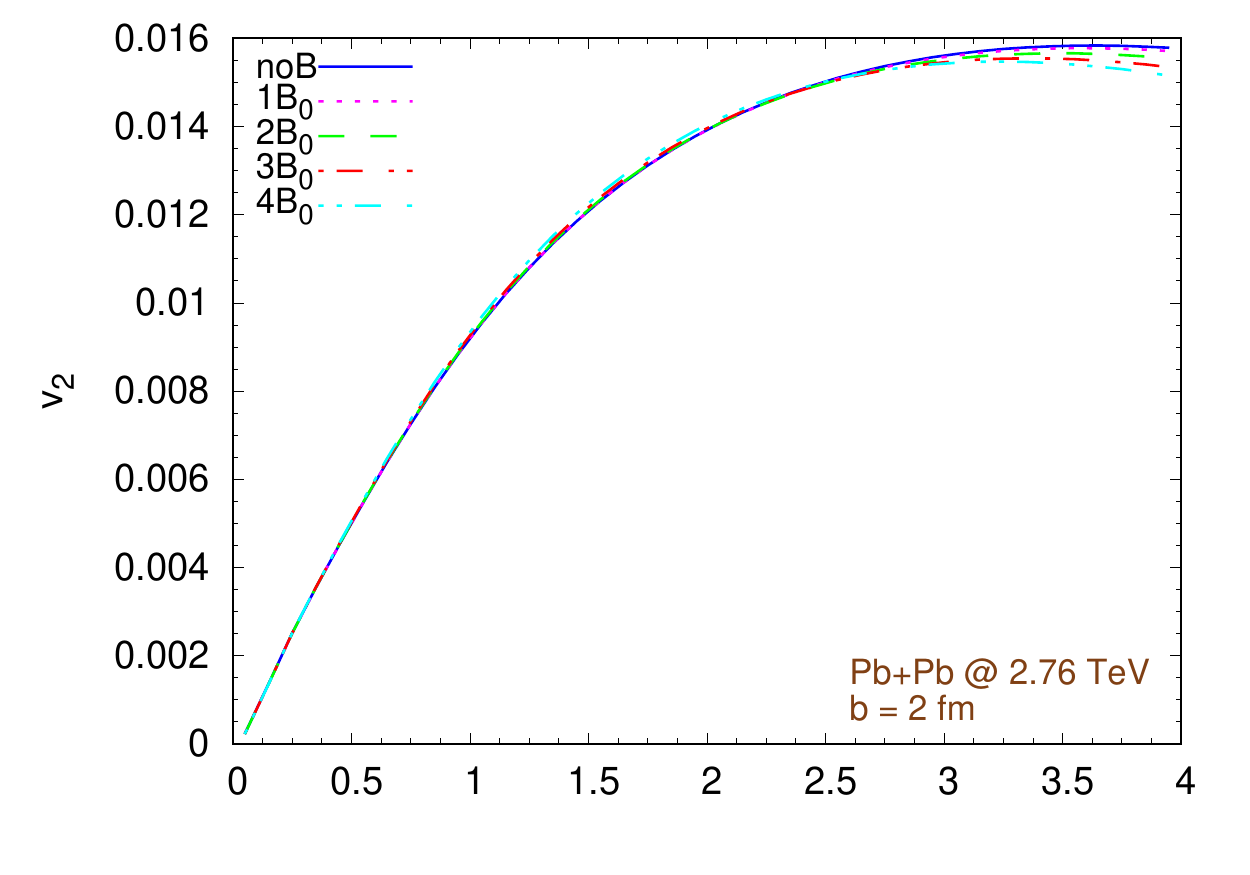}
	\end{minipage}
    \vspace*{-2mm}
	\begin{minipage}[b]{0.37\textwidth}
		\includegraphics[width=1\textwidth]{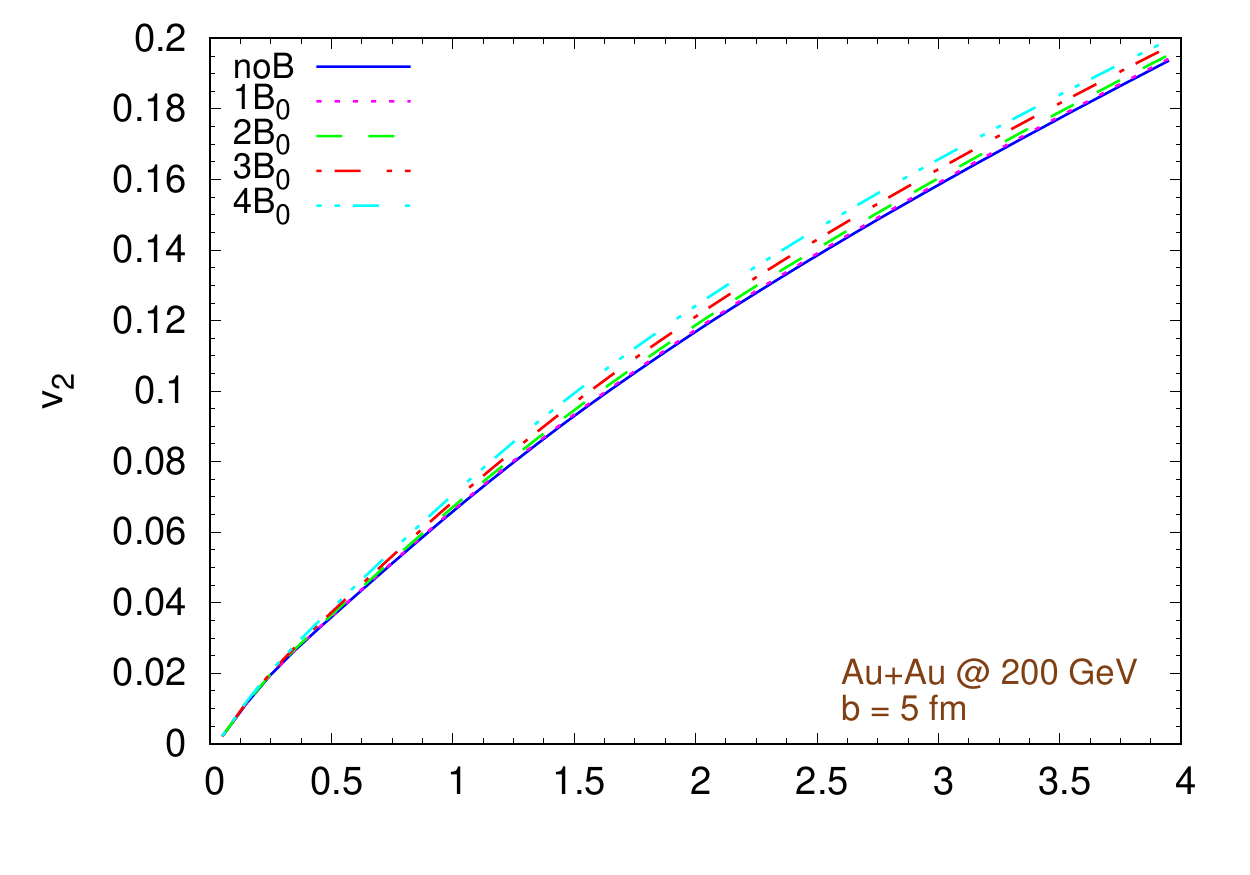}
	\end{minipage}
	\hspace{6mm}
	\begin{minipage}[b]{0.37\textwidth}
		\includegraphics[width=1\textwidth]{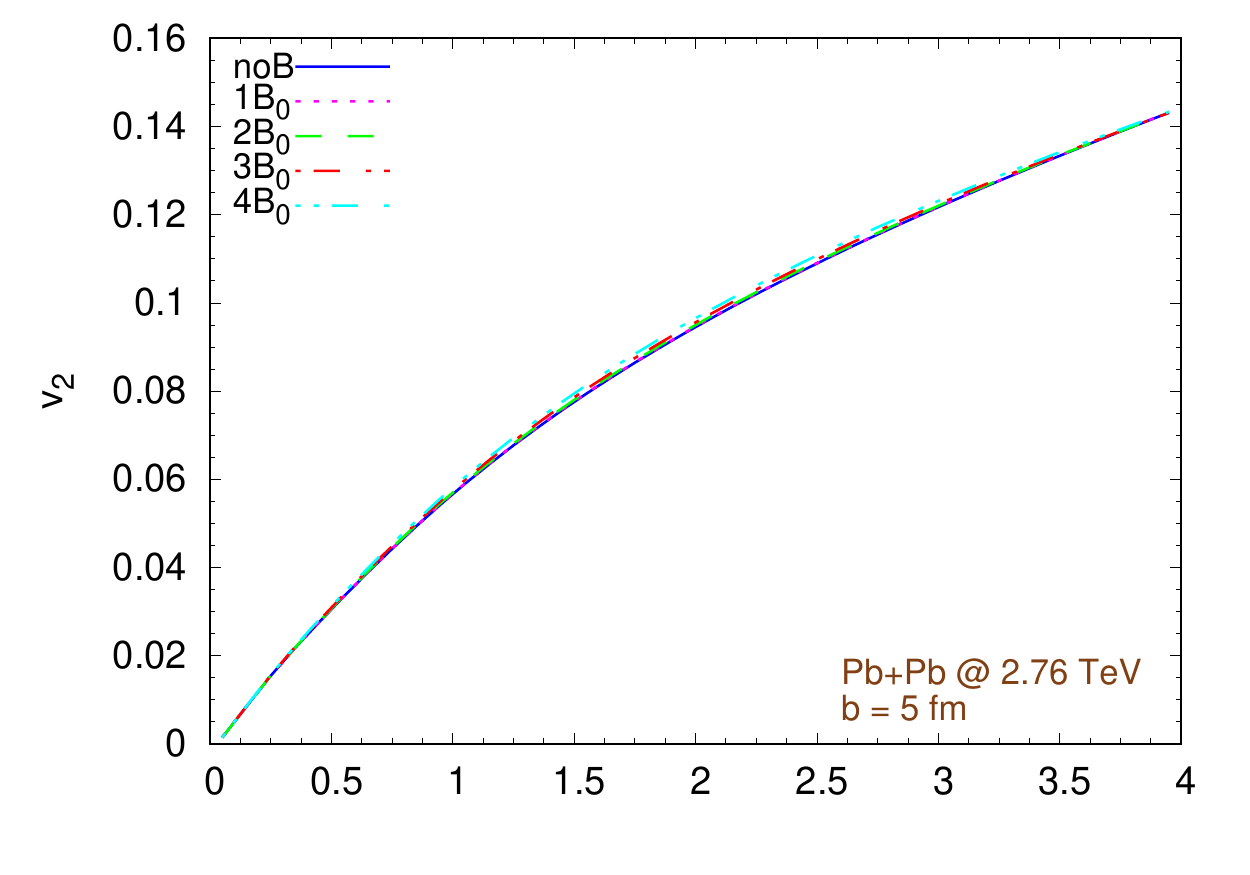}
	\end{minipage}

    \vspace*{-2mm}
    
	\begin{minipage}[b]{0.37\textwidth}
		\includegraphics[width=1\textwidth]{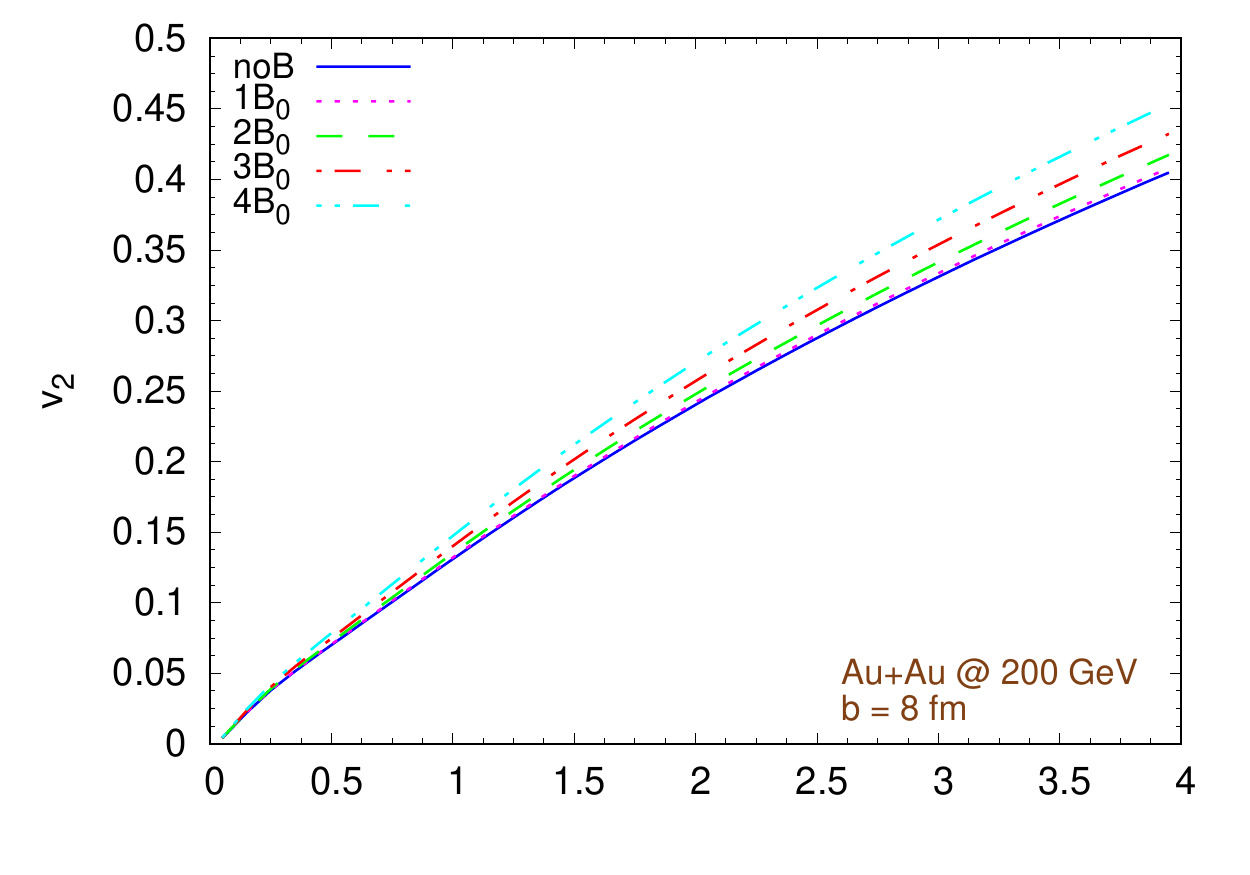}
	\end{minipage}
	\hspace{6mm}
	\begin{minipage}[b]{0.37\textwidth}
		\includegraphics[width=1\textwidth]{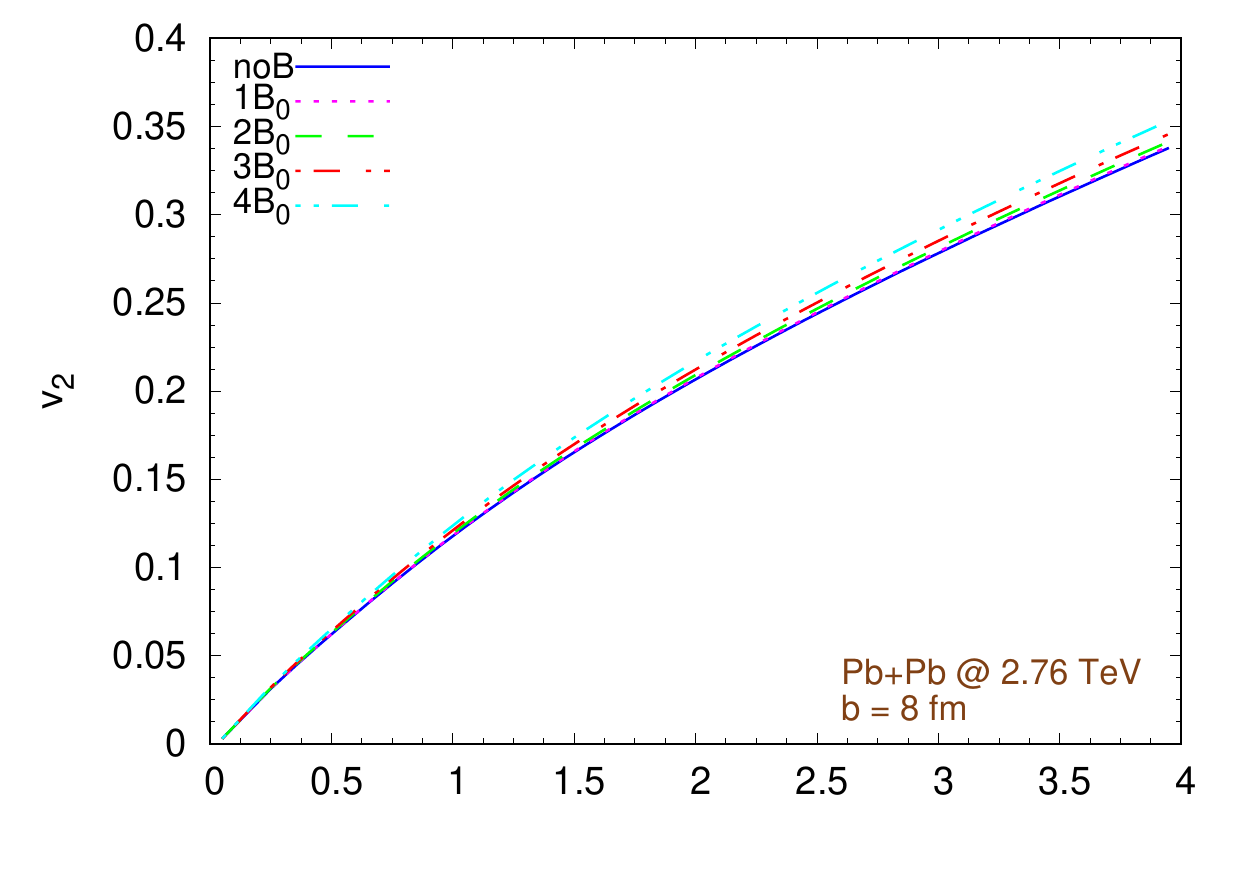}
	\end{minipage}
	
	\vspace*{-2mm}
	
	\begin{minipage}[b]{0.37\textwidth}
		\includegraphics[width=1\textwidth]{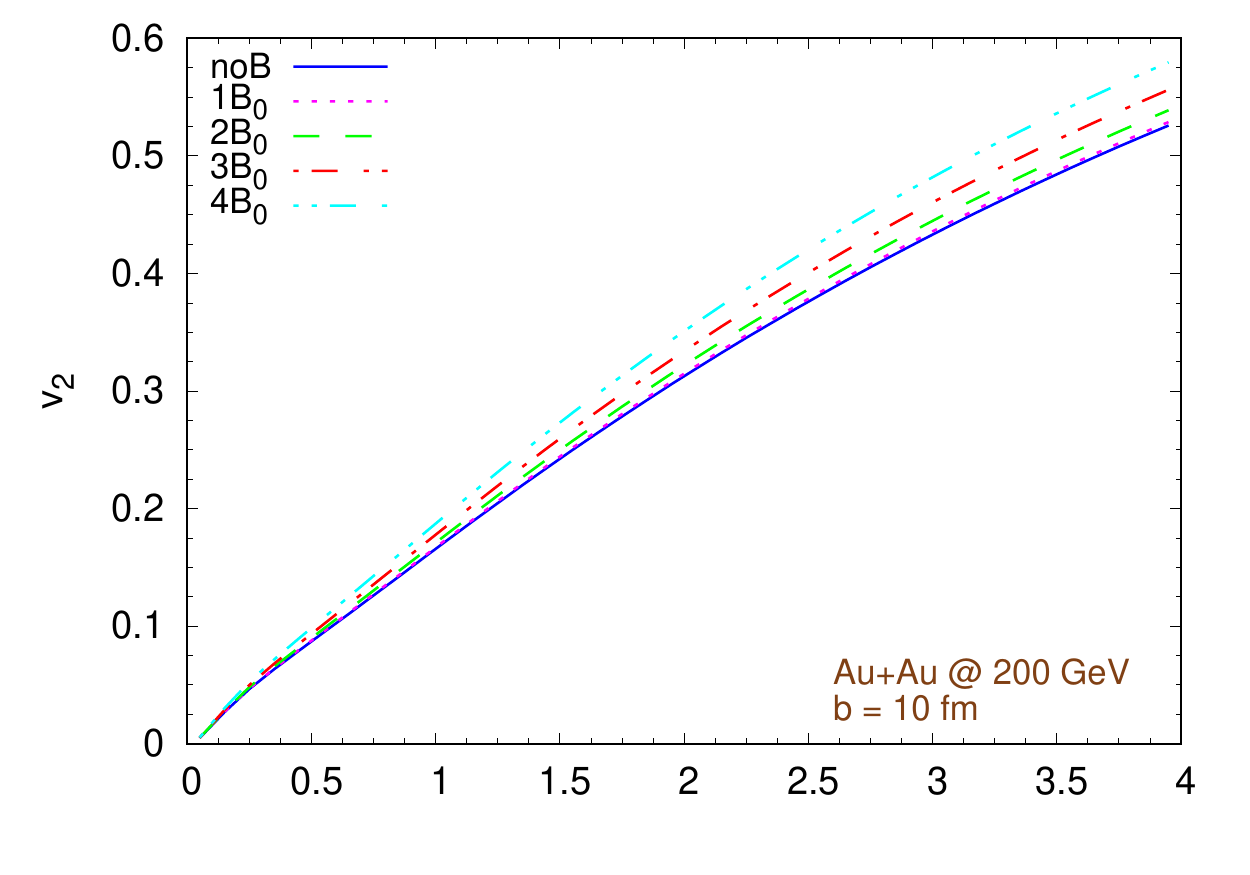}
	\end{minipage}
	\hspace{6mm}
	\begin{minipage}[b]{0.37\textwidth}
		\includegraphics[width=1\textwidth]{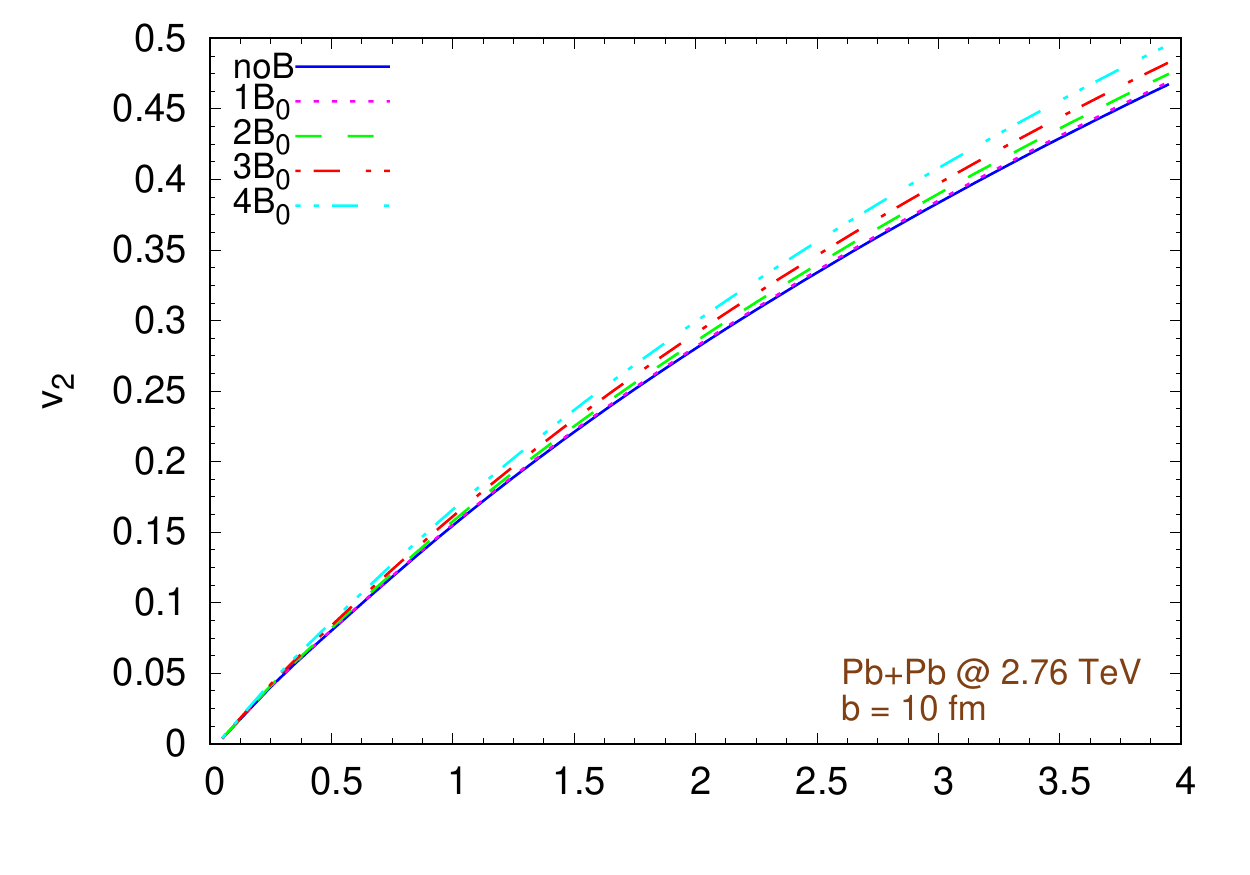}
	\end{minipage}

    \vspace*{-2mm}
	
	\begin{minipage}[b]{0.37\textwidth}
		\includegraphics[width=1\textwidth]{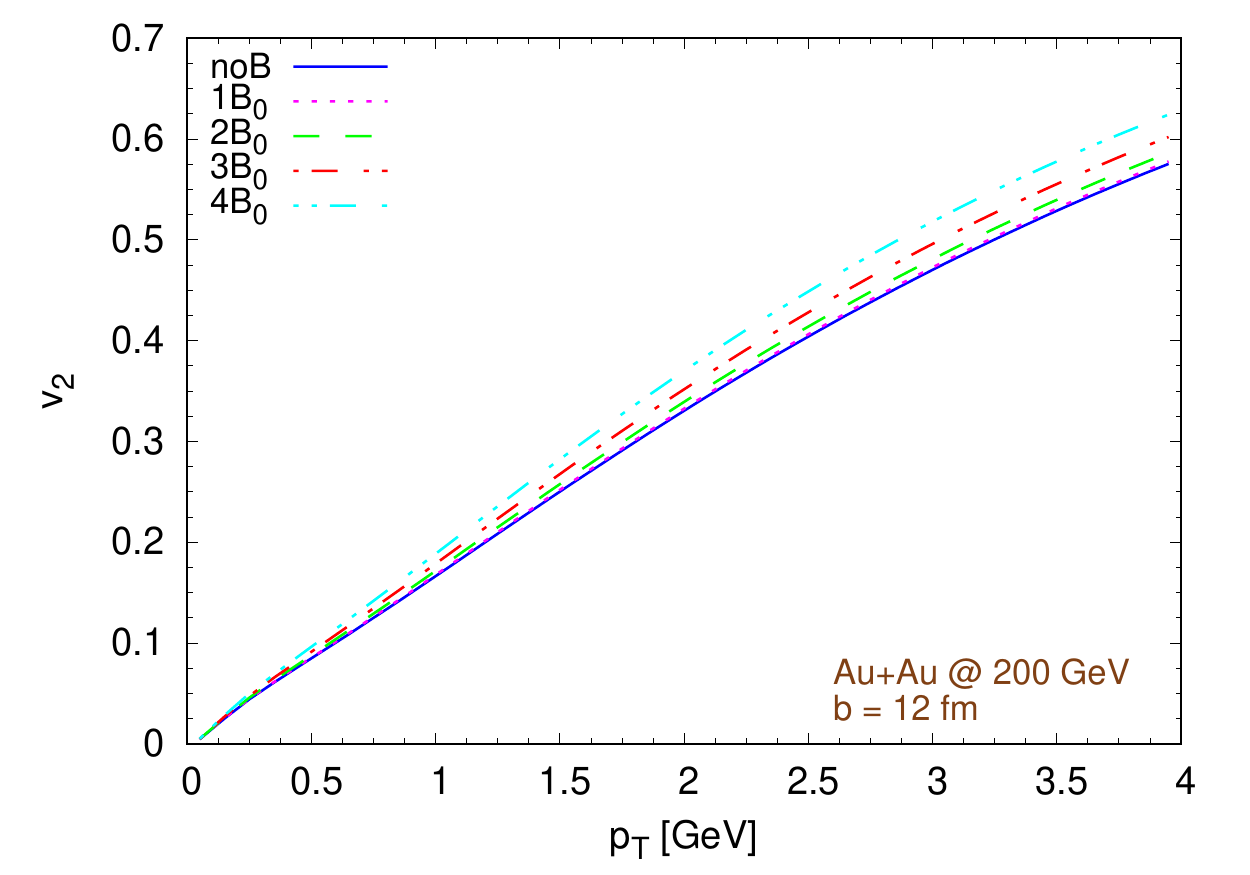}
	\end{minipage}
	\hspace{6mm}
	\begin{minipage}[b]{0.37\textwidth}
		\includegraphics[width=1\textwidth]{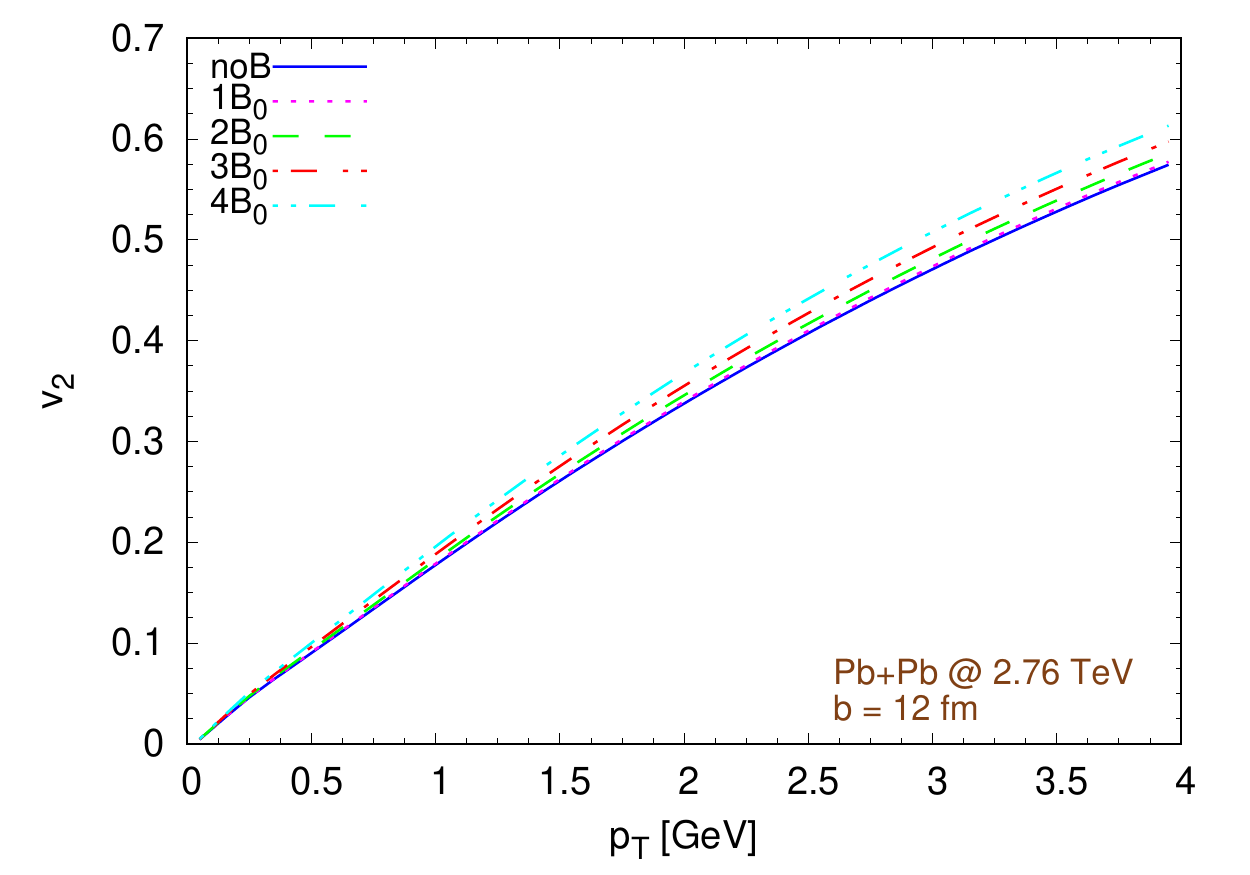}
	\end{minipage}
	\caption{Elliptic flow of pions for various magnetic field strengths $B$ as a function of the transverse momentum $p_T$. The impact parameter $b$ increases from the top to bottom row. Left: Au+Au collisions at $\snn=200\,\textrm{GeV}$. Right: Pb+Pb collisions at $\snn=2.76\,\textrm{TeV}$. }
	\label{fig:B_v_sb}
	\end{center}
\end{figure*}

Fig.~(\ref{fig:B_v_sb}) explores the elliptic flow in more detail.  
The first row of Fig.~(\ref{fig:B_v_sb}) shows that the effect of the magnetic field on $v_2$ for almost central collisions ($b=2\,\textrm{fm}$) is very small, with only a slight suppression for $p_T\gtrsim3\,\textrm{GeV}$, with $B=4B_0$. For more peripheral collisions, however, we observe an enhancement of $v_2$, but only if we increase our initial magnetic field, otherwise its effects are negligible. This is expected, as similar results have already found in simulations with transport models~\cite{Voronyuk:2011jd} and in analytic estimates~\cite{Stewart:2017zsu}.
Although the model, the numbers and many details are quite different, essentially we observe an increasing trend of $v_2$ by increasing the magnitude of the magnetic field as noticed in Ref.~\cite{Roy:2017yvg}. Most likely, this behavior is due to the spatial distribution of the magnetic field along the transverse plane (Fig.~\ref{initial_B_RHIC_b12}), which is rather similar to the spatial distribution of the pressure (Fig.~\ref{initial_pr}). Since the magnetic field produces a pressure ${B^2}/{2}$, it gives an additional contribution to the initial pressure anisotropy of the fluid, whose gradient induces a correspondent momentum anisotropy, of which the elliptic flow is just the second momentum of the Fourier decomposition with respect to the azimuthal angle.\\
Going on with the examination of Fig.~(\ref{fig:B_v_sb}), we note how in the intermediate centrality class ($b=5\,\textrm{fm})$ for RHIC energies we can already observe an enhancement of the elliptic flow which is almost absent at LHC energies. For peripheral collisions, shown in the last rows of Fig.~(\ref{fig:B_v_sb})), we observe an enhancement of $v_2$ both at RHIC and LHC energies, however the effect is stronger at $\snn=200\,\textrm{GeV}$. 

\begin{figure*}[ht!]
	\begin{minipage}[b]{0.48\textwidth}
		\includegraphics[width=1\textwidth]{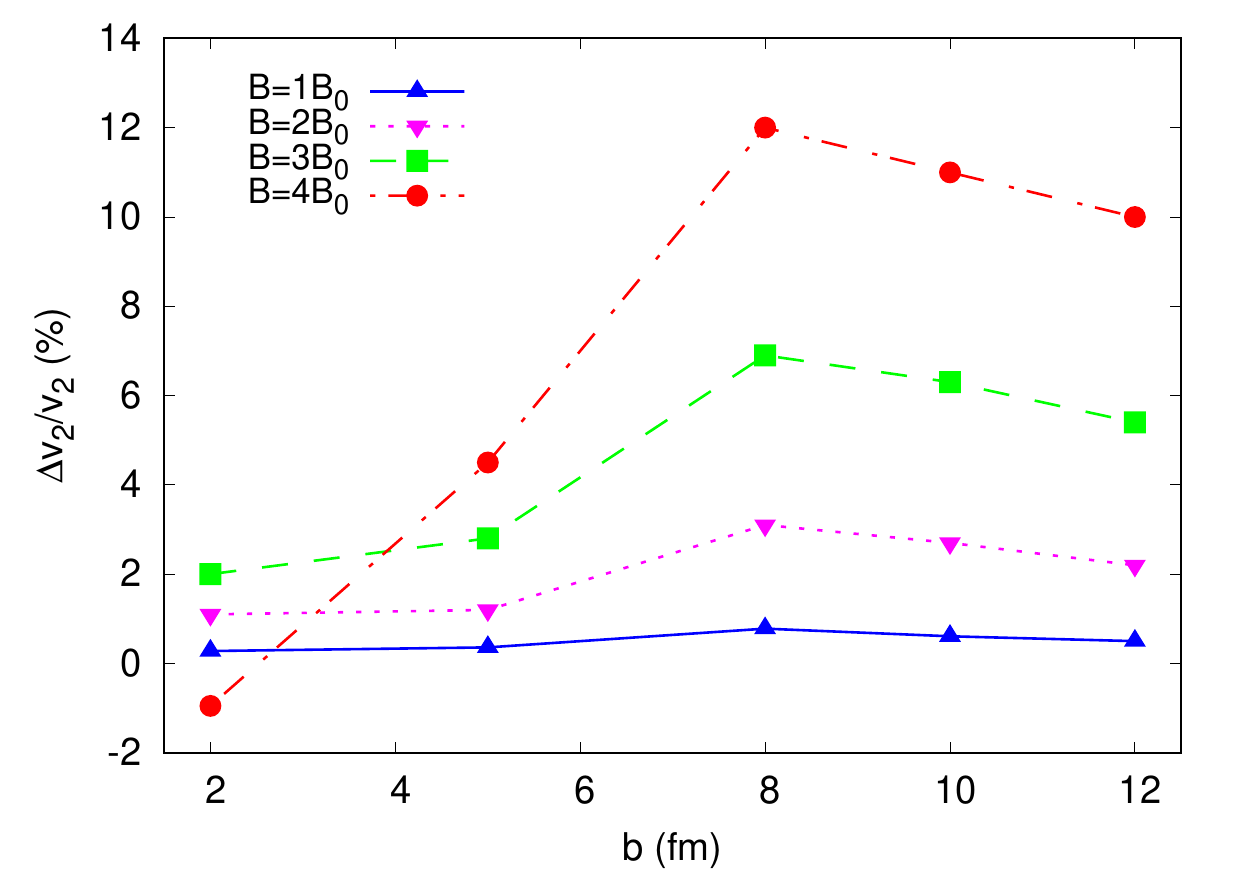}
	\end{minipage}
	\hspace{3mm}
	\begin{minipage}[b]{0.48\textwidth}
		\includegraphics[width=1\textwidth]{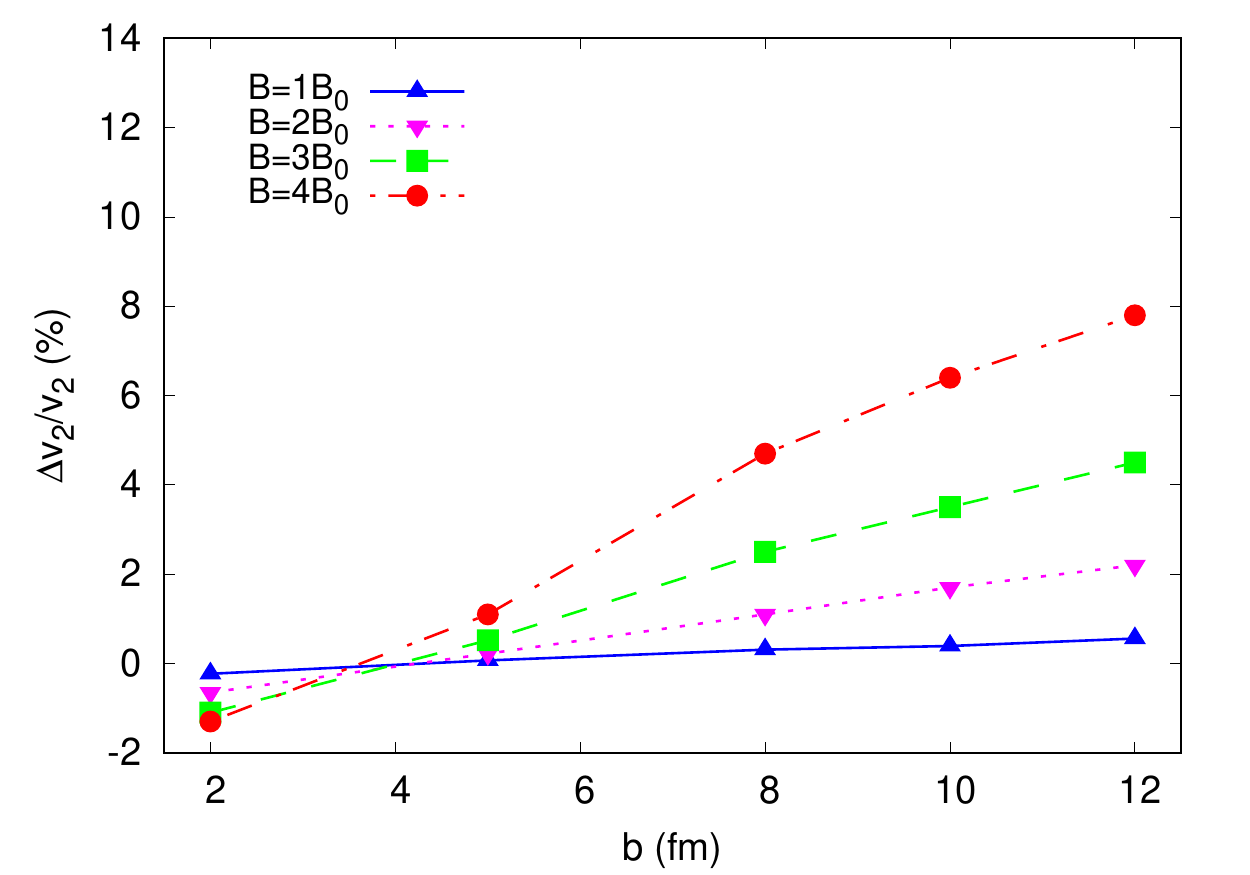}
	\end{minipage}
	\caption{Percentage of enhancement of $v_2(y=0,p_T\simeq3\textrm{GeV})$ compared with the case without magnetic field, $(v_2(B)-v_2(0))/v_2(0)\cdot100$, with respect to the impact parameter $b$. The different lines correspond to the case in which the initial magnetic field, $B_0$, has been multiplied by a factor 1,2,3 or 4. Left: Au+Au collisions at $\snn=200\,\textrm{GeV}$. Right: Pb+Pb collisions at $\snn=2.76\,\textrm{TeV}$.}
	\label{fig:deltav2_vs_b}
\end{figure*}

Fig.~(\ref{fig:deltav2_vs_b}) shows the percentage of enhancement of $v_2(y=0,p_T\simeq3\textrm{GeV})$ compared with the case without magnetic field. If we call $v_2(B)$ the elliptic flow in the presence of a magnetic field and $v_2(0)$ the elliptic flow in pure hydro simulations, the lines display the ratio $(v_2(B)-v_2(0))/v_2(0)\cdot100$ with respect to the impact parameter $b$. The different lines correspond to the case in which the initial magnetic field has been multiplied by a factor 1,2,3, or 4. In the left figure we consider Au+Au collisions at $\snn=200\,\textrm{GeV}$, while in the right figure Pb+Pb collisions at $\snn=2.76\,\textrm{TeV}$.
We observe that at RHIC energies the maximum enhancement is obtained for $b=8\,\textrm{fm}$ and it tends to decrease for more peripheral collisions, while at LHC energies the enhancement steadily increases up to $b=12\,\textrm{fm}$.
It is important to recall that these results are for a specific choice of parameters which we have not tuned to experimental data.

\begin{figure*}[!ht]
	\begin{minipage}[b]{0.48\textwidth}
		\includegraphics[width=1\textwidth]{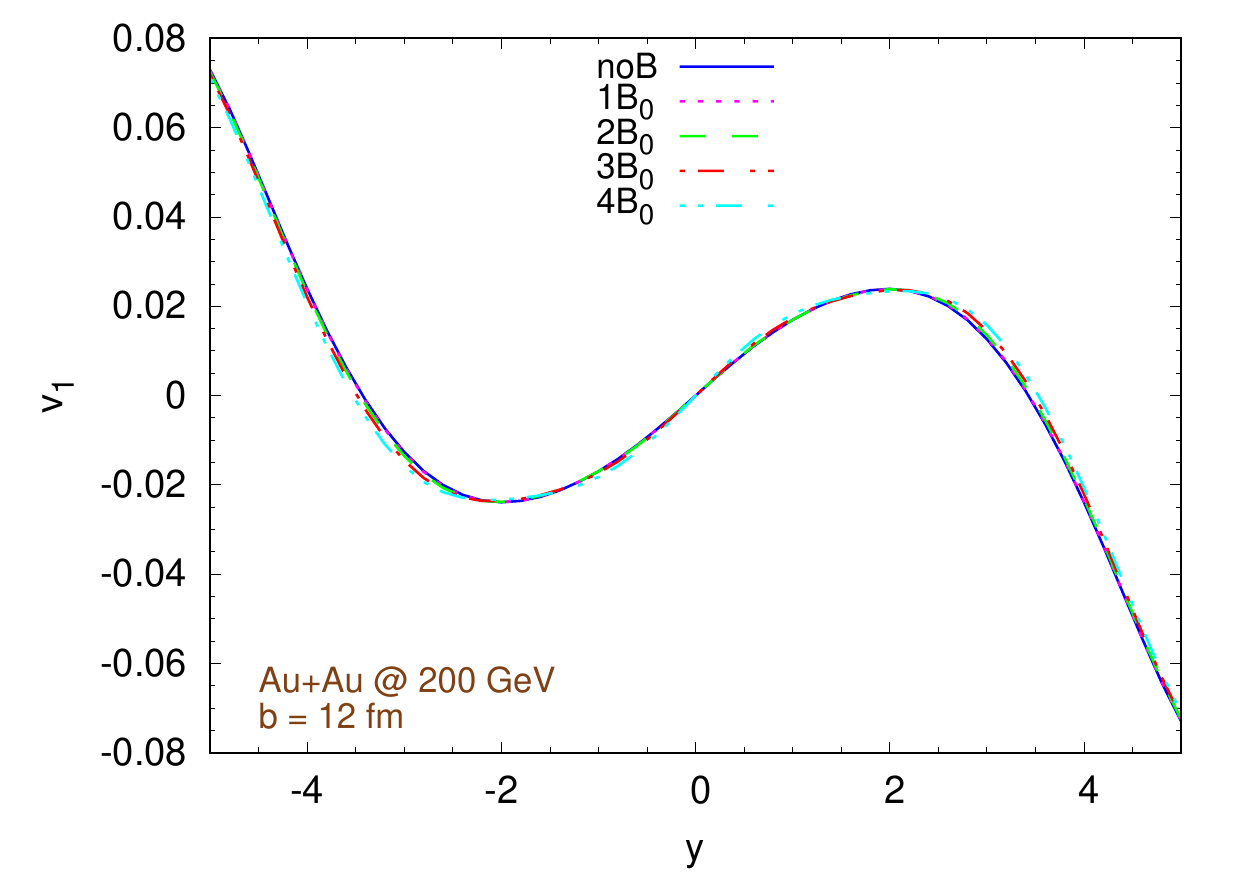}
	\end{minipage}
	\hspace{3mm}
	\begin{minipage}[b]{0.48\textwidth}
		\includegraphics[width=1\textwidth]{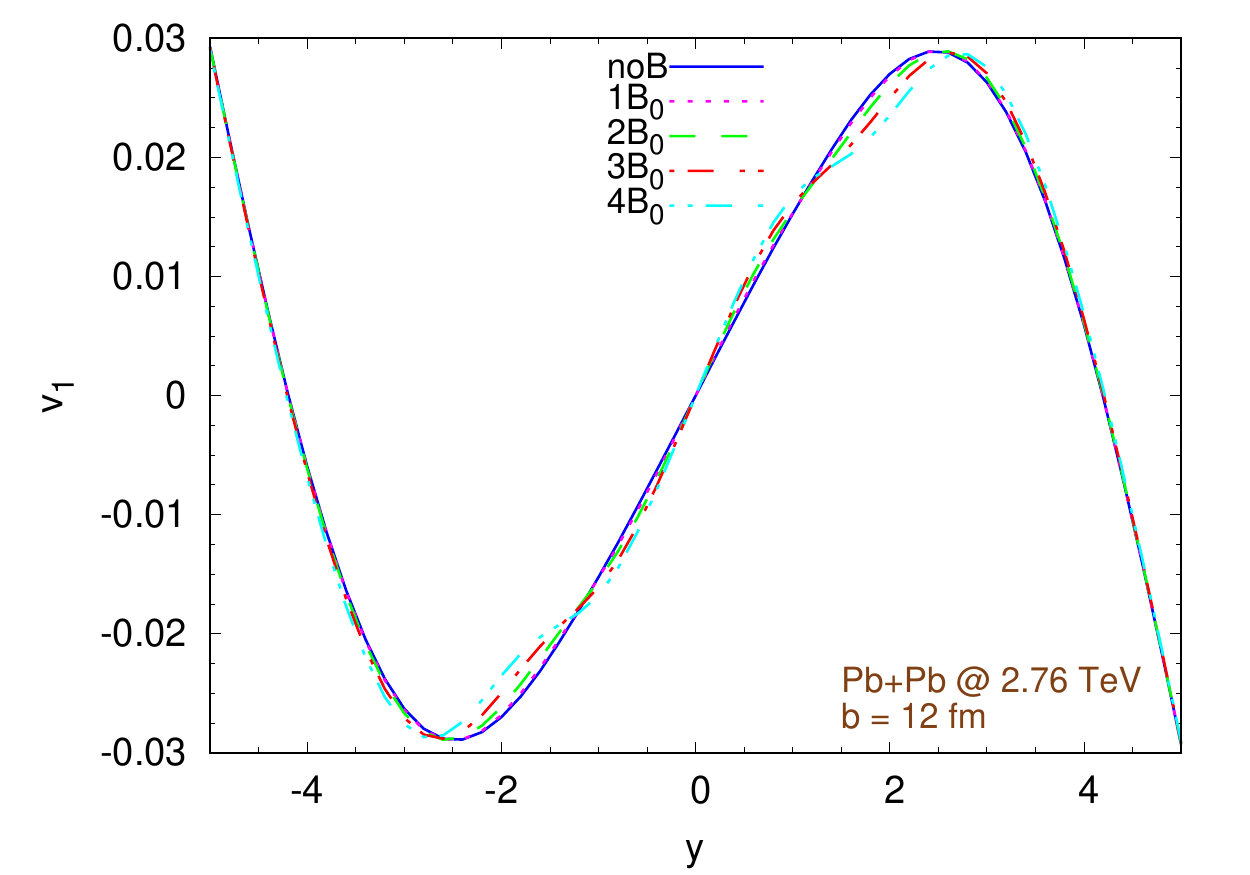}
	\end{minipage}
	\caption{Effect of the magnitude of the initial magnetic field on the directed flow $v_1$ vs rapidity $y$ for peripheral collisions with $b=12\,\fm$. Left: Au+Au collisions at $\snn=200\,\textrm{GeV}$. Right: Pb+Pb collisions at $\snn=2.76\,\textrm{TeV}$.  }
	\label{fig:v1}
\end{figure*}

\begin{figure*}[!ht]
	\begin{minipage}[b]{0.48\textwidth}
		\includegraphics[width=1\textwidth]{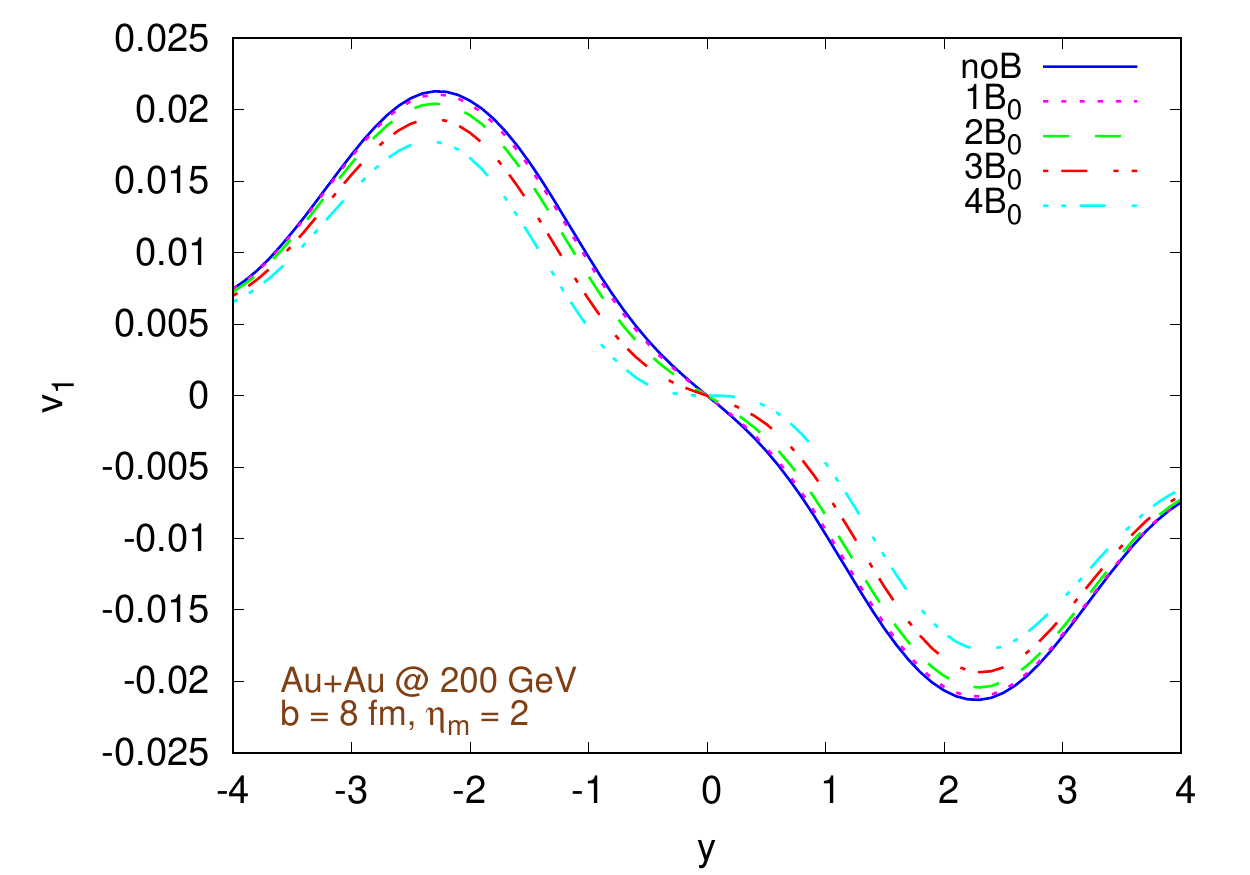}
	\end{minipage}
	\hspace{3mm}
	\begin{minipage}[b]{0.48\textwidth}
		\includegraphics[width=1\textwidth]{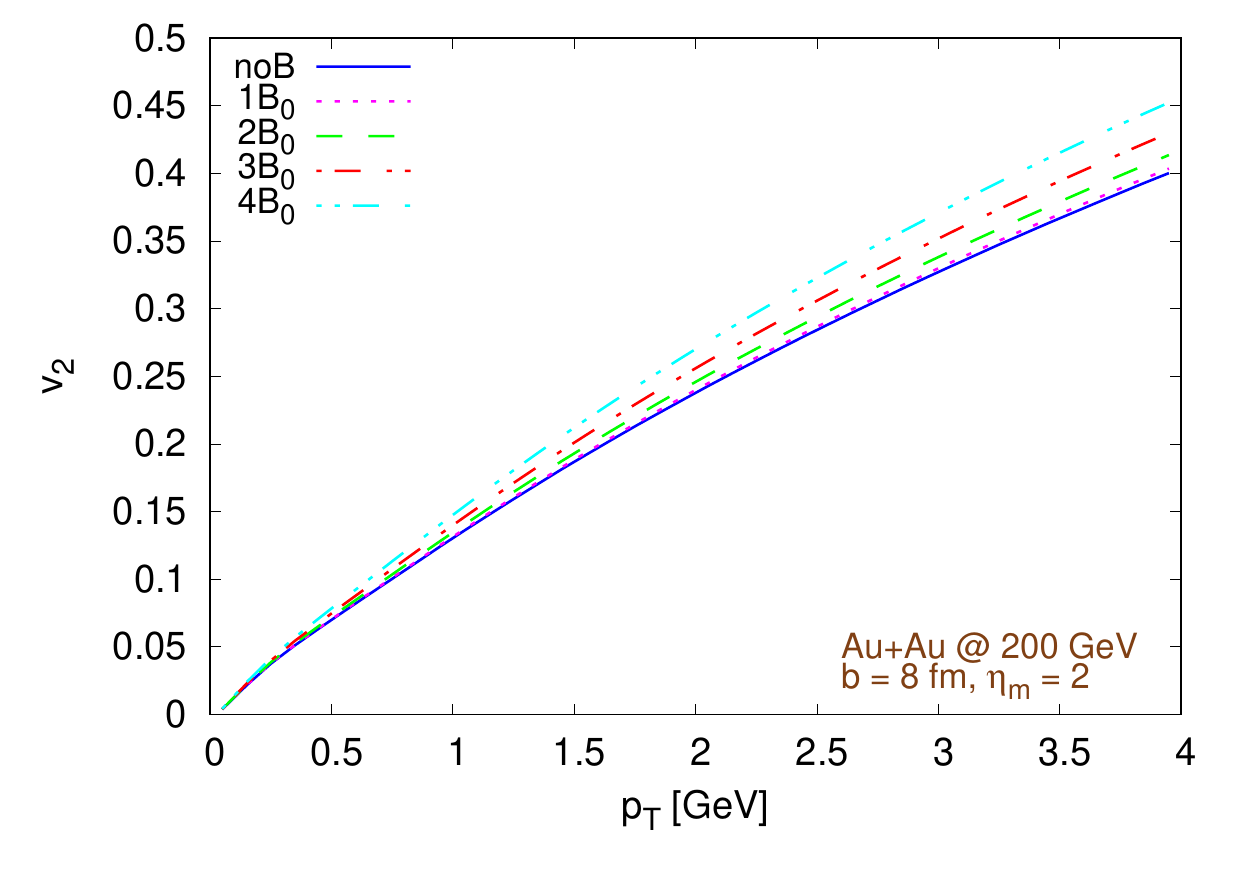}
	\end{minipage}
	\caption{Effect of the magnitude of the initial magnetic field on $v_1(y)$ (left) and $v_2(p_T,y=0)$ (right) in  Au+Au collisions at $\snn=200\,\textrm{GeV}$, with $b=8\,\fm$, when considering a tilted initial energy density distribution (see App.~\ref{tilt-desc}) with parameter $\eta_m=2$.}
	\label{fig:v1-tilted}
\end{figure*}

Fig.~(\ref{fig:v1}) depicts the directed flow vs rapidity for collisions with impact parameter $b=12\textrm{fm}$. In this case the effects of the magnetic field seem to be very modest both at RHIC and LHC energies, albeit in the second case we observe a kind of ripple at $|y|\approx2$. However, if we modify the initial conditions by introducing a tilting in the initial energy density distribution~\cite{Bozek:2010bi,Becattini:2015ska} as described in App.~\ref{tilt-desc}, so to obtain a directed flow with a negative slope at mid-rapidity $\left( \frac{dN}{dy}|_{y=0}<0 \right)$, more similar to the experimental results~\cite{Tang:2010mz,Adamczyk:2011aa}, then the situation changes. As shown in Fig.~(\ref{fig:v1-tilted}), for Au+Au collisions at $\snn=200\,\textrm{GeV}$ with $b=8\,\textrm{fm}$ and tilted initial energy density distribution ($\eta_m=2$), the directed flow exhibits a clear dependence on the magnitude of the magnetic field.

\subsection{Impact of the magnetic field of chiral origin}
\begin{figure*}[ht]
	\begin{minipage}[b]{0.48\textwidth}
		\includegraphics[width=1\textwidth]{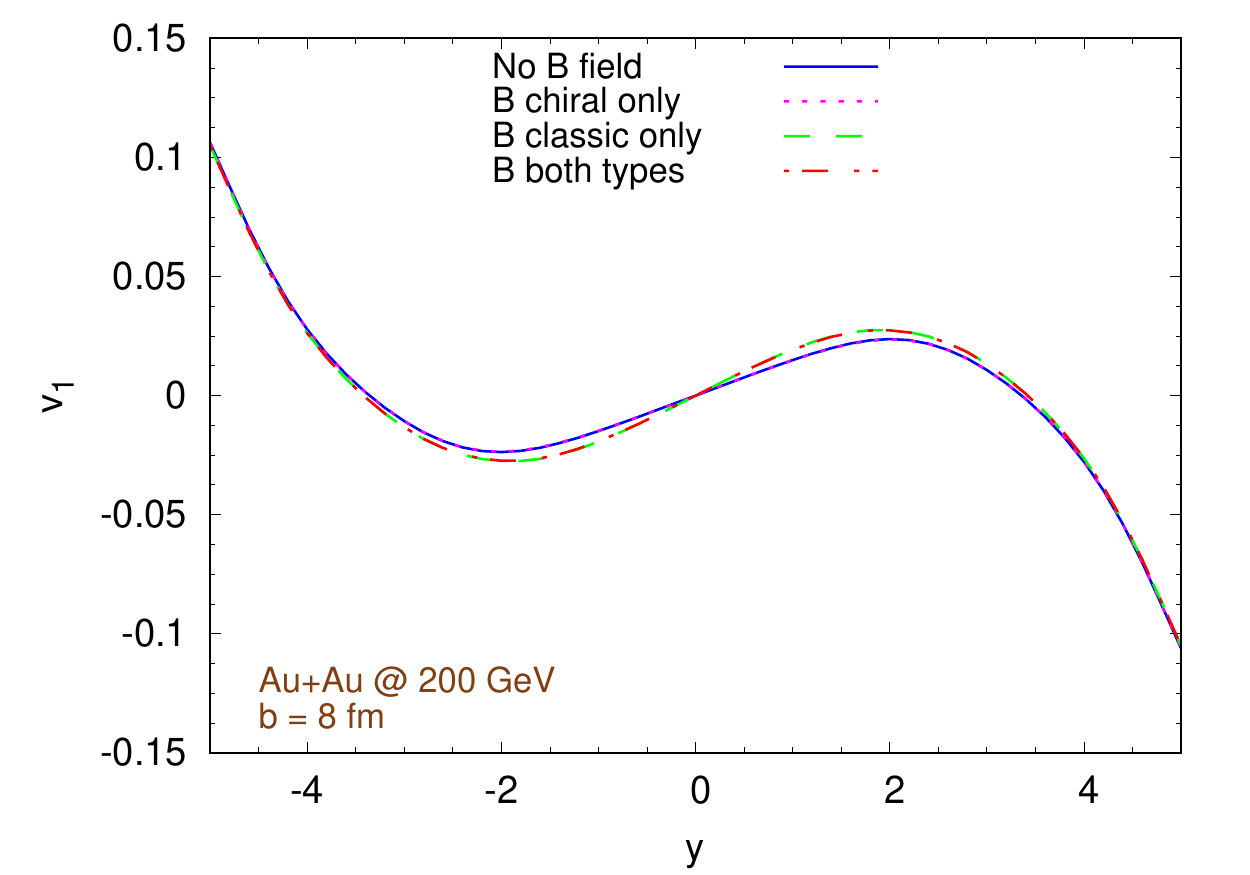}
	\end{minipage}
	\hspace{3mm}
	\begin{minipage}[b]{0.48\textwidth}
		\includegraphics[width=1\textwidth]{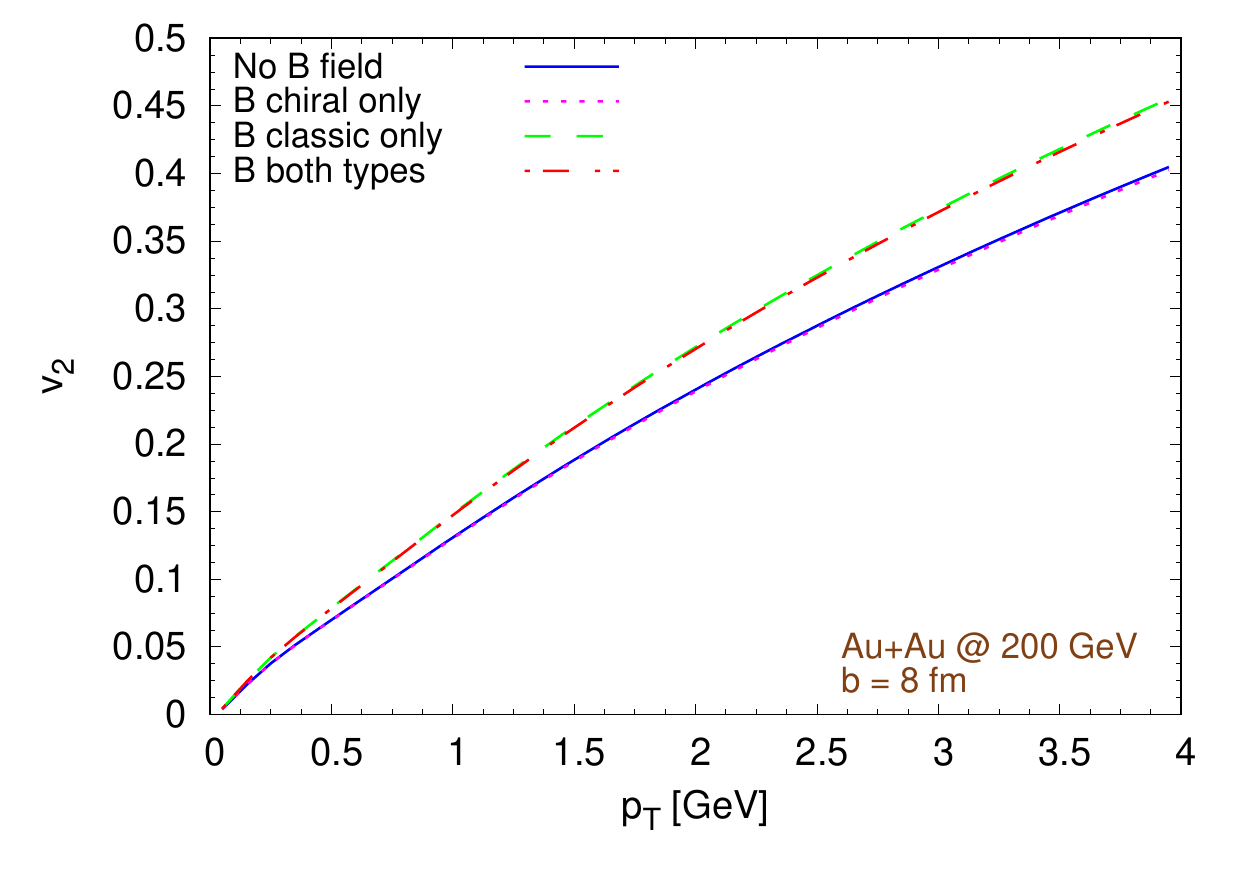}
	\end{minipage}
	\begin{minipage}[b]{0.48\textwidth}
		\includegraphics[width=1\textwidth]{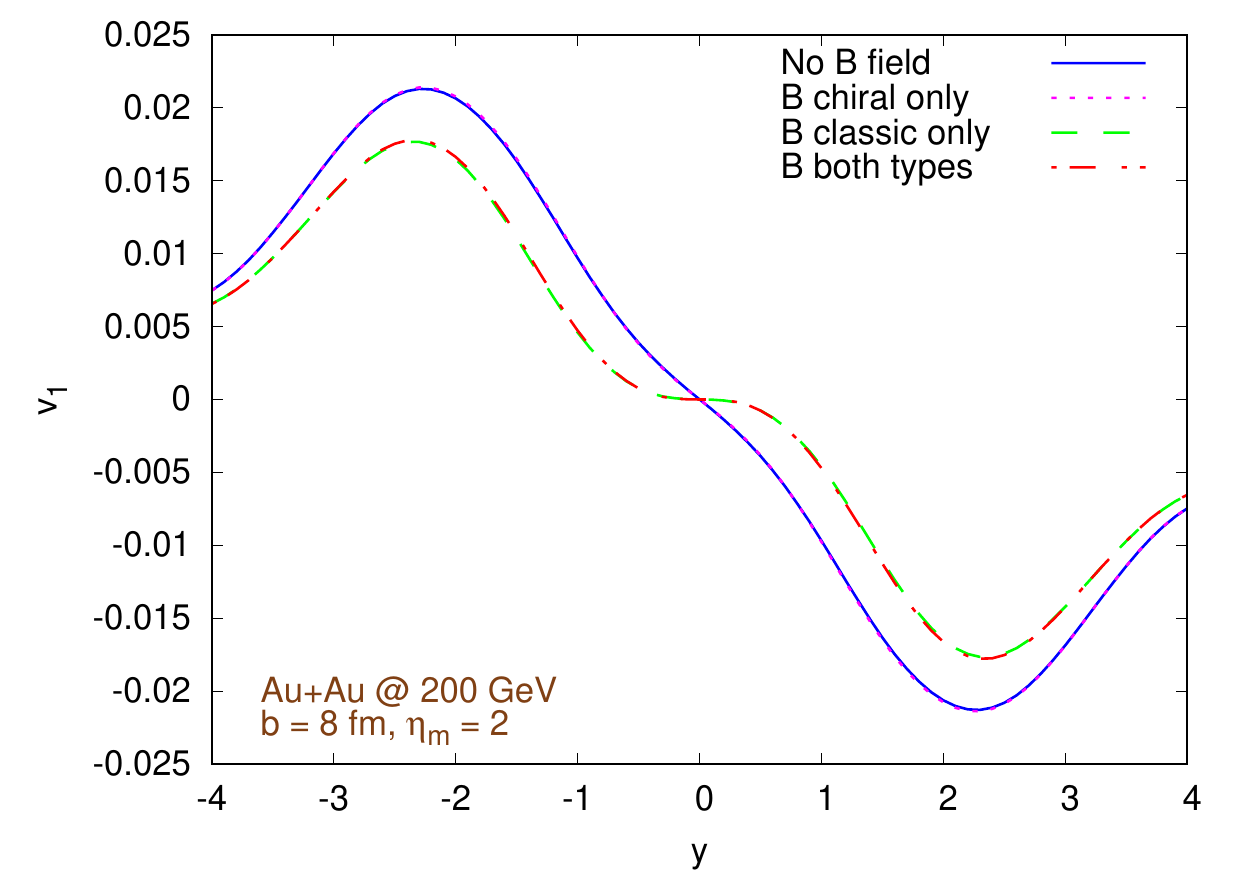}
	\end{minipage}
	\hspace{3mm}
	\begin{minipage}[b]{0.48\textwidth}
		\includegraphics[width=1\textwidth]{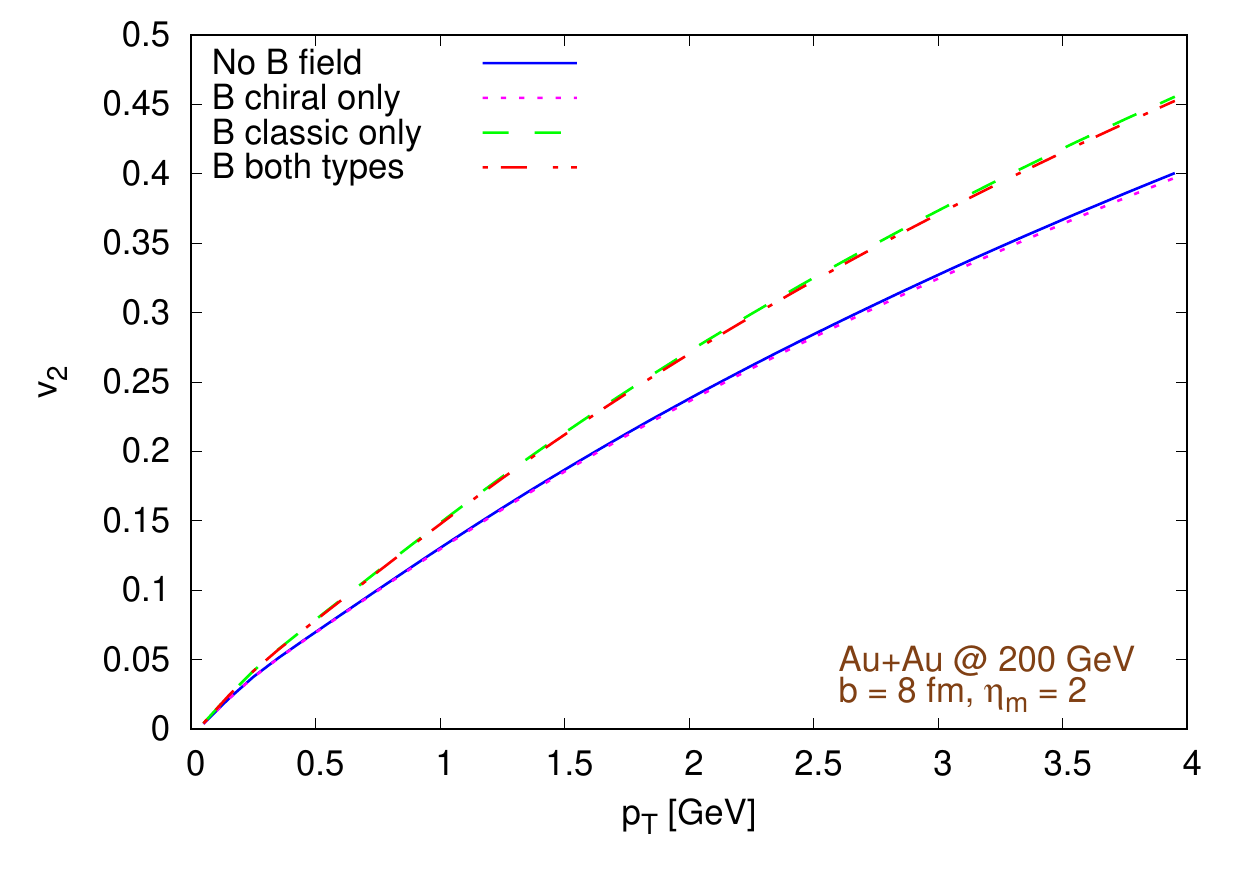}
	\end{minipage}
	\caption{Evaluation of the effects of the magnetic field having classic or chiral origin on $v_1(y)$ (left) and $v_2(p_T,y=0)$ (right), for Au+Au collisions at $\snn=200\,\textrm{GeV}$ with $b=8\,\textrm{fm}$. Top: Simulations with no initial energy density profile tilting (see App.~\ref{app:initial_edens}). Bottom: Simulations with initial energy density distribution tilting ($\eta_m=2$, see App.~\ref{tilt-desc}).}
	\label{kind_B}
\end{figure*}

\begin{figure*}[ht]
	\begin{minipage}[b]{0.48\textwidth}
		\includegraphics[width=1\textwidth]{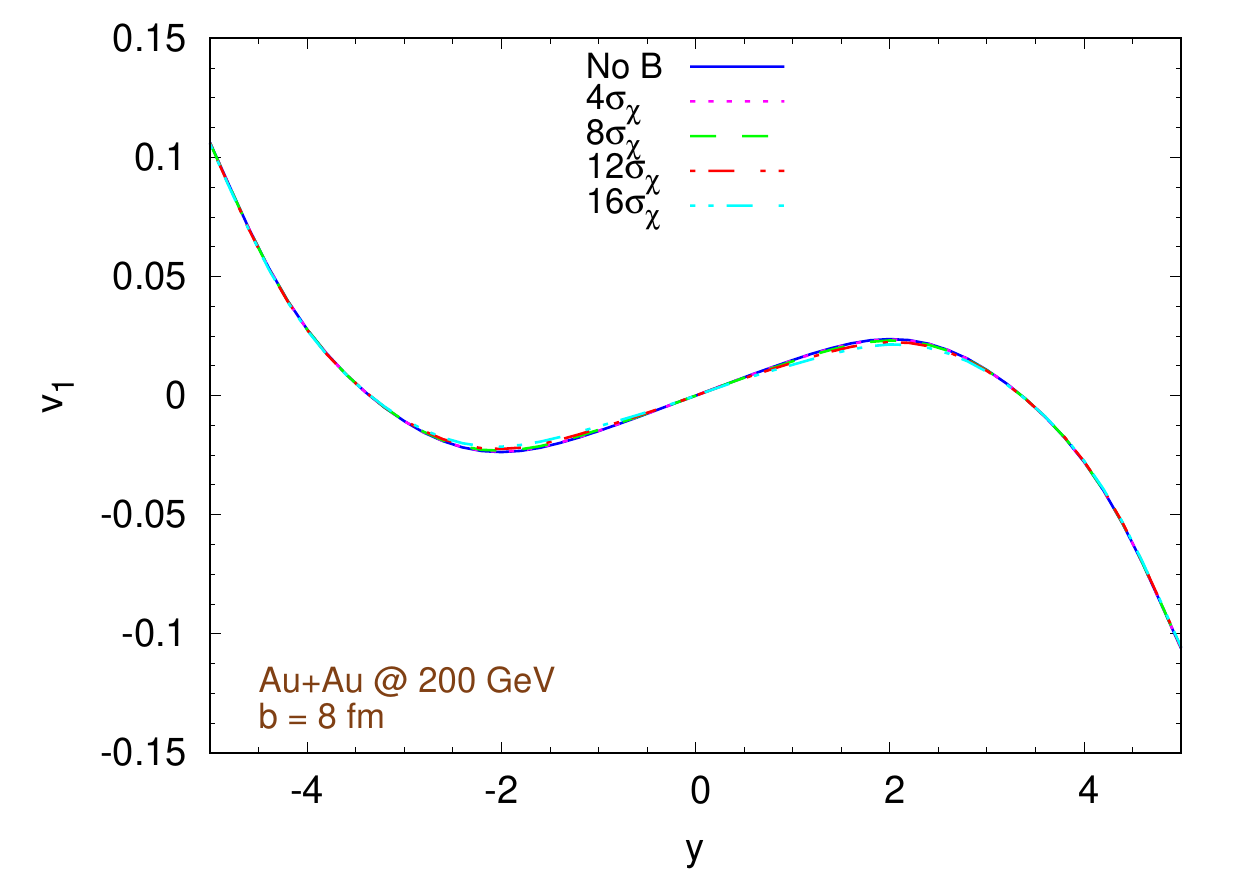}
	\end{minipage}
	\hspace{3mm}
	\begin{minipage}[b]{0.48\textwidth}
		\includegraphics[width=1\textwidth]{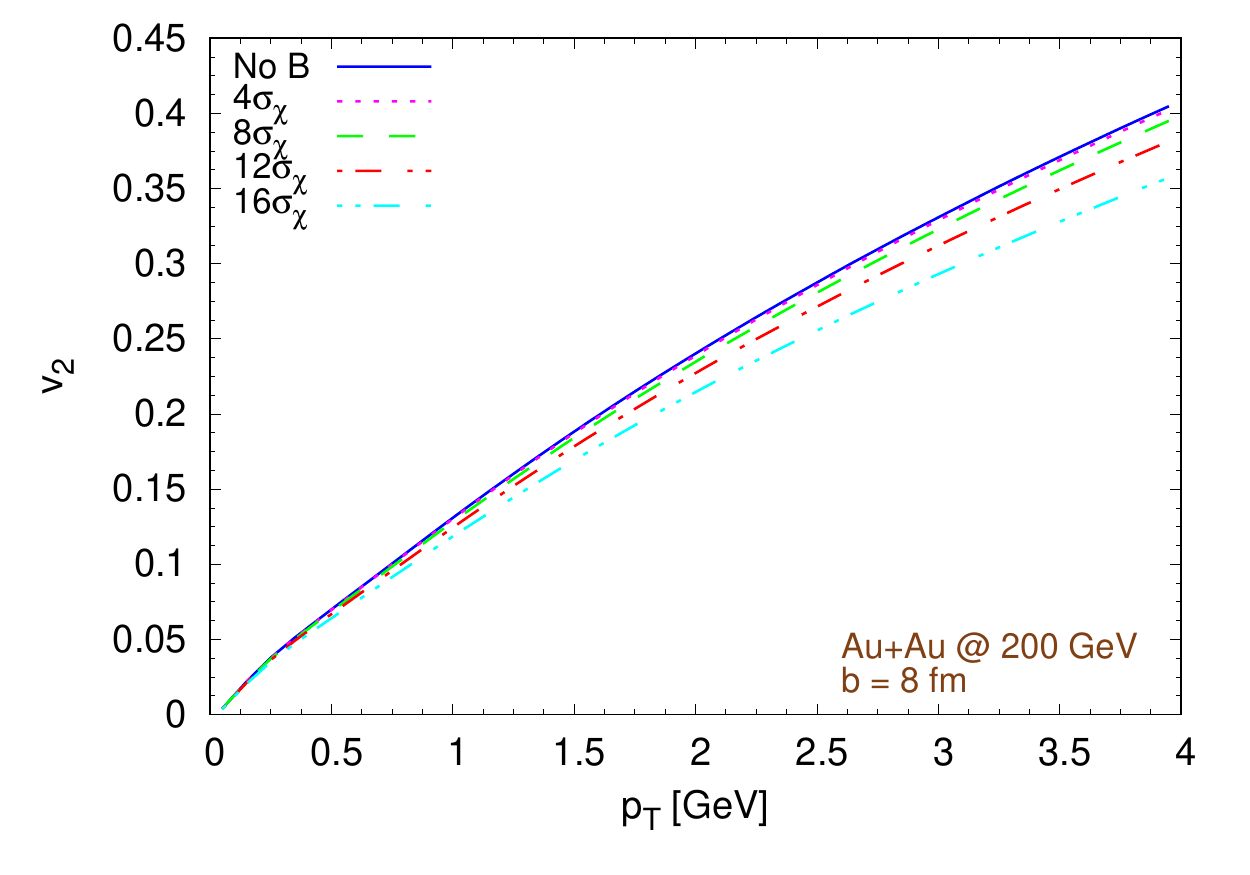}
	\end{minipage}
	\caption{Evaluation of the effects of a magnetic field of chiral origin only, amplified by factors 4, 8, 12, and 16, on $v_1(y)$ (left) and $v_2(p_T,y=0)$ (right), in Au+Au collisions at $\snn=200\,\textrm{GeV}$ with a fixed $b=8\,\textrm{fm}$.}
	\label{ch_extreme_RHIC}
\end{figure*}

In this section, we examine separately the contributions of magnetic fields of classic and chiral origin, as given in Eqs.~(\ref{Eq:inBp}-\ref{Eq:inBz}). To this aim, we perform simulations for Au+Au collisions at $\snn=200\,\textrm{GeV}$ with impact parameter $b=8\,\textrm{fm}$. We consider only an initial $B$ field of chiral or classic origin, in both cases multiplied by a factor of 4 from the na\"{i}ve values. Fig.~(\ref{kind_B}) shows the $v_{1}$ and $v_2$ of charged pions at mid-rapidity. The magnetic field of chiral origin generically produces only small modifications of the Fourier harmonics. The magnetic field of chiral origin does not significantly affect $v_{1}$ and $v_2$ even in the case of an initialization with a tilted energy density distribution (with $\eta_m=2$, see App.~\ref{tilt-desc}), as shown in the bottom part of Fig.~(\ref{kind_B}). However, in the tilted geometry, the impact of the magnetic field of classic origin on $v_1$ is indeed noticeable (top vs bottom left side of the figure). On the other hand, the $v_n$'s produced by the simulations with only the magnetic field of classic origin overlap quite well with the $v_n$ of the simulations with magnetic field of both classic and chiral origin. To obtain varying results, we evaluate the case of chiral magnetic fields which are 4, 8, 12, and 16 times larger than the estimates given by Eqs.~(\ref{Eq:inBp}-\ref{Eq:inBz}). In this situation, illustrated in Fig.~(\ref{ch_extreme_RHIC}) for collision at $\snn=200\,\textrm{GeV}$ with impact parameter $b=8\,\textrm{fm}$ and no initial energy density tilting, the effect of the magnetic field on $v_1$ (left figure) is still negligible, but $v_2$ is suppressed, an effect opposite to magnetic field of classic origin. It is likely that the different behavior is a consequence of the different orientation of the initial fields. 
It is important to note that this situation is one where only the magnetic field of chiral origin is present; physically this is an unlikely situation, as the magnitude of the standard magnetic field is expected to be an order of magnitude larger.
Moreover, we are treating the evolution of a magnetic field of chiral origin as a classic one, completely neglecting the evolution of axial charges. It is reasonable to expect that any axial charge produced will cascade into magnetic fields~\cite{Hirono:2015rla}, further increasing the magnetic field of chiral origin. However, in a more realistic scenario, it remains unclear how strong the contribution of magnetic fields of chiral origin on $v_2$ is. 

\subsection{Impact of the electrical conductivity in the pre-equilibrium phase}

\begin{figure*}
	\begin{minipage}[b]{0.48\textwidth}
		\includegraphics[width=1\textwidth]{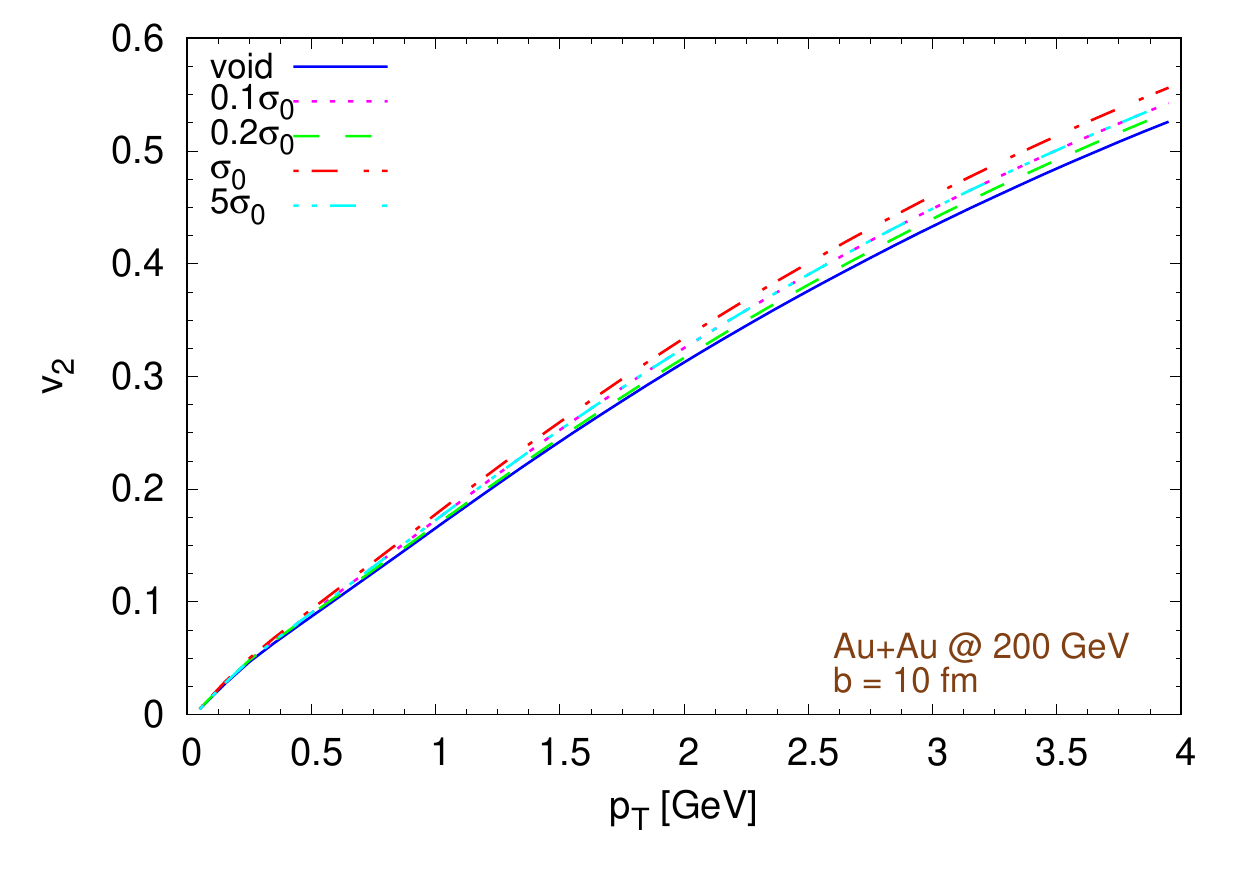}
	\end{minipage}
	\hspace{3mm}
	\begin{minipage}[b]{0.48\textwidth}
		\includegraphics[width=1\textwidth]{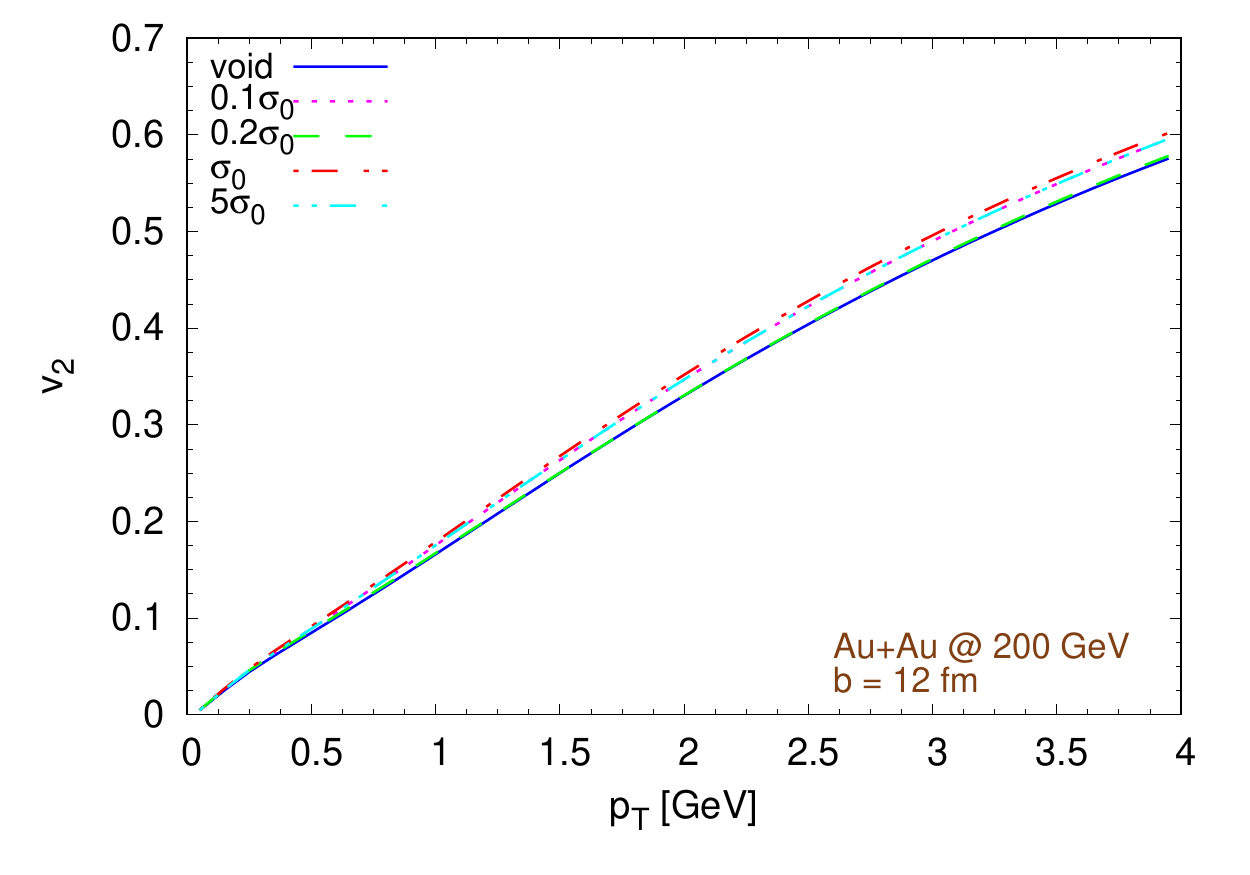}
	\end{minipage}
	\caption{Evaluation of the impact of different initial magnetic fields on $v_2(p_T,y=0)$, computed by assuming different electrical and magnetic chiral conductivities of the medium in the pre-equilibrium phase and multiplied by a factor 3. Both figures refer to Au+Au collisions at $\snn=200\,\textrm{GeV}$, with $b=10\,\textrm{fm}$ (left) and $b=12\,\textrm{fm}$ (right).} 
	\label{sigmas_x3_rhic}
\end{figure*}

Next, we explore the impact of the electrical and chiral magnetic conductivities of the medium in the pre-equilibrium phase, which enter in Eqs.~(\ref{Eq:inBp}-\ref{Eq:inBz}) and determine the magnitude and the spatial distribution of the initial magnetic field. In the context of our model, these are the only parameters which are directly related to the intrinsic properties of the medium and not to the collision energy or geometry. We consider the cases in which the conductivities of the medium in the pre-hydro phase ($\sigma_0$) are both scaled by a factor 0.1, 0.2, and 5. We consider the case of the void, as well. We find that simple variations of $\sigma_0$ are not sufficient to trigger significant modifications in $v_2$ at mid-rapidity, compared to the case with no magnetic field, both at RHIC and LHC energies, not even for peripheral collisions. However, if we enlarge the initial B fields by a factor 3, we notice that the enhancement of the elliptic flow depends on the chosen values of $\sigma$ and $\sigma_{\chi}$ to initialize B, with a dependence neither linear nor monotonic, as Fig.~(\ref{sigmas_x3_rhic}) shows for Au+Au collisions at $\snn=200\,\textrm{GeV}$ with $b=10\,\textrm{fm}$ (left) and $b=12\,\textrm{fm}$ (right). 
With this point, we will end our examination of the consequences of the different choices of the initial conductivities, as the basic hypothesis underlying Eqs.~(\ref{Eq:inBp}-\ref{Eq:inBz}), i.e. the propagation of charged particles with constant velocities in a medium with electrical and magnetic chiral conductivities constant in time and space, is clearly only schematic and the estimates of the initial $B$ field should be taken with care. 
However, these preliminary results suggest a non-trivial relationship between the intrinsic properties of the medium in the very early stages of the collision and the elliptic flow of the final pions, which should be more thoroughly investigated in the future.

\section{Charge production and the CME}
\label{sec:charge_cme}
The basic mechanism behind the Chiral Magnetic Effect (CME) has long been understood, and has been confirmed in a variety of theory calculations, such as with lattice studies~\cite{Buividovich:2009wi,Yamamoto:2011gk,Mace:2016shq}, and observed in condensed matter experiments~\cite{Li:2014bha,Huang:2015eia,arnold-shekar16,Xiong413}. 
In astrophysics, anomalous transport phenomena may explain certain observations, e.g. neutron star kicks~\cite{Kaminski:2014jda}.
However, a clear signal at current heavy-ion collision experiment results remains elusive~\cite{Abelev:2009ac,Abelev:2012pa,Adamczyk:2014mzf,Adamczyk:2015eqo}. From the chiral anomaly, it has long been known that topological configurations of the gauge fields will produce domains of axial charge (chiral fermions); this axial charge is then the seed for the CME current when an external magnetic field is present. Recent progress has been made however towards understanding sources of axial charge production in heavy-ion collisions. At the earliest times, the initially very strong gauge fields feature a longitudinal `flux-tube'-like structures which may source axial charge~\cite{Kharzeev:2001ev,Lappi:2006fp,Lappi:2017skr}. These configurations quickly break up, whereby an over-occupied non-Abelian plasma known as the Glasma is formed~\cite{Lappi:2006fp}. In this regime, real-time topological transition, called sphaleron transitions, can produce axial charge. While this stage is very short, as the system is rapidly thermalizing, it has been determined that a few sphaleron transitions can be expected in the average collision at RHIC or the LHC~\cite{Mace:2016svc}. As opposed to field strength fluctuations, these domains of topological charge will produce chiral charge coherently, and may allow for the growth of a sizeable CME current. During the QGP stage, it is also possible that chiral fermions are anomalously produced via sphaleron transitions~\cite{Moore:2010jd}. However, as the magnetic field during the QGP stage of the evolution is expected to be small compared to that of the pre-equilibrium stage, it is likely that CME production after the pre-equilibrium stage is negligible. It was also argued in Ref.~\cite{Mace:2016svc} that the sphaleron rate out of equilibrium is parametrically larger than in equilibrium, reinforcing the assumption of the dominance of pre-equilibrium CME generation. However, a full chiral-MHD description is needed to study the dynamics of chirality production during the hydrodynamic evolution. 

The pre-equilibrium axial charge production can be encoded, for simplicity, in terms of a uniform chiral chemical potential, $\mu_5$. By virtue of MHD being a long-wavelength effective description, it is natural to encode this initial axial charge as a chiral magnetic conductivity, $\sigma_\chi$.
In essence, the pre-equilibrium axial charge is cast in macroscopic language of MHD in terms of a helical (chiral) magnetic field sourced by $\sigma_\chi$. This however neglects any pre-equilibrium CME current generation; previous real-time lattice studies have demonstrated that these may be sizeable~\cite{Mace:2016shq}. The motivation for this effective description of microscopic axial charge in terms of macroscopic magnetic helicity can also be understood due to a self-similar cascade~\cite{Hirono:2015rla}, and may be driven by a chiral plasma instability~\cite{Akamatsu:2013pjd}. 

\begin{figure*}[ht]
	\begin{minipage}[b]{0.485\textwidth}
		\includegraphics[width=1\textwidth]{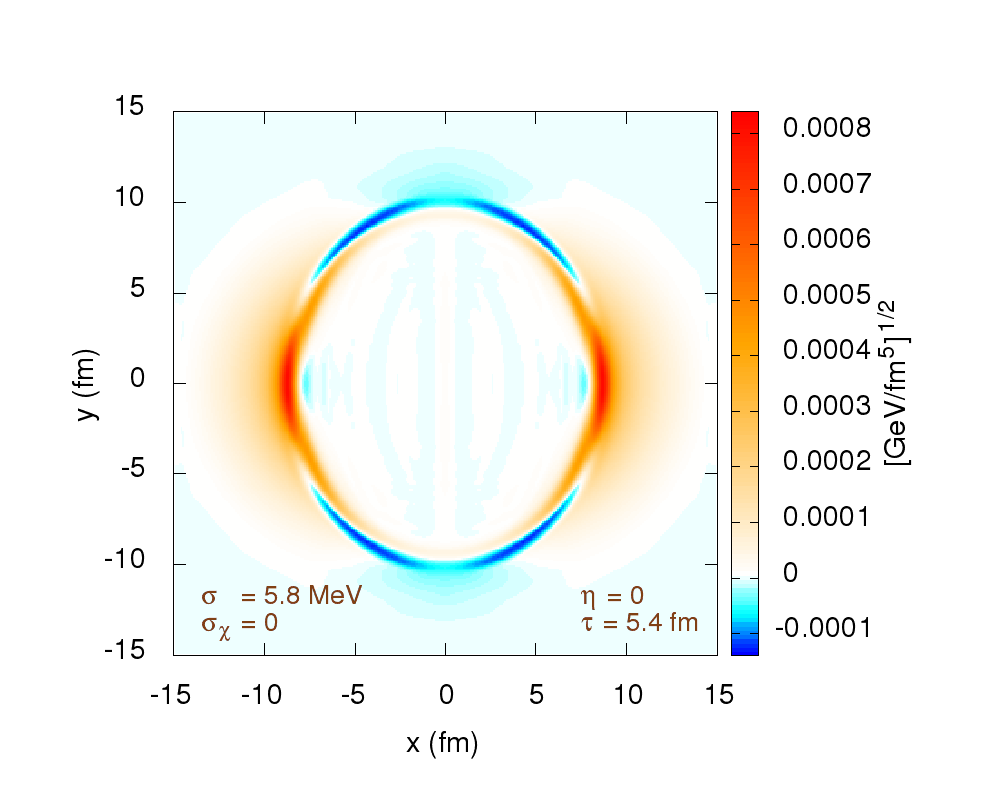}
	\end{minipage}
    	\hspace{0.1mm}
	\begin{minipage}[b]{0.485\textwidth}
		\includegraphics[width=1\textwidth]{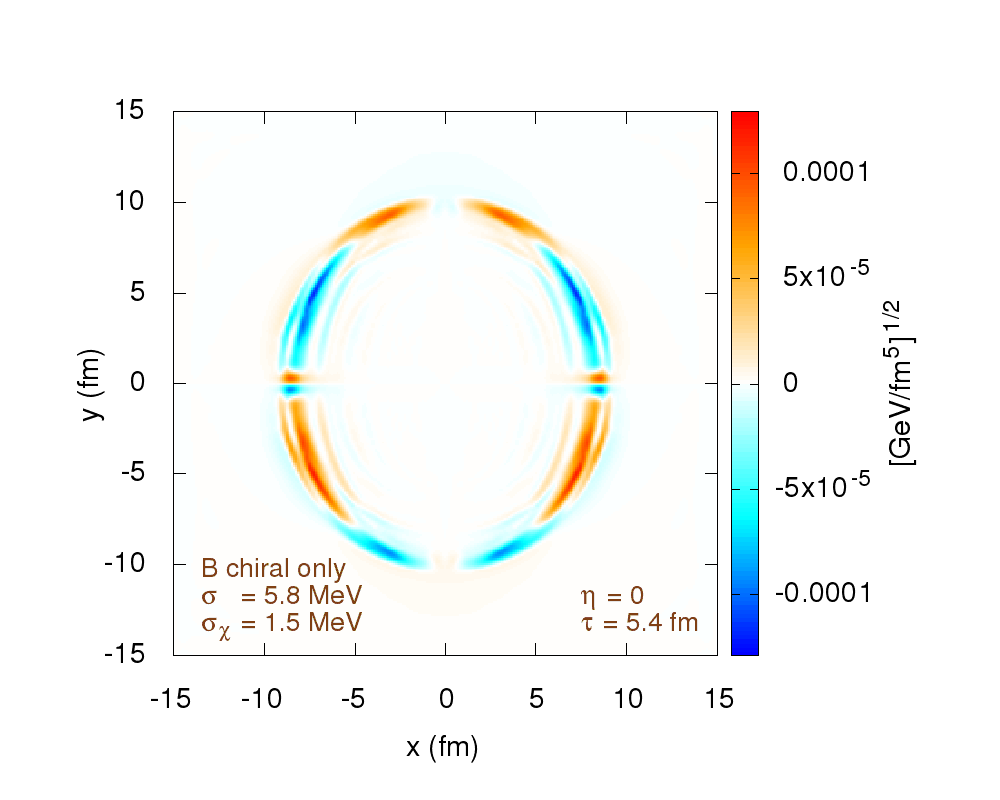}
	\end{minipage}
    	\hspace{0.5mm}
	\begin{minipage}[b]{0.485\textwidth}
	\includegraphics[width=1\textwidth]{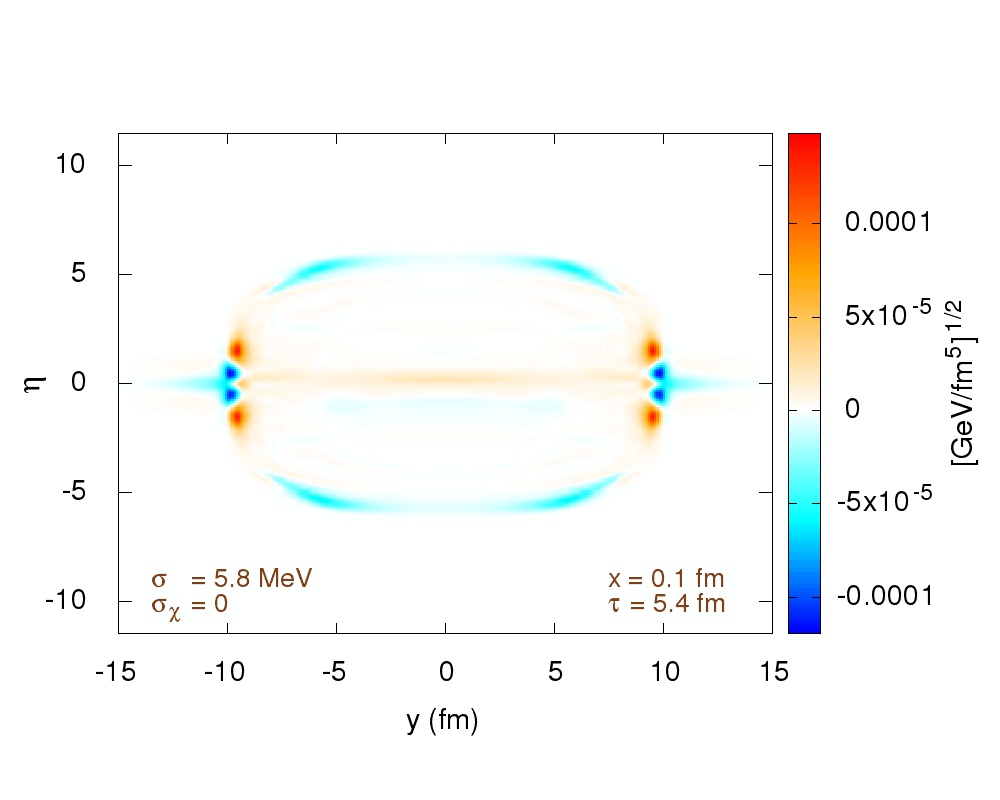}
\end{minipage}
\hspace{0.1mm}
\begin{minipage}[b]{0.485\textwidth}
	\includegraphics[width=1\textwidth]{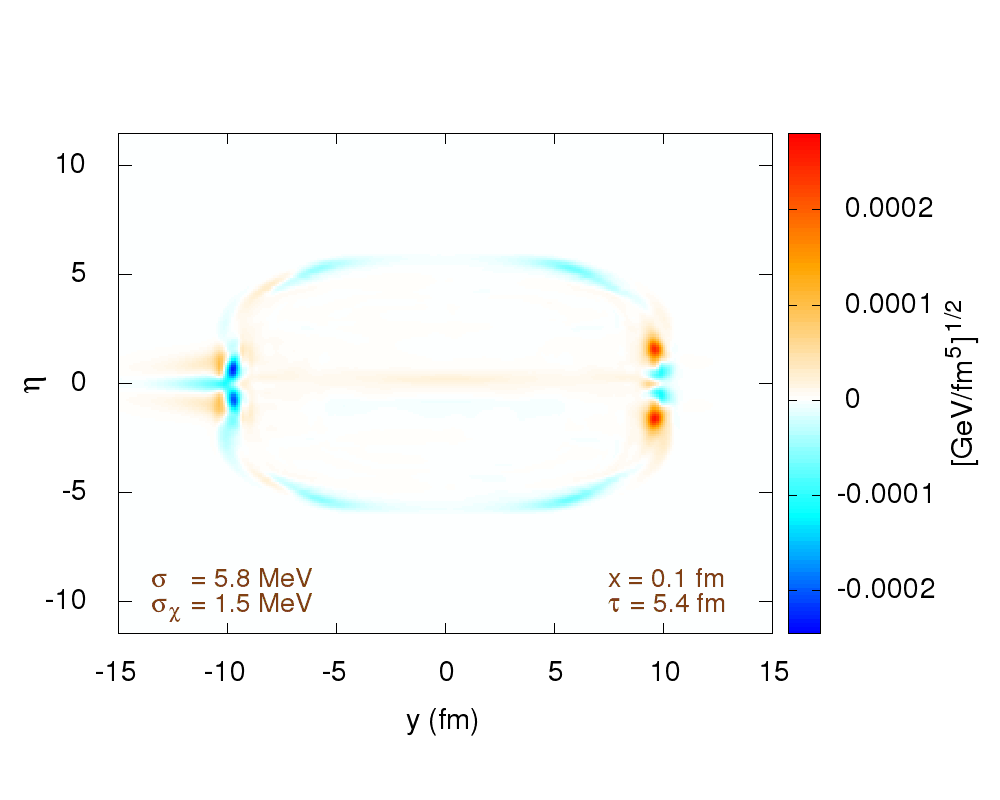}
\end{minipage}
	\caption{Electric charge density distribution in the fluid comoving frame at $\tau=5.4\,\fm$. Top left: Magnetic field of classical origin in transverse plane at mid-rapidity. Top right: Magnetic field of chiral origin only in transverse plane at mid-rapidity. Bottom left: Magnetic field of classical origin only in the $y-\eta$ plane for $x=0.1~\fm$. Bottom right: Magnetic field of classical and chiral origin in the $y-\eta$ plane for $x=0.1~\fm$. Here $\sigma=5.8\,\textrm{MeV}$ and $\sigma_\chi=1.5\,\textrm{MeV}$.}
	\label{charge_sep_comp}
\end{figure*}

\begin{figure*}[ht]
	\begin{minipage}[b]{0.485\textwidth}
		\includegraphics[width=1\textwidth]{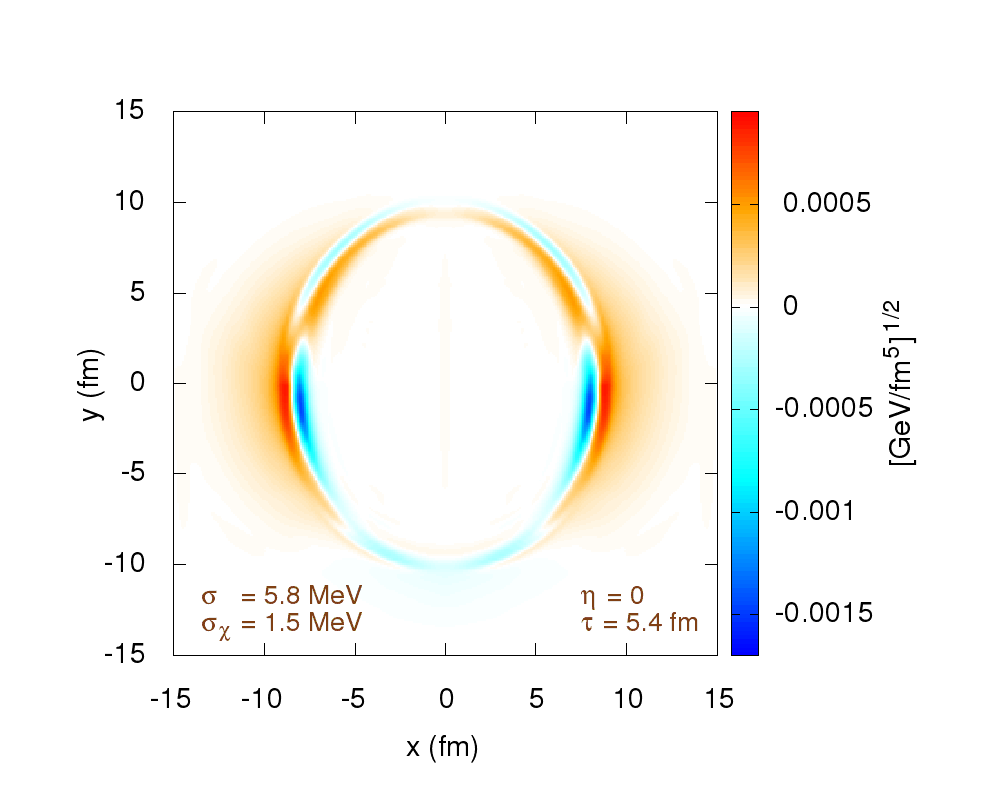}
	\end{minipage}
	\hspace{0.1mm}
	\begin{minipage}[b]{0.485\textwidth}
		\includegraphics[width=1\textwidth]{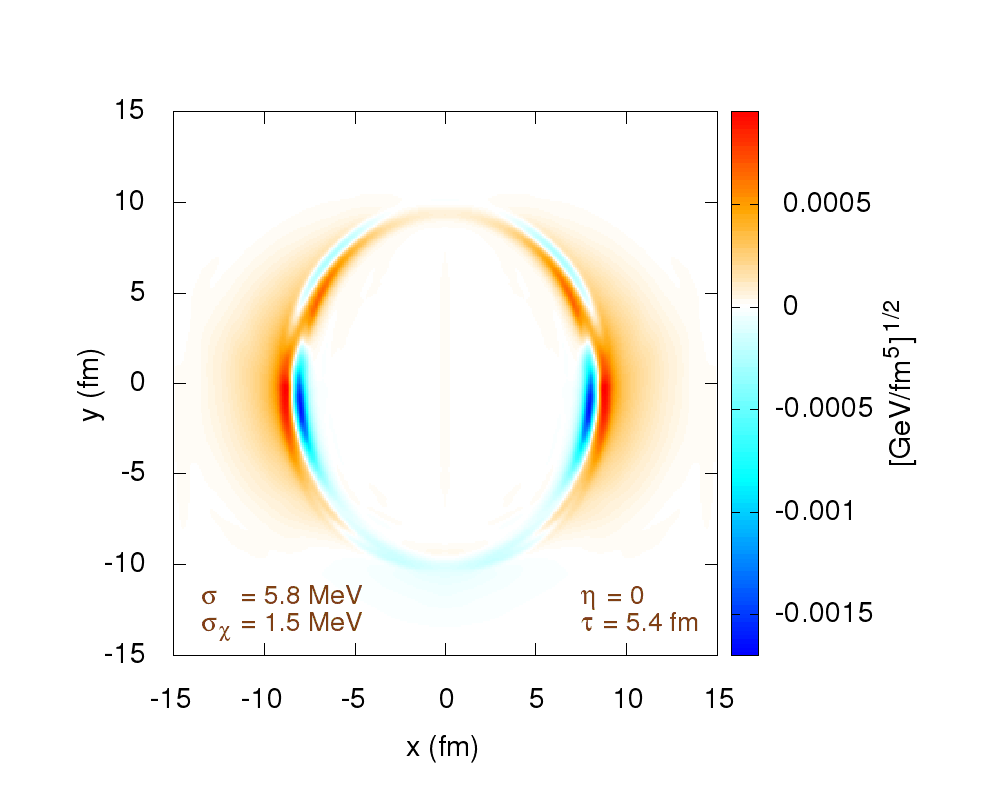}
	\end{minipage}
\caption{Electric charge density distribution in the fluid comoving frame from magnetic fields of both classical and chiral origin in the transverse plane at mid-rapidity. Left: Maxwell equations charge determination method. Right: Vorticity method. Here $\sigma=5.8\,\textrm{MeV}$ and $\sigma_\chi=1.5\,\textrm{MeV}$. }
\label{charge_sep_method}
\end{figure*}

At present, there are no quantitative calculations of the amount of axial charge produced during the pre-equilibrium stage of a heavy-ion collision.
In light of this, we consider a range of reasonable values for the chiral magnetic conductivity, from $\sigma_\chi=1.5~\text{MeV}$ (as before; this corresponds to $\mu_5\approxeq 322~\text{MeV}$) to $\sigma_\chi=15~\text{MeV}$. We then can study the local electric charge density generated in the fluid by the magnetic field. 
In Figures~\ref{charge_sep_comp}, we show this charge density in the transverse plane at mid-rapidity (top row) and in the $y-\eta$ plane for $x=0.1~\fm$ (bottom row) at $\tau=5.4\,\fm$. In the left column, only a magnetic field of classical origin is considered. We see that the charge densities accumulated at the edges of the fireball along the magnetic field (y) direction are manifestly symmetric. Thus no charge separation is seen. However, in the top right column, we consider only a magnetic field of chiral origin and can clearly see an electric charge dipole along the magnetic field direction. In the bottom right figure, we consider a magnetic field of both classical and chiral origin, and observe that an electric charge dipole still remains manifestly present. Results for the electric charge from magnetic fields of both classical and chiral origin in the transverse plane at mid-rapidity are shown in Fig.~\ref{charge_sep_method}. While the dipole structure is less apparent than the top right figure of Fig.~\ref{charge_sep_comp}, it is nevertheless still present.

Furthermore, in Fig.~\ref{charge_sep_method}, the left and the right panels present results obtained using two different methods to compute the electric charge density in the fluid comoving frame. In the bottom left, the charges are determined directly from the Maxwell equations, while in the bottom right they are determined from the vorticity (Fig.~\ref{charge_sep_comp} uses this method). These methods, explained in App.~\ref{app:chargedepcomp}, involve the computation of different partial derivatives. In principle the resulting charge density should be the same, however, in practice they can disagree, especially in regions with strong gradients. It is important to stress two points about the charge densities shown in Figs.~\ref{charge_sep_comp} and~\ref{charge_sep_method}. First, they are located in the region (in Bjorken coordinates) close to the expansion wave, under the assumption that the whole computation domain is filled by a fluid, but in general the region where the QGP transforms into particles, i.e. the freezeout hypersurface, is different. 
Second, the detectors measure a particle distribution in the momentum space, therefore it is crucial to understand how to translate this spatial charge distribution into some experimental observables.

\begin{figure*}[h!]
	\begin{minipage}[b]{0.48\textwidth}
		\includegraphics[width=1\textwidth]{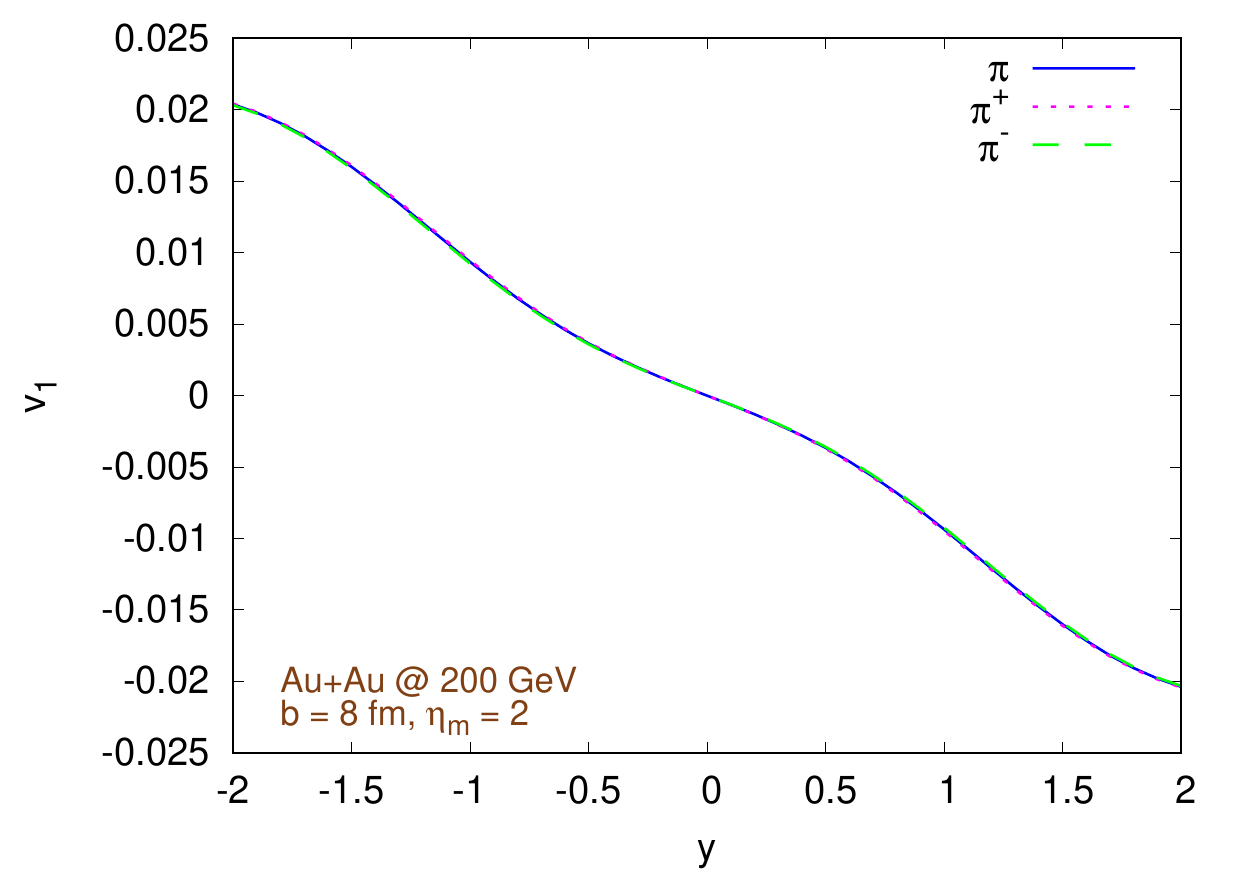}
	\end{minipage}
	\hspace{3mm}
	\begin{minipage}[b]{0.48\textwidth}
		\includegraphics[width=1\textwidth]{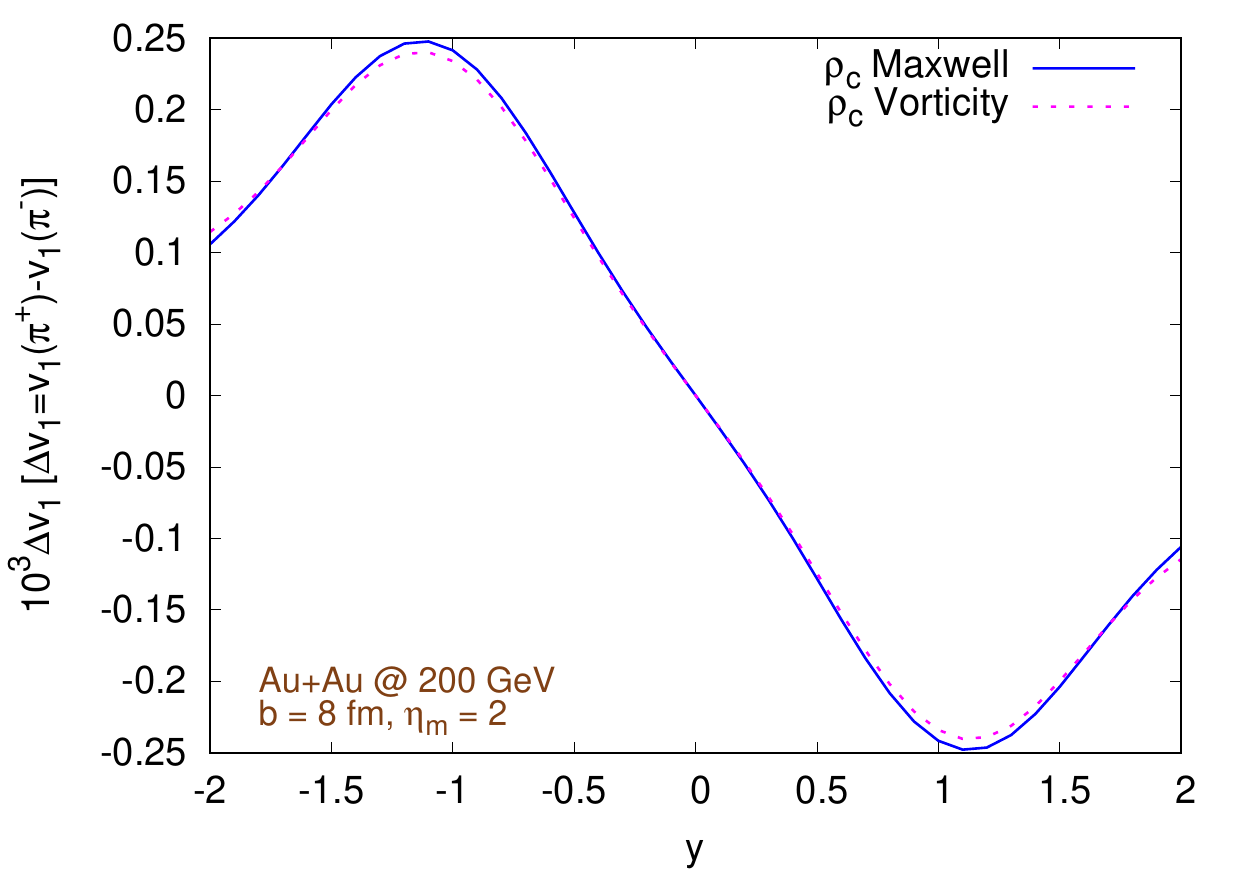}
	\end{minipage}
	\caption{In this figure we evaluate the effects of a charge dependent freezeout (see App.~\ref{app:chargedepfo}) on the directed flow of pions. Left: $v_1(y)$ for pions neglecting their electric charge and for positive and negative pions. Right: difference between $v_1(y)$ of positive and negative pions ($\times 10^3$), computed by using either the Maxwell equations or the vorticity. Both plots refers to Au+Au collisions at $\snn=200\,\textrm{GeV}$ with impact parameter $b=8\,\textrm{fm}$ and tilted initial energy density distribution with parameter $\eta_m=2$.}
	\label{v1_charged}
\end{figure*}

\begin{figure*}[h!]
	\begin{minipage}[b]{0.48\textwidth}
		\includegraphics[width=1\textwidth]{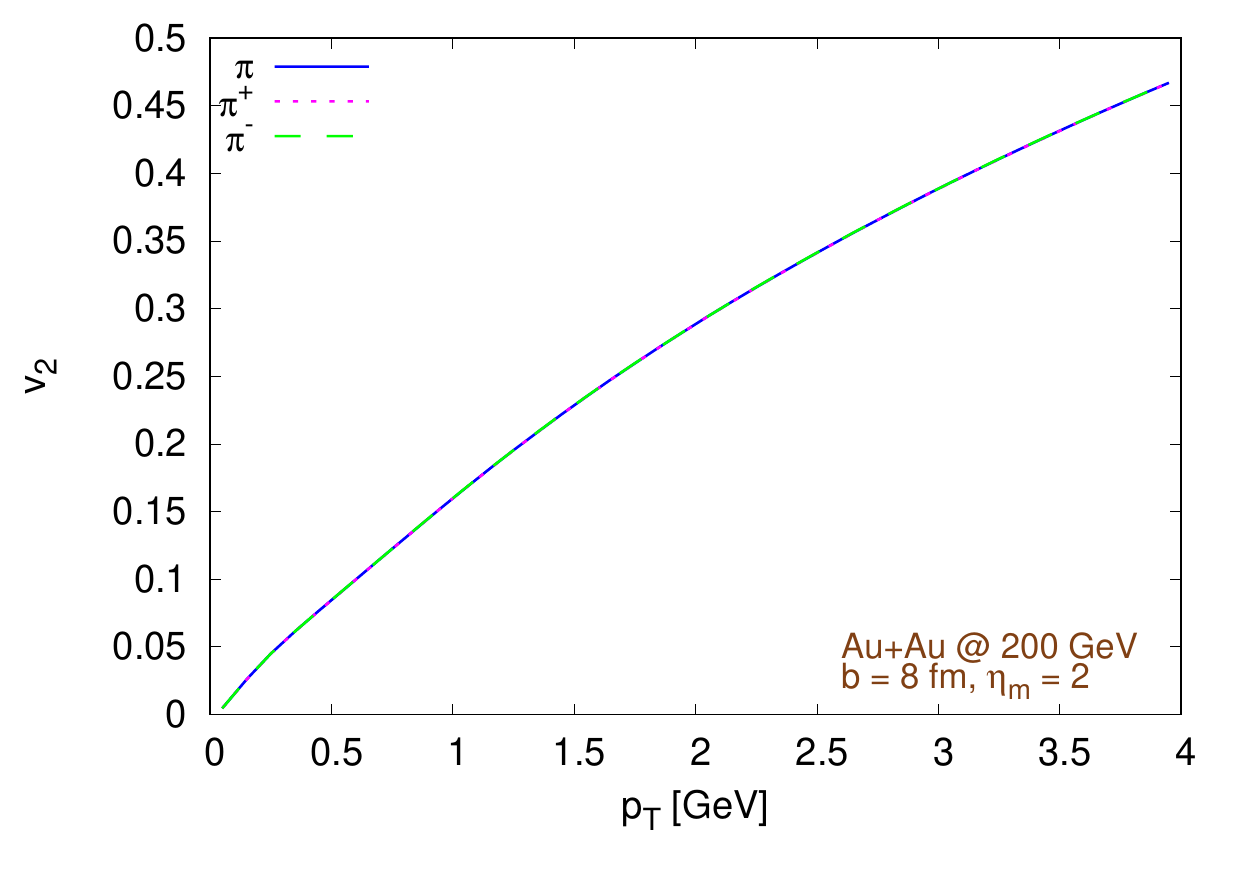}
	\end{minipage}
	\hspace{3mm}
	\begin{minipage}[b]{0.48\textwidth}
		\includegraphics[width=1\textwidth]{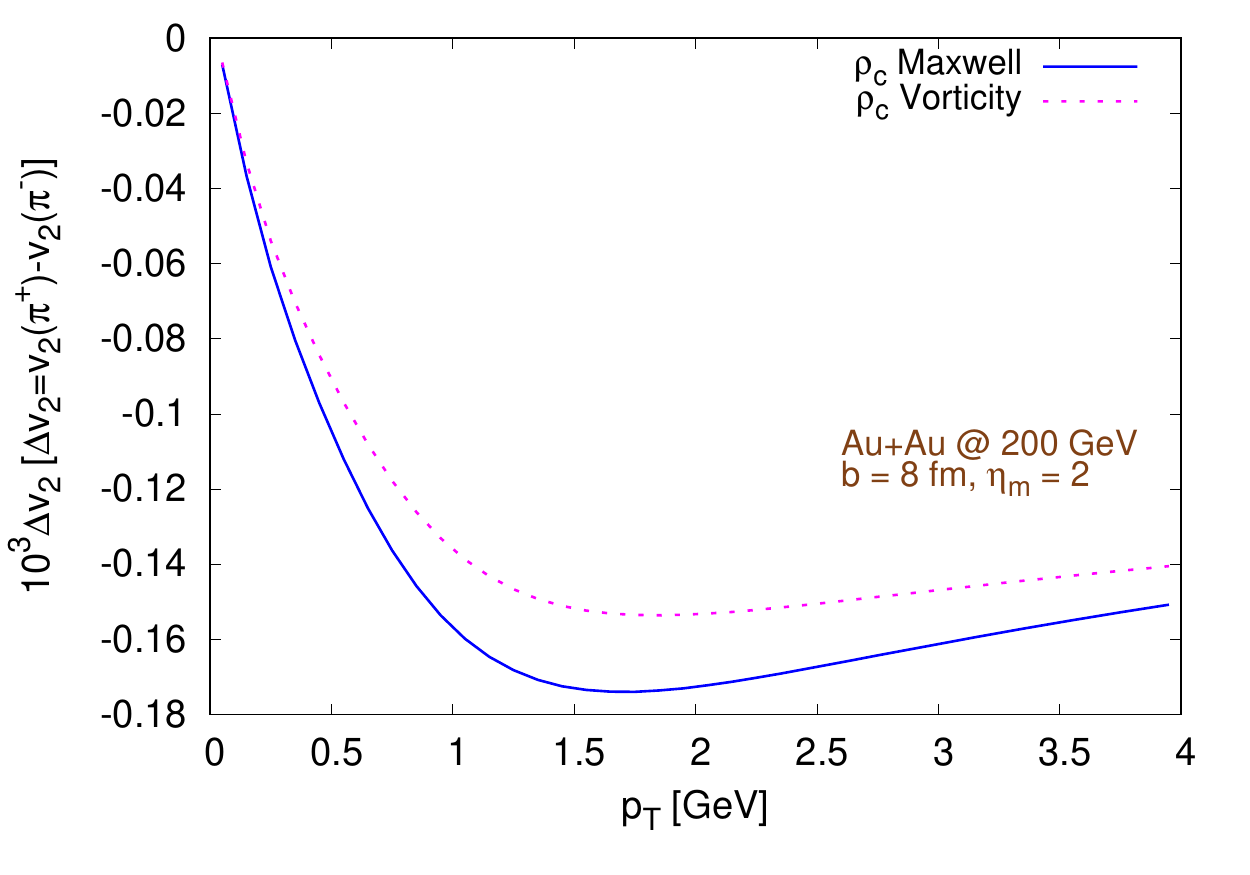}
	\end{minipage}
	\caption{Evaluation the effects of a charge dependent freezeout (see App.~\ref{app:chargedepfo}) on the elliptic flow of pions. Left: $v_2(p_T,y=0)$ for pions without considering their electric charge and for positive and negative pions. Right: Difference between $v_2(p_T,y=0)$ of positive and negative pions ($\times 10^3$), computed by using either the Maxwell equations or the vorticity. Both plots refers to Au+Au collisions at $\snn=200\,\textrm{GeV}$ with $b=8\,\textrm{fm}$ and tilted initial energy density distribution ($\eta_m=2$).}
	\label{v2_charged}
\end{figure*}

In our model we compute the spectra of the emitted particles by using the Cooper-Frye prescription~\cite{Cooper:1974mv}, so we should modify this recipe in such a way to take into account a contribution due to the electromagnetic fields.
Unfortunately, the underlying assumption of the ideal MHD approach that the electric field in the comoving frame of the fluid is always zero, with an infinitely small relaxation time, makes modifying the distribution function in the denominator of the Cooper-Frye equation in a consistent way very difficult~\cite{Feng:2017giy,Gursoy:2018yai}. It will be possible to overcome this limitation with resistive MHD. However, in this work we appeal to a simple, physically motivated solution based on the charge density in the comoving frame of the fluid. Essentially, we compute the electric charge density, as measured in the comoving frame of the fluid, for each cell of the freeze-out hypersurface. We then modify the number of the produced positive and negative pions in such a way to obtain the same net charge density. With this procedure, the distribution of the momenta of the emitted hadrons in each cell is still insensitive to the electric charge. Details for this procedure are discussed in App.~\ref{app:chargedepfo}. While physically motivated, we acknowledge that this method is imperfect. Nevertheless, our aim is to perform an exploratory study on the effects of the dynamics of magnetic fields on a fluid and not quantitative phenomenology, so such an approximation seems justified.

We evaluate the effects of our charge dependent freezeout procedure for charged pions in Au+Au collisions at $\snn=200\,\textrm{GeV}$ with impact parameter $b=8\,\textrm{fm}$, initializing the energy density distribution with tilting parameter $\eta_m=2$. We look at the directed and elliptic flow and at the  $ \langle \sin \phi \rangle $ observables, where $\phi$ is the angle between the direction of transverse momentum of the pion and the reaction plane. 

On the left side of Fig.~\ref{v1_charged} we show the directed flow $v_1$ of pions versus the rapidity, while on the right side we show the difference $\Delta v_1=v_1(\pi^+)-v_1(\pi^-)$ between the directed flow of positive and negative pions (multiplied by a factor $10^3$). The right side of Fig.~\ref{v1_charged} demonstrates a fair agreement between the results obtained by computing the electric charge density using the Maxwell equations or vorticity method.

We note that the sign of $\Delta v_1$ with respect to rapidity seems to be the opposite of what the preliminary experimental measurements suggest~\cite{Margutti:2017lup}, albeit these experimental data refer to unidentified charged particles (so, not only pions) with respect to pseudorapidity at a different collision energy ($\sqrt{s_{NN}}=5.02\,\mathrm{TeV}$). We will discuss this further in Sec.~\ref{sec:issues}. 
In the left side of Fig.~\ref{v2_charged} we show the elliptic flow $v_2$ of pions versus rapidity, while in the right side we show the difference $\Delta v_2$ between the elliptic flow of positive and negative pions (multiplied by a factor $10^3$). The right side of Fig.~\ref{v2_charged} shows some clear discrepancies between the results obtained by computing the electric charge density with the Maxwell equations or with the vorticity method, probably due to the numerical errors in summing over two different sets of gradients and to the limited accuracy (second order) in solving the evolution equations. However, in any case the difference between the $v_2$ of positive and negative pions seems to be rather small.

The $\langle \sin \phi \rangle$~\footnote{ $\langle \sin \phi \rangle=\dfrac{\int_0^{p_T(max)} \int_0^{2\pi}\sin(\phi) p_T \d N/(p_T \d p_T \d\phi \d y)_{|y=0} \d \phi \d p_T}{\int_0^{p_T(max)} \int_0^{2\pi}p_T \d N/(p_T \d p_T \d\phi \d y)_{|y=0} \d \phi \d p_T}$, with $p_T(max)=4\,\GeV$.} observable (with $\phi$ the angle between the direction of transverse momentum of the pion and the reaction plane) aims to isolate particle production in the direction of the magnetic field, perpendicular to the reaction plane. Therefore, it can act as a simple signal of magnetic field driven effect like the CME. If the average of $\sin \phi $ is different from 0, it means that there is an unbalance between the number of particles emitted up or down along the direction orthogonal to the reaction plane. The results of this calculation for different choices of magnetic field are reported in Table~\ref{t:sinphi}. We observe that when considering an initial magnetic field which includes a contribution from chiral charges (second row), there is a clearly non-zero $\langle \sin \phi \rangle$. However, if the magnetic field is generated by classical vector currents only, $ \langle \sin \phi \rangle \simeq 0$ (first row). Thus, we can clearly demonstrate the conversion of initial chiral magnetic fields in final state charge asymmetry. The change in the sign of $ \langle \sin \phi \rangle$ might be explained by the following considerations. The CME current is proportional to $\mu_5$ and magnetic field. The chiral magnetic field is equivalent locally to some value of $\mu_5$ -
the chiral charge of the chiral knot of magnetic field corresponds locally to a certain chiral charge, and thus to a chiral chemical potential $\mu_5$. So the CME current for the chiral magnetic field only is $J^{chiral} \sim \mu_5^{chiral} B^{chiral}$.
Now, for chiral + classical, the total magnetic field changes by about an order of magnitude, while $\mu_5$ stays the same, so the CME
current is $J^{chiral+classical} \sim \mu_5^{chiral} B^{chiral+classical}$. Since $B^{chiral+classical}$ is an order of magnitude larger than $B^{chiral}$, the CME current and the resulting charge separation should be an order of magnitude larger as well.\\
Moreover, if the direction of $B^{chiral+classical}$ is different from the direction of $B^{chiral}$, the sign of separation can be different as well. In Ref. 40 it is shown that, in the transverse plane, the components of classic magnetic fields have an azimuthal orientation, while chiral magnetic fields have a radial orientation. Going beyond the aims of the current work, it would be interesting  to study more elaborate magnetic field driven charge dependent observables, like those of Refs.~\cite{Voloshin:2004vk,Ajitanand:2010rc,Magdy:2017yje}, but probably it is wiser to dive into this project only after the development of a more refined version of our model.

\begin{center}
	\begin{table}
\begin{tabular}{|l|c|c|c|}
	\hline
	& $ \langle \sin \phi \rangle \,(\pi^+)$ & $ \langle \sin \phi \rangle \,(\pi^-)$ & $\Delta \langle \sin \phi \rangle$\\
	\hline
Classic $\vec{B}$ & $-1.49\cdot10^{-15}$  &  $-1.67\cdot10^{-15}$& $1.8\cdot10^{-16}\approx 0$\\
Chiral  $\vec{B}$ & $-2.82\cdot10^{-6}$ & $+2.82\cdot10^{-6}$ & $-5.64\cdot10^{-6}$\\
Classic + chiral  $\vec{B}$ & $8.16\cdot10^{-5}$ & $-5.54\cdot10^{-5}$ & $1.37\cdot10^{-4}$\\
\hline
\end{tabular}
\caption{First two columns: $\langle \sin \phi \rangle$  at $|y|\simeq 0$ for positive and negative pions, third column: $\Delta \langle \sin \phi \rangle= \langle \sin \phi \rangle \,(\pi^+)-\langle \sin \phi \rangle \,(\pi^-)$. The results refer to numerical simulations of Au+Au collisions at $\snn=200\,\textrm{GeV}$ with impact parameter $b=8\,\textrm{fm}$. In these computations the electric charge density was computed by using the vorticity method (App. ~\ref{vorticity_method}). When including an initial magnetic field with a contribution from chiral charges, the results turn from a roughly null value (first row) to a non zero signal (second row)}
    \label{t:sinphi}
	\end{table}
\end{center}

\section{Known issues}
\label{sec:issues}

In the previous section, we noted that there is tension between our results in right panel of Fig.~\ref{v1_charged} for the rapidity dependence of $\Delta v_1(y)=v_1(\pi^+)-v_1(\pi^-)$ compared to recent ALICE results~\cite{Margutti:2017lup}.
A similar discrepancy between theoretical and experimental results has been observed in a complementary framework in Refs.~\cite{Gursoy:2018yai,Gursoy:2014aka}, whereby the electromagnetic fields are added to the fluid evolution perturbatively (see also~\cite{Das:2016cwd,Chatterjee:2018lsx,Chatterjee:2019xtg,Coci:2019nyr} for related calculations for heavy flavors). There, the authors considered finite conductivity of the medium and therefore have both electric and magnetic fields. They identified two classical magnetic field driven mechanisms which resulted in charge- and rapidity-odd contributions to the Fourier harmonics: I) the Faraday effect, resulting from changing magnetic fields, II) the Lorentz force, resulting from a charged particle moving in a magnetic field. They observed that these effects contribute with opposite signs to charge-dependent $v_1(y)$. Using this as a basis to speculate on the nature of the apparent disagreement between theory and experiment on the sign of $\Delta v_1(y)$, it is possible that in our formalism, and potentially others, there is a delicate interplay between the properties of the magnetic field in the medium, related to the rate of expansion in comparison to the decrease in the magnetic field with time, which the inclusion of a temperature dependent conductivity and viscosity might drastically alter. 
There also several other possible explanations for the behavior of $\Delta v_1(y)$ obtained in the present simulations. A possible source of error might be the prescription to determine charge dependent spectra, which assumes a kind of chemical potential to modify the particle species abundances, without taking into account any modification of the momentum distribution due to the electromagnetic field. Moreover, in this study the local charge imbalance is attributed only to pions, neglecting the contributions of other less abundant hadrons. The computation of the electric charge density itself shows some evidence of a limited accuracy, e.g. in Fig.~(\ref{v2_charged}), which might be improved either adopting constrained transport schemes~\cite{lond04,Mignone:2019ebw} or other frameworks in which this quantity is explicitly evolved~\cite{Denicol:2018rbw,Denicol:2019iyh}. There are also notable improvements possible for the initialization procedure. To begin, by virtue of the ideal MHD framework, the electric field is only a derived quantity and therefore, by setting the initial velocity field to zero, the initial electric field is also set to zero; we thus neglect its contribution. However, in Ref.~\cite{Gursoy:2018yai} the electric field has been shown to be of the same order of magnitude of the magnetic field. In addition to that, the magnetic field generated by the protons is assumed to be instantaneously transferred to the QGP fluid. This instant quench should be replaced with a characteristic interaction time profile. Another contribution, albeit probably less important, might come from nucleon and sub-nucleon fluctuations in the initial conditions, which we neglect in the current implementation. This may produce non-zero localized charge densities, inducing non-linear effects, which are not captured by the present approach. Finally, the solution utilized to reproduce the experimental $v_1$ slope, tilting the initial energy density distribution~\cite{Bozek:2010bi,Becattini:2015ska}, deserves further study and improvement. If we do not utilize this tilting, for LHC energies we obtain the same sign of the slope of $\Delta v_1(y)$ as seen in experiment, while for RHIC energies the slope remains unchanged.

Given the considerable effort required in refining this model and extensively testing a wider set of alternative initial conditions, we postpone this task to a future work. We believe that this current framework is an instructive first step towards the ultimate goal of a dissipative resistive chiral MHD formulation, which will require the development of a more advanced numerical framework.

\section{Summary and outlook}
\label{sec:concl}
In this study, we performed (3+1)D one fluid ideal MHD simulations of relativistic collisions of gold nuclei at $\snn=200\,\textrm{GeV}$, for RHIC, and of lead nuclei at $\snn=2.76\,\textrm{TeV}$, for the LHC. 
We adopted a somewhat simplified approach to make possible a first systematic study of the evolution and influence of backreacting magnetic fields on common heavy-ion observables. 
Taking into account the possibility of an underestimation of the initial magnetic field, we studied the effects of amplifying the magnitude of the magnetic field components by up to a factor four. Compared to ``pure'' hydrodynamics, we found an enhancement of $v_2$ in peripheral collisions, with the greatest change found for the largest values of the magnetic field. Otherwise, this enhancement is almost negligible, in overall general agreement with Refs.~\cite{Voronyuk:2011jd,Roy:2017yvg,Stewart:2017zsu,Das:2017qfi}. At RHIC energies, this enhancement tends to be stronger, and with a sufficiently large magnetic field, it is evident already in semi-peripheral collisions. In the initialization of the magnetic field $\vec{B}$, we took into account also a contribution of chiral origin, although we neglected any fluctuations or dissipation of axial charge. We studied how different values of the electric and chiral magnetic conductivities of the medium in the pre-equilibrium phase can impact the final $v_n$. We found that a simple modification of these parameters is not sufficient to produce significant changes in $v_{1}$ and $v_{2}$. However, the initialization of the electromagnetic fields, based on the simple, but unrealistic assumption of constant scalar conductivities for all time~\cite{Amato:2013naa,Aarts:2014nba,Steinert:2013fza,Puglisi:2014sha,Greif:2014oia,Greif:2016skc,Hattori:2016cnt,Hammelmann:2018ath}, should be replaced by more sophisticated modeling. 

We then computed the electric charge density in the comoving frame of the fluid. When initializing the magnetic field with a contribution originating from a pre-equilibrium CME, represented in terms of a chiral conductivity $\sigma_\chi$, we detected an electric charge separation in the direction orthogonal to the reaction plane. However, since this is not a quantity that can be directly measured in the experiments, it is necessary to translate this charge imbalance into an observable which can be experimentally evaluated. The ideal MHD assumption prevents a simple modification of the Cooper-Frye prescription to take into account the different electric charges of the hadrons, therefore we have presented an alternative method based on electric charge conservation, whereby an effective electric charge chemical potential is determined and used to ascertain the charge dependent pion spectra. We observed splitting between the directed flow of positive and negative pions that is of the same order of magnitude as recent experimental results~\cite{Margutti:2017lup}, however with a different ordering of $\pi^+$ compared to $\pi^-$. We note however that our results are in qualitative agreement with a number of recent theoretical calculations ~\cite{Gursoy:2018yai,Gursoy:2014aka,Das:2016cwd,Chatterjee:2018lsx,Chatterjee:2019xtg,Coci:2019nyr}. It is therefore a challenge to all of these descriptions to understand the physical mechanism which may result in this different ordering. Finally, within our simplified setup we were able to compute a non-zero average value of the sine of the emission angle of the particles with respect to the reaction plane, a necessary signature of the CME. This motivates further refinements of our model, in hopes of making more quantitative predictions of the CME.

The most obvious improvement to our approach is the inclusion of resistive effects, already present in many relativistic MagnetoHydroDynamic codes for astrophysics ~\cite{Komissarov:2007wk,Bucciantini:2012sm,Tomei:2019zpj,DelZanna:2016uiq,Dionysopoulou:2012zv,Mignone:2019ebw,Porth:2019wxk}, together with the viscous hydrodynamic corrections (these exist already in ECHO-QGP~\cite{DelZanna:2013eua}, but are presently kept separate from the MHD module). From the theory side, there are recent advancements towards a deeper understanding of the connections between the kinetic theory and MHD~\cite{Denicol:2018rbw,Denicol:2019iyh} that can help in improving the consistency of the formalism.
For a proper quantitative investigation of the CME within the chiral-MHD framework, it will be necessary to consider the evolution of axial charges coupled with the fluid~\cite{Hirono:2014oda,Shi:2017cpu,Hattori:2017usa}, which one may be able to derive from chiral kinetic theory~\cite{Son:2012wh,Chen:2012ca,Stephanov:2012ki,Son:2012zy,Manuel:2014dza,Chen:2015gta,Hidaka:2016yjf,Mueller:2017lzw,Huang:2018wdl,Mueller:2019gjj}; a possible intermediate step could be the simplified, but still interesting, formalism illustrated in Ref.~\cite{DelZanna:2018dyb}.

\section*{Acknowledgments}
We thank F.~Becattini, A.~Beraudo, H.~Elfner, K.~Eskola, U. Gursoy, M.~Kaminski, I.~Karpenko, E.~Molnar, H.~Niemi, D.~Oliinychenko, H.~Olivares, K. Rajagopal, L.~Rezzolla, D.~Rischke, C. Shen, and L.~Tinti for fruitful discussions and useful suggestions. We thank H. Niemi for comments on the manuscript and I.~Karpenko for sharing his notes on E.M. field transformations. G. Inghirami is supported by the Academy of Finland, Project no. 297058. MM is supported by the European Research Council, grant ERC-2015-CoG-681707. This material is based on work supported by the U.S. Department of Energy, Office of Science, Office of Nuclear Physics, under Contracts No. DE-SC0012704 (D.K, Y.H, and M.M.) and No. DE-FG02-88ER40388 (D.K., M.M.), and within the framework of the Beam Energy Scan Theory (BEST) Topical Collaboration. The work of Y.H. was supported in part by the Korean Ministry of Education, Science and Technology, Gyeongsangbuk-do and Pohang City for Independent Junior Research Groups at the Asia Pacific Center for Theoretical Physics. During the initial parts of this work, M. Mace was supported by the BEST Topical Collaboration. During the initial part of this work, G. Inghirami was supported by a GSI scholarship in cooperation with the John von Neumann Institute for Computing. G. Inghirami also gratefully acknowledges past support from the HIC for FAIR and from Helmholtz Graduate School for Hadron and Ion Research. This project was supported by COST Action CA15213 ``THOR''. The computational resources were provided by the by the Center for Scientific Computing (CSC) of the Goethe University, the National Energy Research Scientific Computing Center, a DOE Office of Science User Facility supported by the Office of Science of the U.S. Department of Energy under Contract No. DE-AC02-05CH11231, by the INFN - Sezione di Firenze and by the Frankfurt Institute for Advanced Studies. Moreover, we acknowledge grants of computer capacity from the Finnish Grid and Cloud Infrastructure (persistent identifier urn:nbn:fi:research-infras-2016072533 ). 

\appendix
\section{Ideal relativistic MagnetoHydroDynamics}
\label{app:ideal_mhd}
We follow the relativistic MagnetoHydroDynamics (MHD)~\cite{bekenstein78,an89} approach, based on the conservation laws for a one-fluid current $N^\mu$ and for the \emph{total} (matter and electromagnetic fields) energy-momentum tensor of the fluid $T^{\mu\nu}$, namely
\begin{align}
& d_\mu N^\mu  =  0, \label{eq:mass} \\
& d_\mu T^{\mu\nu}  =  0 \label{eq:enmom} ,
\end{align}
with $d_{\mu}$ being the covariant derivative. 
In addition, we consider the second law of thermodynamics 
\be
d_\mu s^\mu \ge 0,
\label{eq:entropy}
\ee
where $s^\mu$ is the entropy current, and, for the electromagnetic fields, the Maxwell's equations 
\begin{align}
& d_\mu F^{\mu\nu} = - I^\nu \quad (d_\mu I^\mu = 0),  \label{eq:maxwell1} \\
& d_\mu F^{\star\mu\nu} = 0,     \label{eq:maxwell2}
\end{align}
where $F^{\mu\nu}$ is the Faraday tensor and $F^{\star\mu\nu}=\textstyle{\frac{1}{2}}\epsilon^{\mu\nu\lambda\kappa}F_{\lambda\kappa}$ is its Hodge dual. When polarization and magnetization effects are neglected, the electromagnetic contribution to the energy-momentum tensor can be expressed as:
\be
T^{\mu\nu} _\mathrm{f} = F^{\mu\lambda}F^\nu_{\,\lambda} - 
\tfrac{1}{4}g^{\mu\nu}F^{\lambda\kappa}F_{\lambda\kappa}.
\ee
In the ideal limit all dissipative fluxes can be neglected, local equilibrium is assumed and we can write a single \emph{fluid four-velocity} $u^\mu$ ($u_\mu u^\mu = -1$). After introducing the electric and magnetic fields measured in the comoving frame of the fluid:
\begin{align}
e^\mu  = F^{\mu\nu}u_\nu,  \quad & (e^\mu u_\mu =0), \label{eq:e_4vec}\\
b^\mu  = F^{\star\mu\nu}u_\nu, \quad &  (b^\mu u_\mu =0), \label{eq:b_4vec}
\end{align}
the Faraday tensor and its dual can be decomposed with respect to $u^\mu$ as
\begin{align}
F^{\mu\nu} & = u^\mu e^\nu - u^\nu e^\mu + \epsilon^{\mu\nu\lambda\kappa} b_\lambda u_\kappa, \\
F^{\star\mu\nu} & = u^\mu b^\nu - u^\nu b^\mu - 
\epsilon^{\mu\nu\lambda\kappa} e_\lambda u_\kappa.
\end{align}
Since we are dealing with electromagnetic fields strongly coupled with the fluid, we need a relation (\emph{Ohm's law}) between the electric currents and the fluid, which is adequately modeled, in many cases, by a linear dependence: 
\be
I^\mu = \tilde{\rho_\mathrm{e}} u^\mu + j^{\,\mu}; \quad j^{\,\mu} = \sigma^{\,\mu\nu} e_\nu,
\label{eq:el_current}
\ee
where $\tilde{\rho_\mathrm{e}}$ is the electric charge density in the comoving frame, $j^{\,\mu}$ the conduction current ($j^{\,\mu} u_\mu = 0$), and $\sigma^{\,\mu\nu}$ the fluid conductivity tensor. The ideal MHD approximation, adopted in this work, consists in assuming $\sigma\rightarrow\infty$, so that, to avoid the onset of infinite currents, this requires
\be
e^\mu = 0
\label{eq:ohm}
\ee
must hold, as well.
The adoption of Eq.~(\ref{eq:ohm}) brings many simplifies the Faraday tensor and its dual, as well as the structure of the evolution equations; at the end are given by the following system (see Ref.~\cite{Inghirami:2016iru} for details): 
\begin{align}
& d_\mu ( n u^\mu) = 0,  \label{eq:mhd1} \\
& d_\mu [(e + p + b^2) u^\mu u^\nu +  ( p + \tfrac{1}{2} b^2) g^{\mu\nu} - b^\mu b^\nu] = 0,  \label{eq:mhd2} \\
& d_\mu ( u^\mu b^\nu - u^\nu b^\mu ) = 0  \label{eq:mhd3} , 
\end{align}
in the unknowns $n$, $e$, $p$, $u^\mu$, and $b^\mu$, all defined in the comoving frame of the fluid. In Eqs.~(\ref{eq:mhd1}-\ref{eq:mhd3}), $n= - N^\mu u_\mu$ is the baryon density, $e = T^{\mu\nu}_\mathrm{m}u_\mu u_\nu $ the fluid energy density, and $p = \tfrac{1}{3} \Delta_{\mu\nu} T^{\mu\nu}_\mathrm{m}$ the kinetic pressure, $\frac{1}{2}b^2 = \frac{1}{2} b_\mu b^\mu$ the energy density of the magnetic field, and $g^{\mu\nu}$ is the metric tensor.

However, to solve numerically the system of equations (\ref{eq:mhd1}-\ref{eq:mhd3}), we need to perform a $3+1$ splitting~\cite{GourgoulhonFormalismGeneralRelativity2012,alcubierre2008introduction} of time and spatial components and rewrite the equations in conservative form~\cite{laney_computational_1998}. For more details on this procedure, see Refs.~\cite{ldz03,ldz07,DelZanna:2013eua}; here we summarize the most important steps.
We consider two possible metrics: diag($g_{\mu\nu}$)=$(-1,1,1,1)$ for Minkowski space-time or diag($g_{\mu\nu}$)=$(-1,1,1,\tau^2)$  for Milne (Bjorken) space-time. Given the fluid velocity $v^i$ for an Eulerian observer in the laboratory frame, the fluid four velocity can be expressed as
\be
u^\mu = ( \gamma , \gamma v^i ), 
\ee
where $\gamma = (1-v^2)^{-1/2}$ is the Lorentz factor of the bulk flow and $v^2=v_k v^k$. We now consider the three spatial components of the electric field $E^i$ and magnetic field $B^i$ as measured in the laboratory frame, which are related to the four vectors $e^{\mu}$ and $b^{\mu}$ by:
\be
e^\mu = ( \gamma v_kE^k, \gamma E^i + \gamma \varepsilon^{ijk} v_jB_k),
\ee
\be
b^\mu = ( \gamma v_kB^k, \gamma B^i - \gamma \varepsilon^{ijk} v_jE_k),
\ee
where $\varepsilon_{ijk}$ is the Levi-Civita pseudo-tensor of the spatial three-metric, i.e. $\varepsilon_{ijk} = \vert g \vert^{\frac{1}{2}} [ijk]$, with $g=\mathrm{det}\{g_{\mu\nu}\}=-\mathrm{det}\{g_{ij}\} < 0$ and $[ijk]$ the fully antisymmetric Levi-Civita symbol. It can be easily verified that Eq.~(\ref{eq:ohm}) becomes
\be
E_i = - \varepsilon_{ijk}v^jB^k, \label{eq:ohm2}
\ee
therefore in ideal MHD the electric field is a not an independent quantity, but is derived from $v^i$ and $B^i$. 
We can then rewrite Eqs.~(\ref{eq:mhd1}-\ref{eq:mhd3}) in a form appropriate for numerical integration by clearly separating time and space derivatives and tensor components:
\be
\label{eq:echo}
\partial_0 {\bf U}+ \partial_i {\bf F}^i ={\bf S}, 
\ee
where
\be
{\bf U} \! = \! \detmet \! \left(\begin{array}{c}
	\gamma n \\
	S_j \equiv T^0_{\,j}\\
	\mathcal{E} \equiv - T^0_{\,0} \\
	B^j
\end{array}\right), \,
{\bf F}^i \! = \! \detmet \! \left(\begin{array}{c}
	\gamma n v^i \\
	T^i_{\,j}\\
	S^i \equiv - T^i_{\, 0} \\
	v^iB^j - B^iv^j
\end{array}\right)
\label{eq:fluxes}
\ee
are respectively the set of \emph{conservative variables} and \emph{fluxes}, 
while the source terms are given by
\be
{\bf S}=\detmet \left(\begin{array}{c}
	0 \\
	\tfrac{1}{2} T^{ik}\partial_j g_{ik} \\
	- \tfrac{1}{2} T^{ik}\partial_0 g_{ik} \\
	0
\end{array}\right).
\label{eq:sources}
\ee
The components of energy momentum tensor $T^{\mu\nu}$ are:
\begin{align}
S_i = & (e+p)\gamma^2v_i  + \varepsilon_{ijk}E^jB^k , \label{eq:S} \\
T_{ij} = & (e+p)\gamma^2 v_iv_j + (p + u_\mathrm{em} )g_{ij}\nonumber \\
&-E_iE_j - B_iB_j ,  \\
\mathcal{E} = & (e+p)\gamma^2 - p + u_\mathrm{em} \label{eq:E}, 
\end{align}
in which $u_\mathrm{em}=\tfrac{1}{2}(E^2+B^2)$ is the energy density.

Finally, the solenoidal condition
\be
\partial_i (\vert g \vert^{\frac{1}{2}} B^i) = 0,
\label{eq:solenoidal}
\ee
coming from the time component of Eq.~(\ref{eq:mhd3}), is enforced by using a modified version of the hyperbolic divergence cleaning method~\cite{dedner02,Palenzuela21042009,Mignone20102117,penner11,moesta14,Dionysopoulou:2012zv}. This procedure has been validated through several tests~\cite{Inghirami:2016iru}. However, we verified that using or not the hyperbolic divergence cleaning method does not have a noticeable impact in many results presented in this paper, although its use has a positive influence on the stability of the code. Nevertheless, since this method controls the error on the divergence of $\vec{B}$ by transporting it away with a dumping effect, it might introduce some spurious effects in the computation of the electric charge density and, therefore, we did not use it in the simulations in which this quantity was evaluated. In the future, it may be desirable to use the Constrained Transport approach~\cite{lond04,Mignone:2019ebw}.

\section{Initialization of the energy density distribution}
\label{app:initial_edens}
We define the \emph{thickness} function: 
\bea
T(x,y) &=&\int_{-\infty}^{\infty} \di z \, n(x,y,z)\nonumber \\
&=& \int_{-\infty}^{\infty} \di z \, 
\frac{n_0}{1+\e^{(\sqrt {x^2+y^2+z^2}- R)/\delta}} \eea
where $n_0$ is the nuclear density, $\delta$  the width and $R$ the radius of the nuclear Fermi distribution, which depends on the specific nucleus. Then, we define the following functions:
\bea
T_1({\bf x}_T) &=& T_+({\bf x}_T)\,\left (1-\left ( 1-\dfrac{\sigma T_-({\bf x}_T)}{A}\right )^A \right ) \\
T_2({\bf x}_T) &=& T_-({\bf x}_T)\,\left (1-\left ( 1-\dfrac{\sigma T_+({\bf x}_T)}{A}\right )^A \right )
\eea
where $\sigma$ is the inelastic NN cross section, $A$ the mass number of the colliding 
nuclei, and
\be\label{thick}
T_+({\bf x}_T) = T({\bf x}_T+{\bf b}/2)\,,\,\, T_-({\bf x}_T) = T({\bf x}_T-{\bf b}/2)
\ee
where ${\bf x}_T=(x,y)$ is the vector of the transverse plane coordinates and 
${\bf b}$ is the impact parameter vector, connecting the centers of the two nuclei. 
The function $W_N$, which gives the contribution of the \emph{wounded} nucleons, is then simply:
\be\label{wound}
W_N(x,y,\eta)=T_1(x,y)+T_2(x,y)\,.
\ee
The initial proper energy density distribution is given by:
\be\label{inidens}
\varepsilon(x,y,\eta)=\varepsilon_0 \,W(x,y,\eta)\,H(\eta),
\ee
where the total weight function $W(x,y,\eta)$ is defined as:
\be\label{weight}
W(x,y,\eta)=\dfrac{(1-\alpha)\,W_{N}(x,y,\eta)+\alpha\,n_{BC}(x,y)}{(1-\alpha)
	\,W_{N}(0,0,0)+\alpha\,n_{BC}(0,0)\Big|_{{\bf b}=0}}, 
\ee
where
$n_{BC}(x,y)$ is the mean number of binary collisions:
\be
n_{BC}(x,y)={\sigma}_{in} T_+(x,y)\,T_-(x,y)
\ee
and $\alpha$ is the {\em collision hardness} parameter, which can vary between 
0 and 1. The energy density profile in the longitudinal direction is modulated by 
\be\label{z-profile}
H(\eta)=\exp\left(-\dfrac{\tilde\eta^2}{2\sigma_{\eta}^2}\right)
\qquad  \tilde\eta=|\eta-\eta_0|-\eta_{flat}/2,
\ee
where 
\be \label{eq:eta0_shifted}
\eta_0=\frac{1}{2}\frac{(T_{+} + T_{\_})\gamma_{beam}+ (T_{+} - T_{\_})\gamma_{beam}\beta_{beam}}{(T_{+} + T_{\_})\gamma_{beam}- (T_{+} - T_{\_})\gamma_{beam}\beta_{beam}},
\ee
while $\eta_{flat}$ and $\sigma_{\eta}$ are parameters.

\begin{figure*}
	\begin{minipage}[b]{0.48\textwidth}
		\includegraphics[width=1\textwidth]{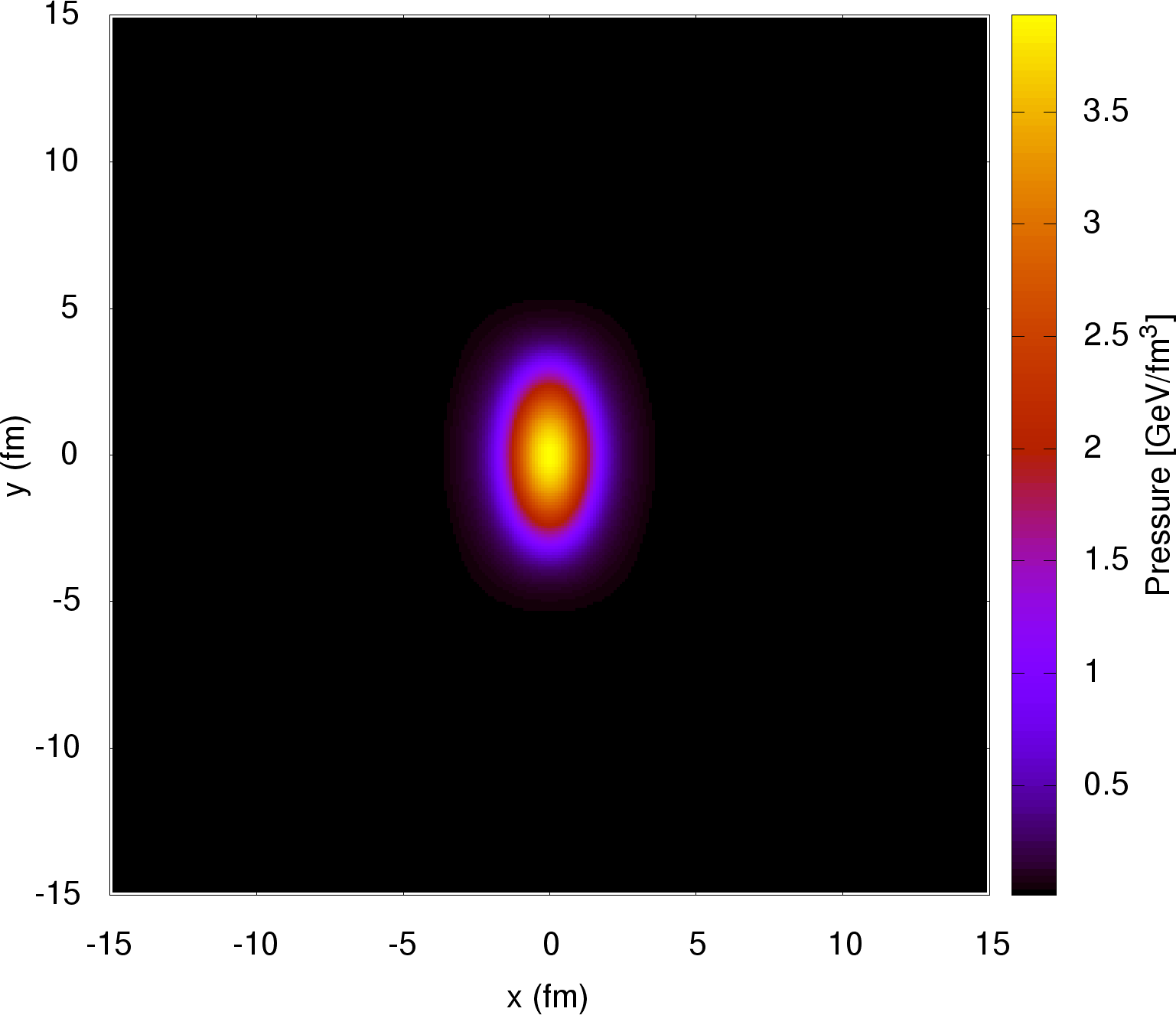}
	\end{minipage}	
	\begin{minipage}[b]{0.48\textwidth}
		\includegraphics[width=1\textwidth]{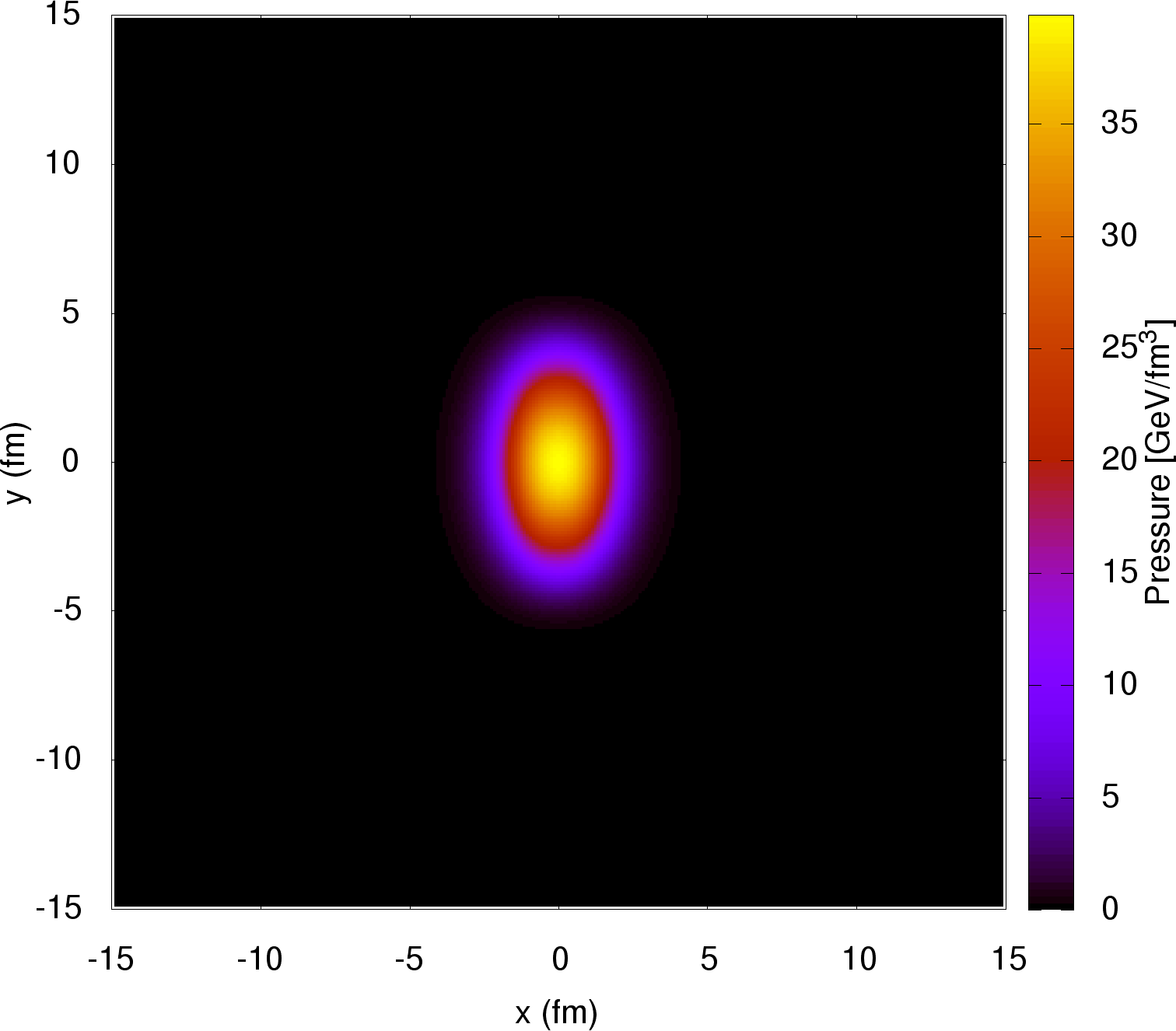}
	\end{minipage}
	\caption{Initial pressure distribution along the transverse plane at $\eta=0$ for collisions with impact parameter $b=12\,\textrm{fm}$. The left figure refers to Au+Au collisions at $\snn=200\,\textrm{GeV}$ at $\tau_0=0.4\,\textrm{fm/c}$, the right figure refers to Pb+Pb collisions at $\snn=2.76\,\textrm{TeV}$ at $\tau_0=0.2\,\textrm{fm/c}$.}
	\label{initial_pr}
\end{figure*}

\subsection{Tilted initial energy density distribution}
\label{tilt-desc}
In some cases, in order to obtain a directed flow in a better agreement with the experimental results~\cite{Bozek:2010bi}, we modify 
Eq.~(\ref{wound}) and redefine the wounded nucleons weight function $W_N$ as:
\be\label{wound_tilted}
W_N(x,y,\eta)=2\,\left( T_1(x,y)f_{-}(\eta)+T_2(x,y)f_{+}(\eta)\right) \nonumber \\
\ee
where:
$$
f_{-}(\eta)=
\begin{cases}
1 & \eta < -\eta_m\\
\dfrac{-\eta+\eta_m}{2\eta_m} & -\eta_m \le \eta \le \eta_m\\
0 & \eta > \eta_m
\end{cases}
$$
and
$$
f_{+}(\eta)=
\begin{cases}
0 & \eta < -\eta_m\\
\dfrac{\eta+\eta_m}{2\eta_m} & -\eta_m \le \eta \le \eta_m\\
1 & \eta > \eta_m
\end{cases}
,
$$

while Eq.~(\ref{z-profile}) becomes:
\be\label{profile_tilted}
H(\eta)=\exp\left(-\dfrac{\tilde\eta^2}{2\sigma_{\eta}^2}\theta(\tilde \eta)\right)
\qquad  \tilde\eta=|\eta|-\eta_{flat}/2.
\ee

\begin{figure*}
	\begin{minipage}[b]{0.48\textwidth}
		\includegraphics[width=1\textwidth]{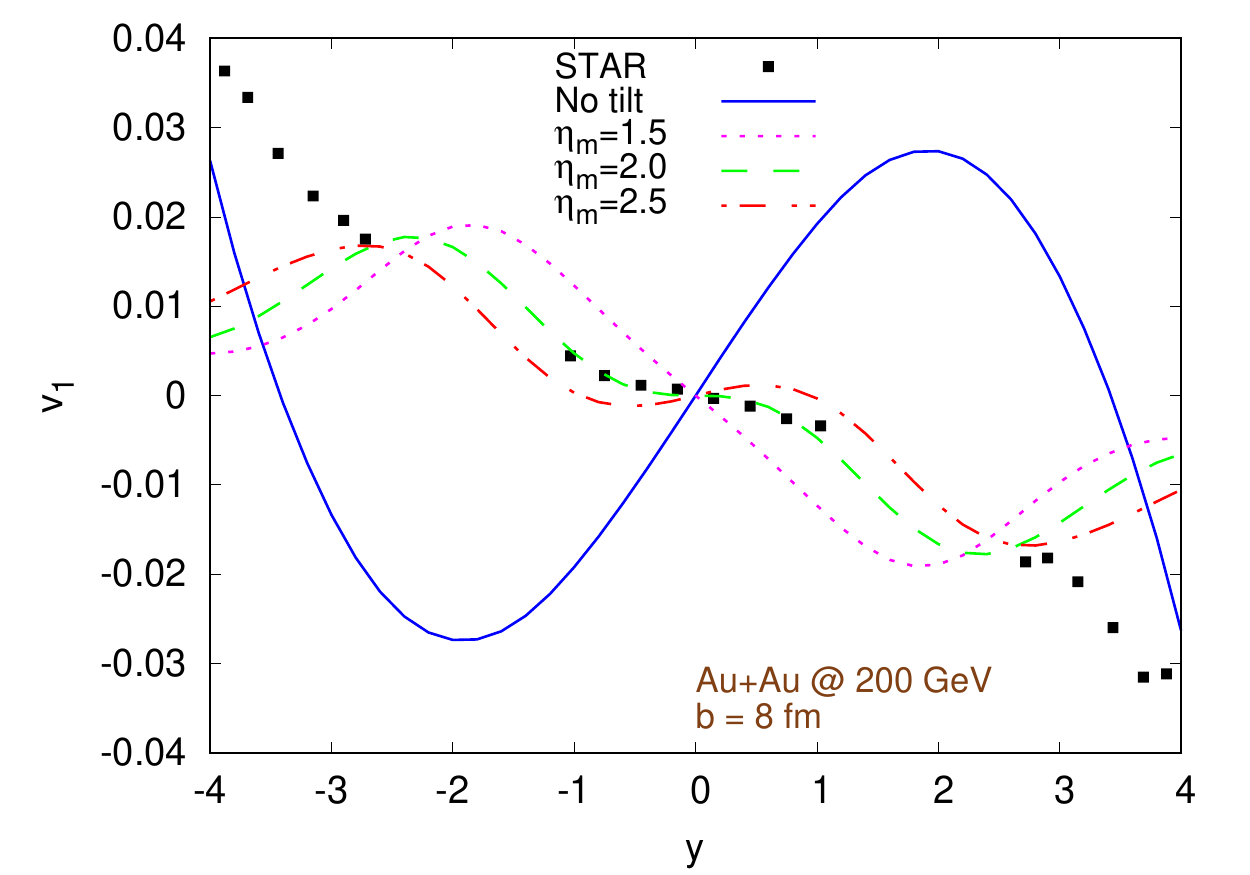}
	\end{minipage}	
	\begin{minipage}[b]{0.48\textwidth}
		\includegraphics[width=1\textwidth]{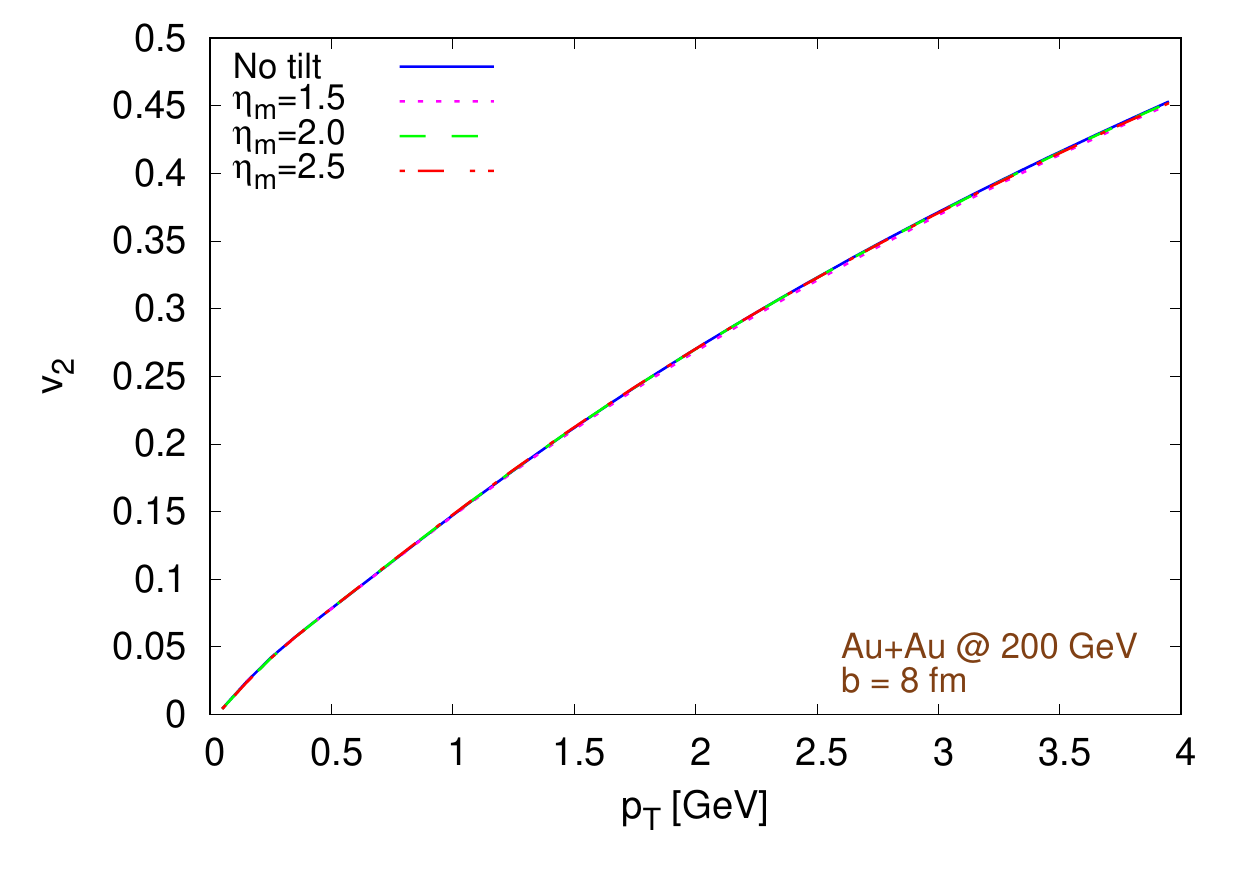}
	\end{minipage}
	\caption{Comparison between the outcomes of simulations with an initial energy density distribution non tilted and tilted by using different values of the parameter $\eta_m$. Left: directed flow $v_1$ versus rapidity $y$. Right: elliptic flow $v_2$ versus transverse momentum $p_T$. Both figures refer to Au+Au collisions at $\snn=200\,\textrm{GeV}$ with impact parameter $b=8\,\fm$.}
	\label{etam_effects}
\end{figure*}

Essentially, this kind of initialization introduces a \emph{tilting} in the initial energy density distribution, regulated by the parameter $\eta_m$ (see Ref.~\cite{Bozek:2010bi} for details). As the left side  of Fig.~(\ref{etam_effects}) shows, this tilting changes significantly the directed flow of pions and, with $\eta_m=2$, it is able to reproduce quite well the experimental data by the STAR collaboration~\cite{Abelev:2008jga} in the central rapidity region, although, of course, for a proper comparison with experimental data viscous effects, hadronic rescattering and resonance decay feed-down should be included in the model as well. However, as shown in the right side of Fig.~(\ref{etam_effects}), the elliptic flow at mid-rapidity seems not to be influenced by the tilting, at least for the values of $\eta_m$ that we tested. Given the weak sensitivity of $v_2$ to the tilting of the initial energy density distribution, we employed this initialization only in a few cases regarding $v_1$, in particular when studying charge dependent spectra, like in Figs.~(\ref{v1_charged},\ref{v2_charged}).

\section{Effects of the energy density pedestal on $v_1$ and $v_2$}
\begin{figure*}
	\begin{minipage}[b]{0.48\textwidth}
		\includegraphics[width=1\textwidth]{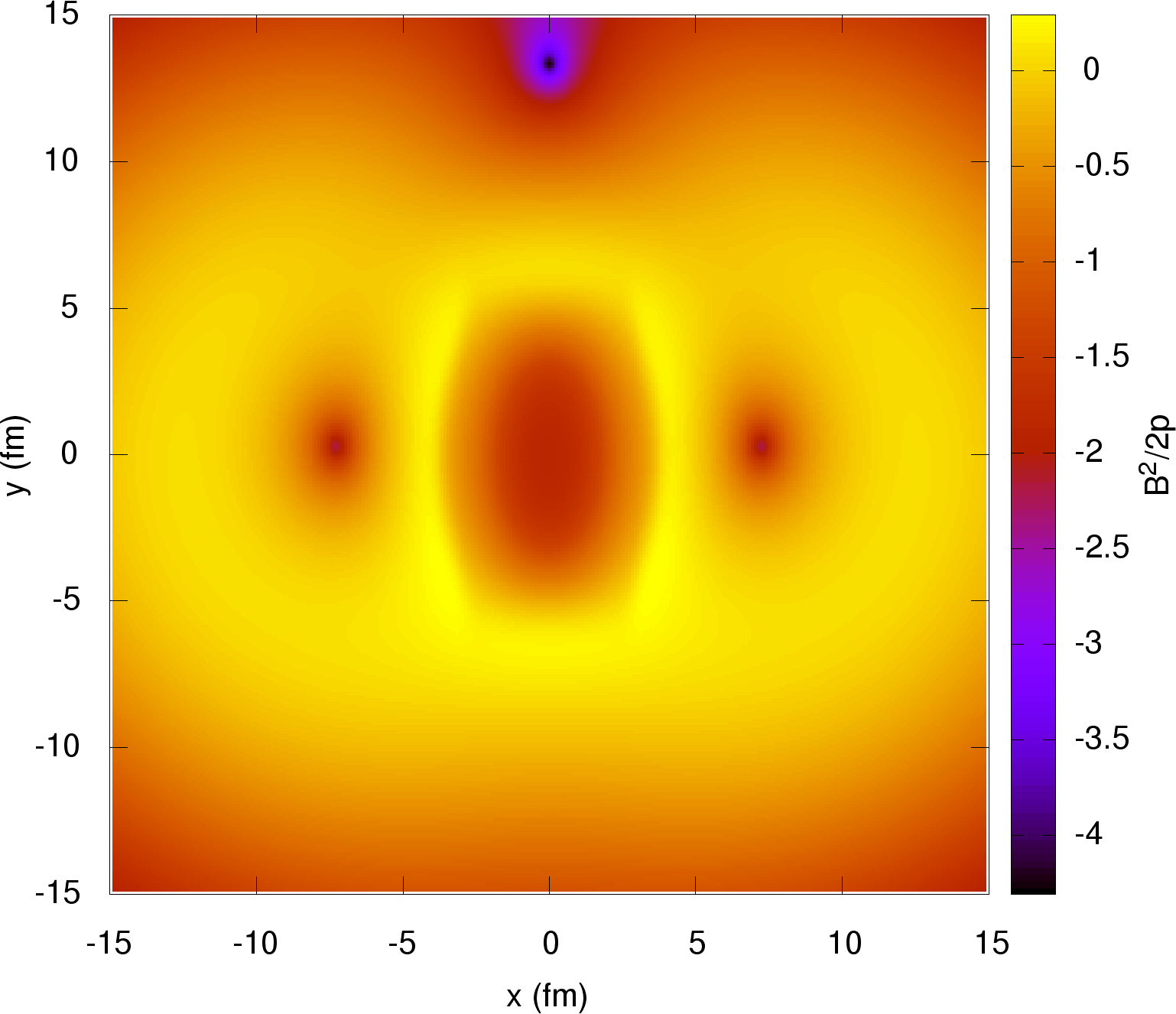}
	\end{minipage}
	\hspace{3mm}
	\begin{minipage}[b]{0.48\textwidth}
		\includegraphics[width=1\textwidth]{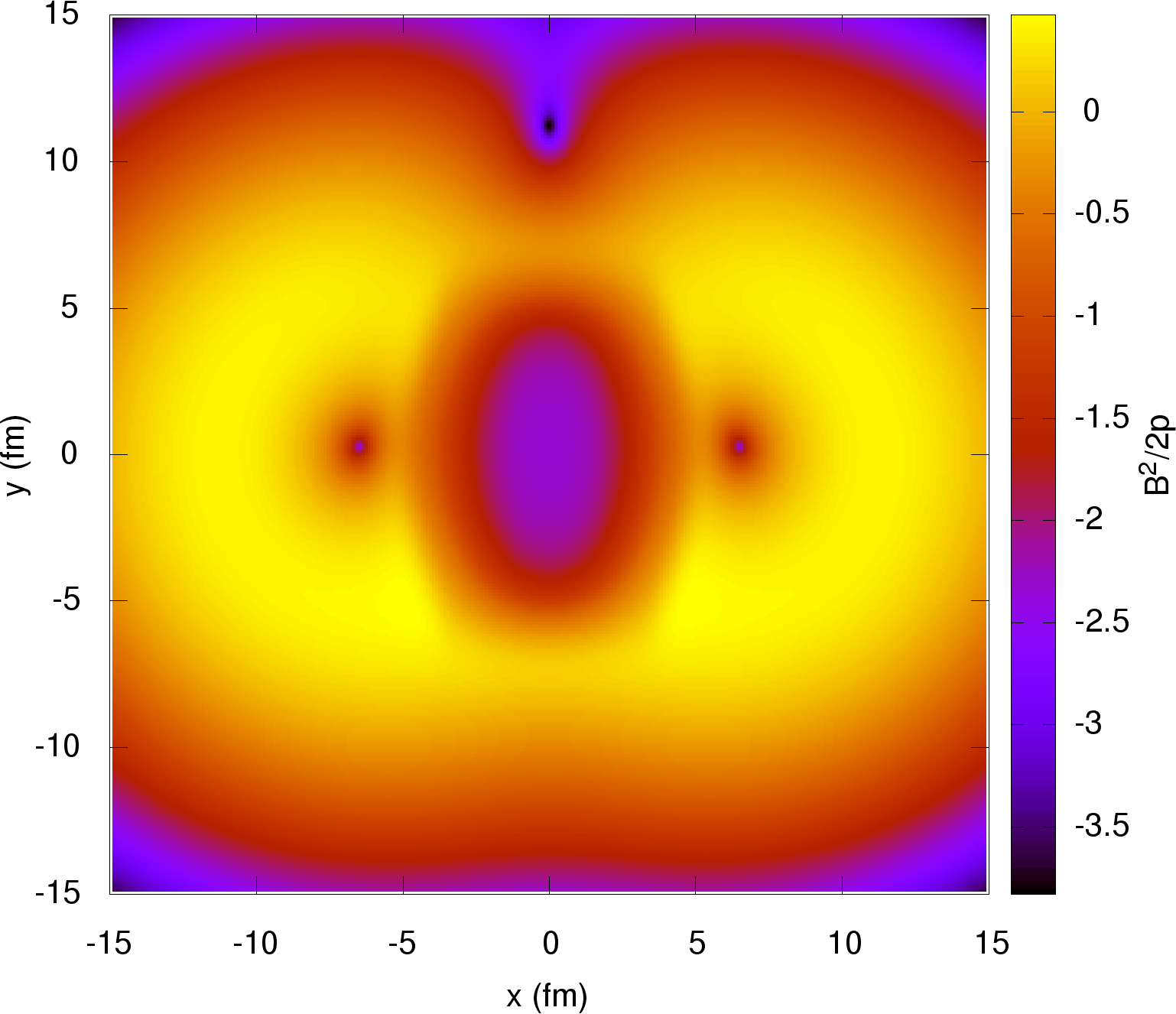}
	\end{minipage}
	\caption{Ratio between the initial magnetic pressure and thermal pressure $B^2/(2p)$. Left figure:  Au+Au collisions at $\snn=200\,\textrm{GeV}$, $\tau_0=0.4,\textrm{fm/c}$. Right figure: Pb+Pb collisions at $\snn=2.76\,\textrm{TeV}$, $\tau_0=0.2,\textrm{fm/c}$. Both figures refer to collisions with an impact parameter $b=12\,\textrm{fm}$, the magnetic field is computed assuming a medium with constant electrical conductivity $\sigma=5.8\,\textrm{MeV}$ and a chiral magnetic conductivity $\sigma_{\chi}=1.5\,\textrm{MeV}$.}
	\label{ratio_mag_th_pressure}
\end{figure*}

\begin{figure*}
\begin{minipage}[b]{0.48\textwidth}
		\includegraphics[width=1\textwidth]{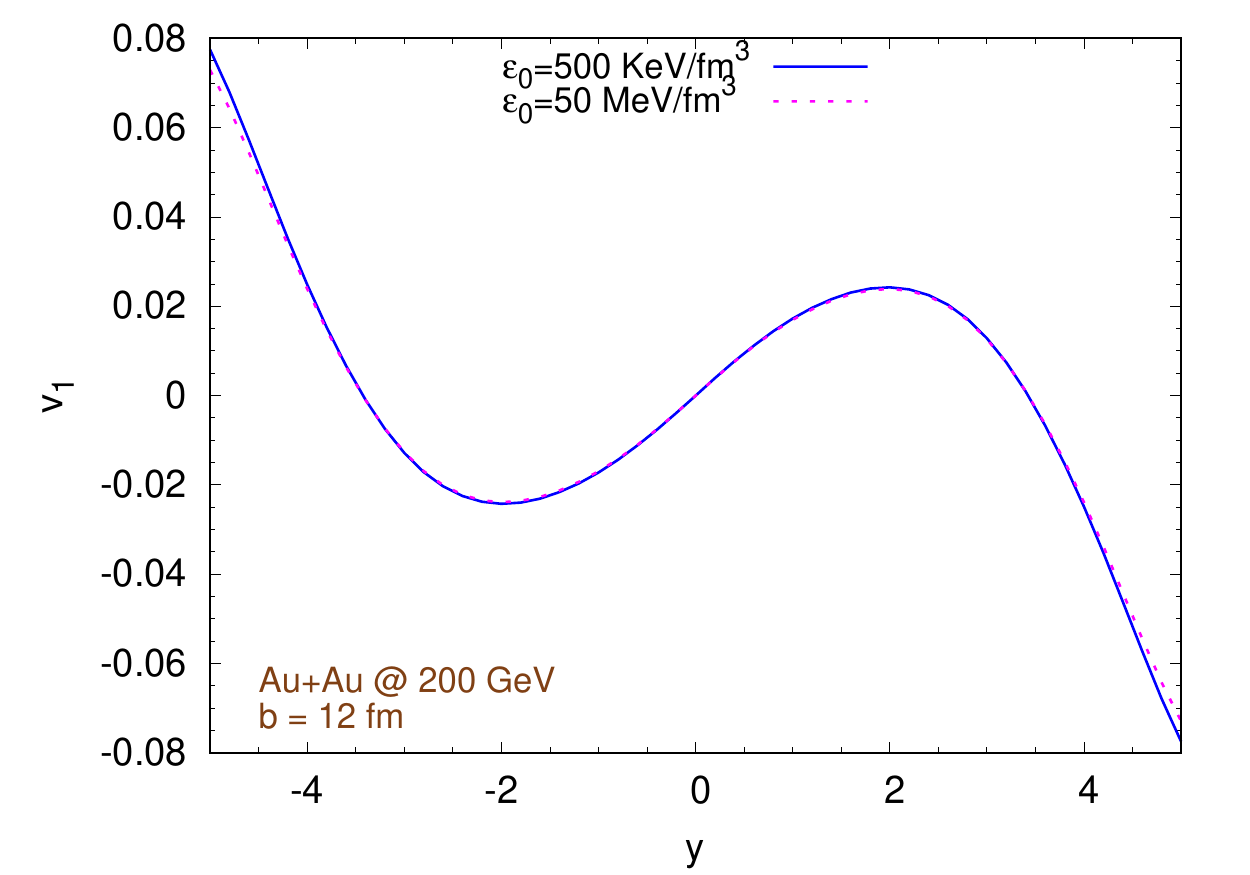}
	\end{minipage}
	\hspace{3mm}
	\begin{minipage}[b]{0.48\textwidth}
		\includegraphics[width=1\textwidth]{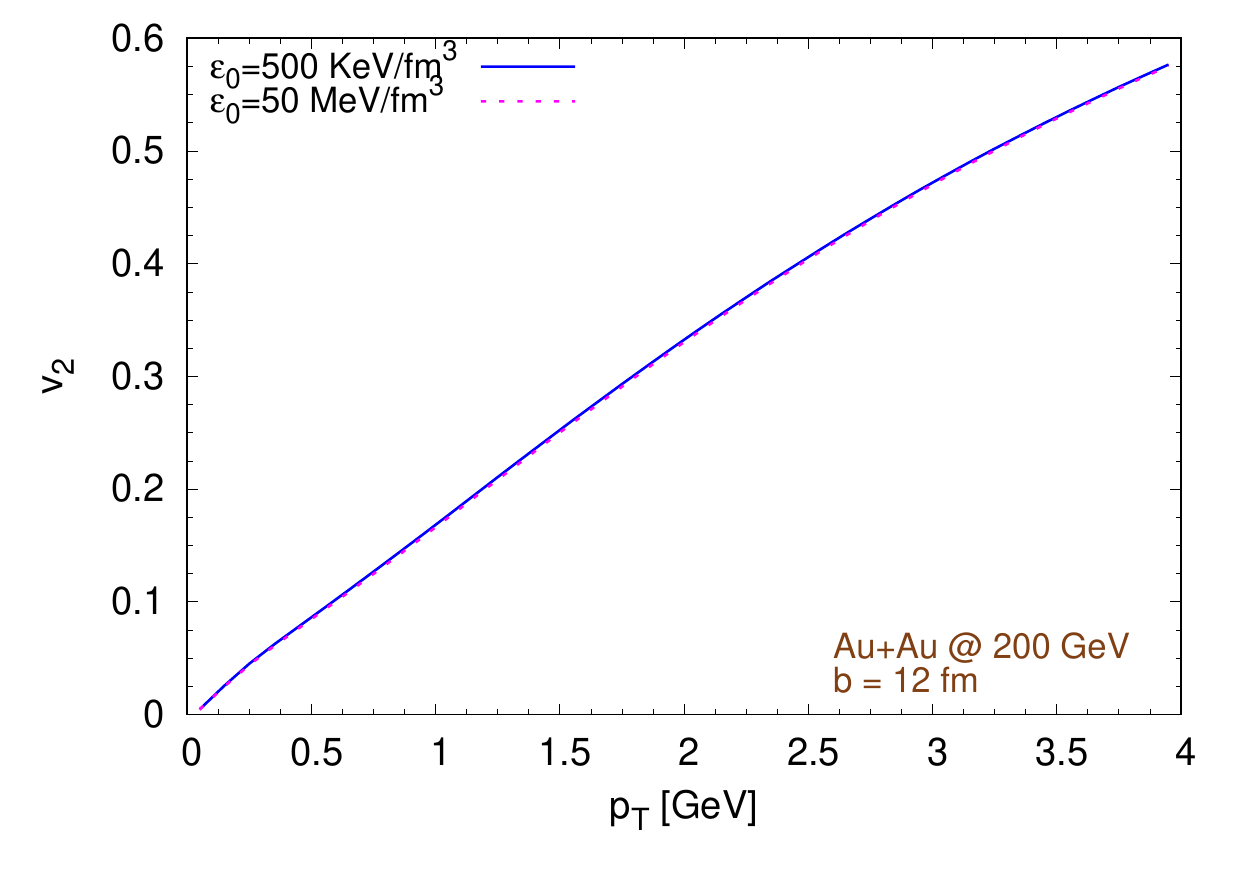}
	\end{minipage}
	\caption{Evaluation of the impact on the energy density pedestal introduced to stabilize the code on the Fourier harmonics $v_{1,2}$. The figures refer to Au+Au collisions at $\snn=200\,\textrm{GeV}$ with an impact parameter $b=12\,\textrm{fm}$ and without magnetic fields. Left figure: integrated directed flow of charged pions vs rapidity. Right figure: elliptic flow of charged pions at mid-rapidity vs transverse momentum. We can notice how at RHIC energies the high energy density background has a very limited effect on $v_2$ at mid-rapidity, while we can notice some differences in $v_1$ for $|y| > 4$.}\label{check_minen_RHIC}
\end{figure*}

\begin{figure*}
	\begin{minipage}[b]{0.48\textwidth}
		\includegraphics[width=1\textwidth]{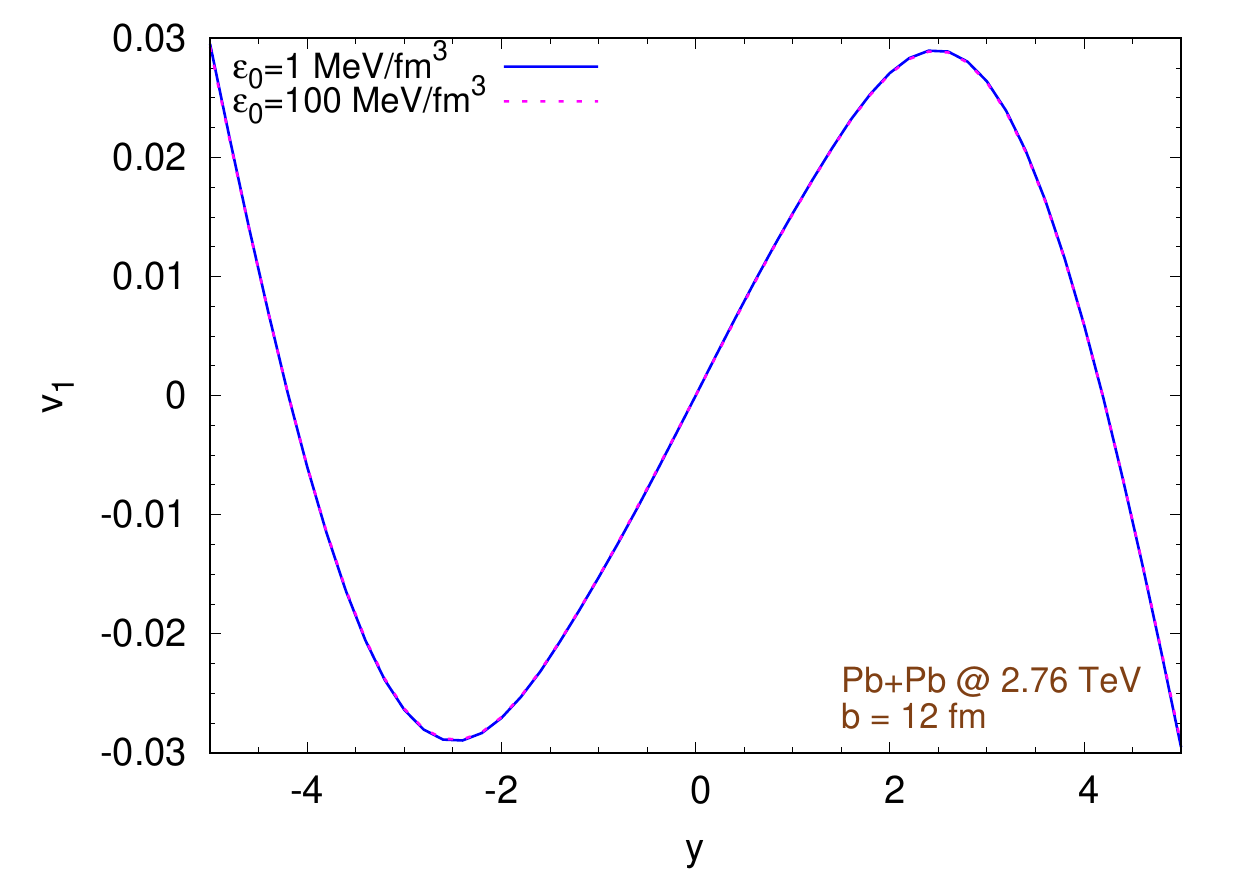}
	\end{minipage}
	\hspace{3mm}
	\begin{minipage}[b]{0.48\textwidth}
		\includegraphics[width=1\textwidth]{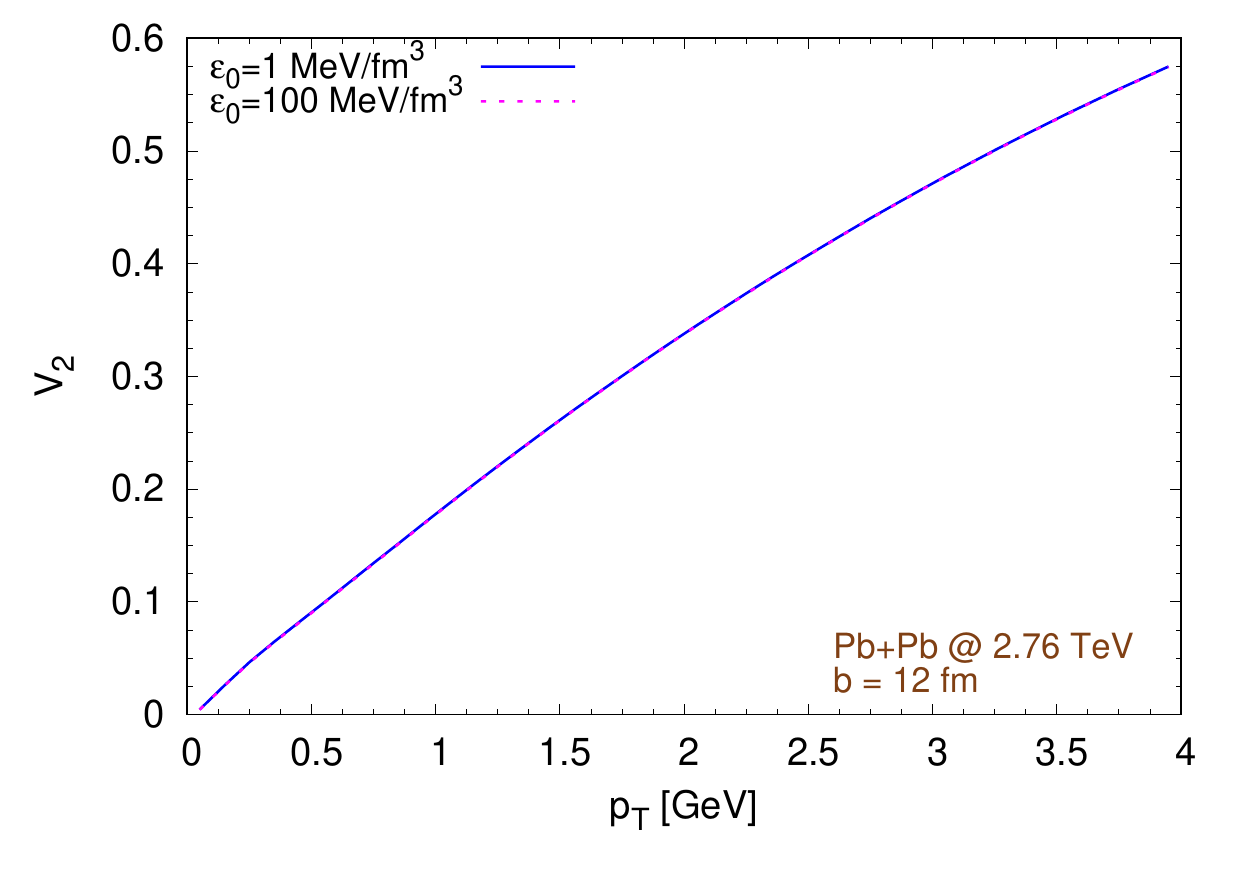}
	\end{minipage}
	\caption{Evaluation of the impact of the energy density pedestal introduced to stabilize the code on the Fourier harmonics $v_{1,2}$. The figures refer to Pb+Pb collisions at $\snn=2.76\,\textrm{TeV}$ with an impact parameter $b=12\,\textrm{fm}$ and without magnetic fields. Left figure: integrated directed flow of charged pions vs rapidity. Right figure: elliptic flow of charged pions at mid-rapidity vs transverse momentum. We can notice how at LHC energies the high energy density background has a very limited effect on the results, at least in the ranges shown in the figures.}
	\label{check_minen_LHC}
\end{figure*}

At the beginning of the simulation, the energy density created by the collision between two nuclei is very high inside the fireball at the center of the grid, but, at least in principle, it should be zero outside of it (see, e.g., Fig.~(\ref{initial_pr})). However, we cannot run hydro simulations in the void, therefore we must add a minimum energy density pedestal. The magnetic field, instead, has a large magnitude even in the space regions outside the fireball, as can be noticed in Fig.~(\ref{initial_B_RHIC_b12}). When the magnetic pressure is much larger than the thermal pressure, the code can easily crash. To avoid this situation, the artificial additional energy density pedestal must be considerably larger than in the pure hydro case. Fig.~(\ref{ratio_mag_th_pressure}) shows the ratio between the initial magnetic pressure and thermal pressure. It is important to verify that this larger minimum energy density layer does not change appreciably the Fourier harmonics $v_n$ compared to the case when a much smaller value is used. We make the test in pure ideal hydrodynamical simulations, assuming that the situation does not significantly change in the MHD case, and we look at the $p_T$ integrated directed flow vs rapidity and at the elliptic flow at mid-rapidity vs transverse momentum of charged pions. At RHIC energies, whose results are shown in Fig.~(\ref{check_minen_RHIC}), we can notice a small, but appreciable deviation in $v_1(y)$ for $|y| > 4$ (left), while the effects are almost negligible for $v_2(p_T,y=0)$. At LHC energies, whose results are shown in Fig.~(\ref{check_minen_LHC}), the effects are almost negligible both for $v_1$ and $v_2$.

\section{Effects of the reconstruction algorithm on $v_1$ and $v_2$}
\begin{figure*}[ht!]
	\begin{minipage}[b]{0.48\textwidth}
		\includegraphics[width=1\textwidth]{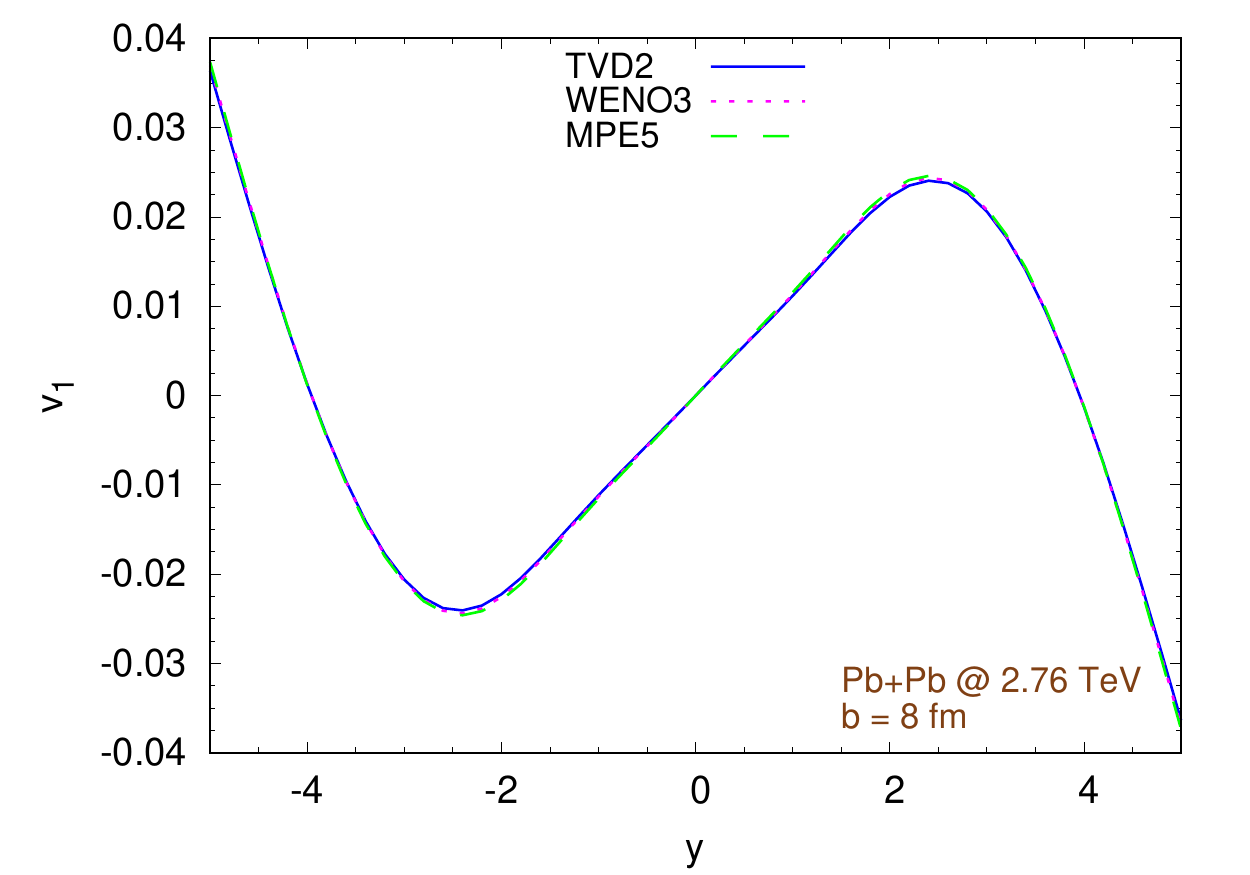}
	\end{minipage}
	\hspace{3mm}
	\begin{minipage}[b]{0.48\textwidth}
		\includegraphics[width=1\textwidth]{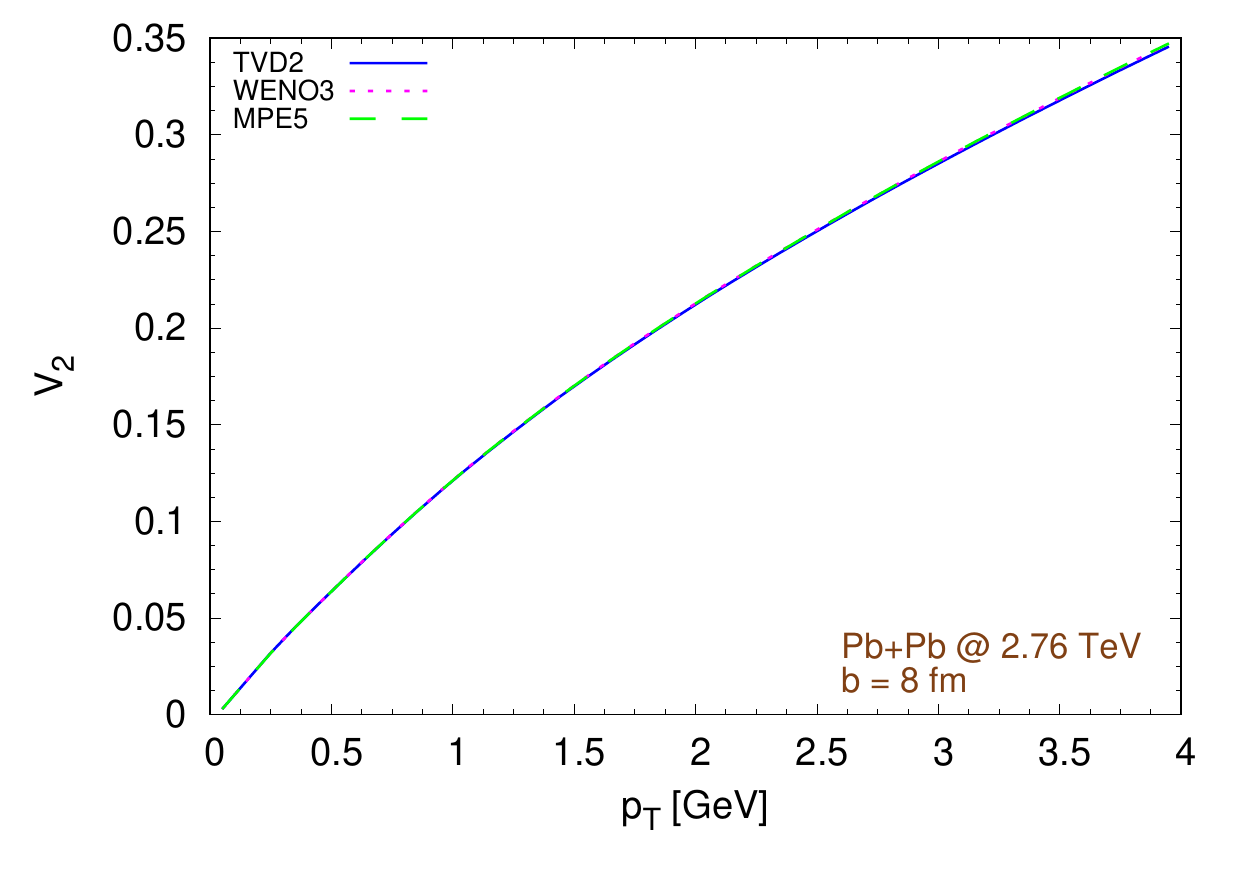}
	\end{minipage}
	\caption{Evaluation of the impact of choice of the reconstruction algorithm on Fourier harmonics $v_{1,2}$. The figures refer to Pb+Pb collisions at $\snn=2.76\,\textrm{TeV}$ with an impact parameter $b=8\,\textrm{fm}$ and without magnetic fields. Left figure: integrated directed flow of charged pions vs rapidity. Right figure: elliptic flow of charged pions at mid-rapidity vs transverse momentum.}
	\label{check_rec_algo_LHC}
\end{figure*}

ECHO-QGP inherits from the ECHO code many high order reconstruction algorithms. In general lower order algorithms tends to be more diffusive, but, when stability is a problem, this weakness becomes an advantage in term of robustness of the code. In this work we use the second order TVD2 algorithm. In this section we check that, for our purposes, the choice of the reconstruction algorithm does affect significantly the final results. This statement is demonstrated by Fig.~(\ref{check_rec_algo_LHC}), showing the integrated directed flow vs rapidity (left) and the elliptic flow at mid-rapidity vs transverse momentum (right) of charged pions produced in Pb+Pb collisions at $\snn=2.76\,\textrm{TeV}$ with an impact parameter $b=8\,\textrm{fm}$, neglecting the effects of magnetic field (pure ideal hydro simulations).

\section{Effects of the freeze-out temperature}
\begin{figure*}[ht!]
	\begin{minipage}[b]{0.48\textwidth}
		\includegraphics[width=1\textwidth]{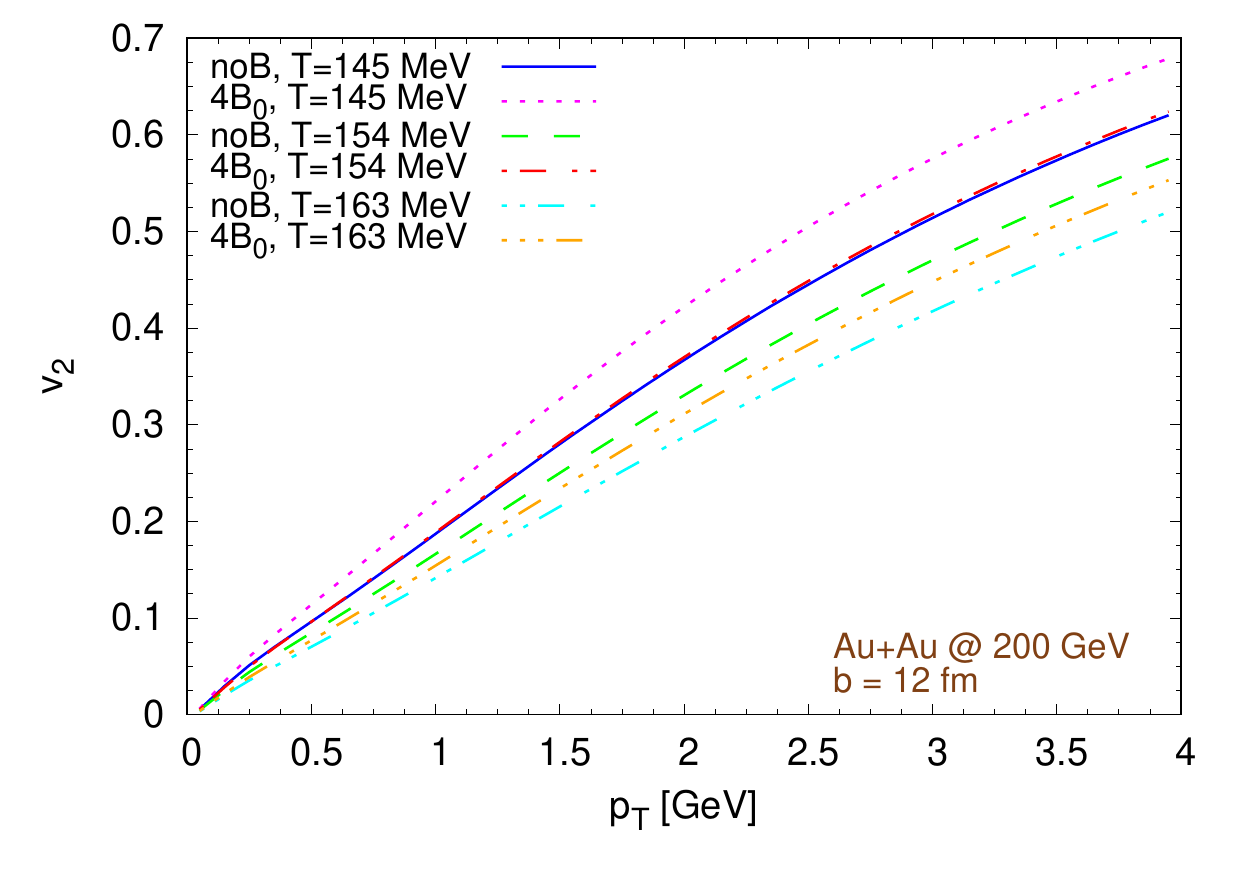}
	\end{minipage}
	\hspace{3mm}
	\begin{minipage}[b]{0.48\textwidth}
		\includegraphics[width=1\textwidth]{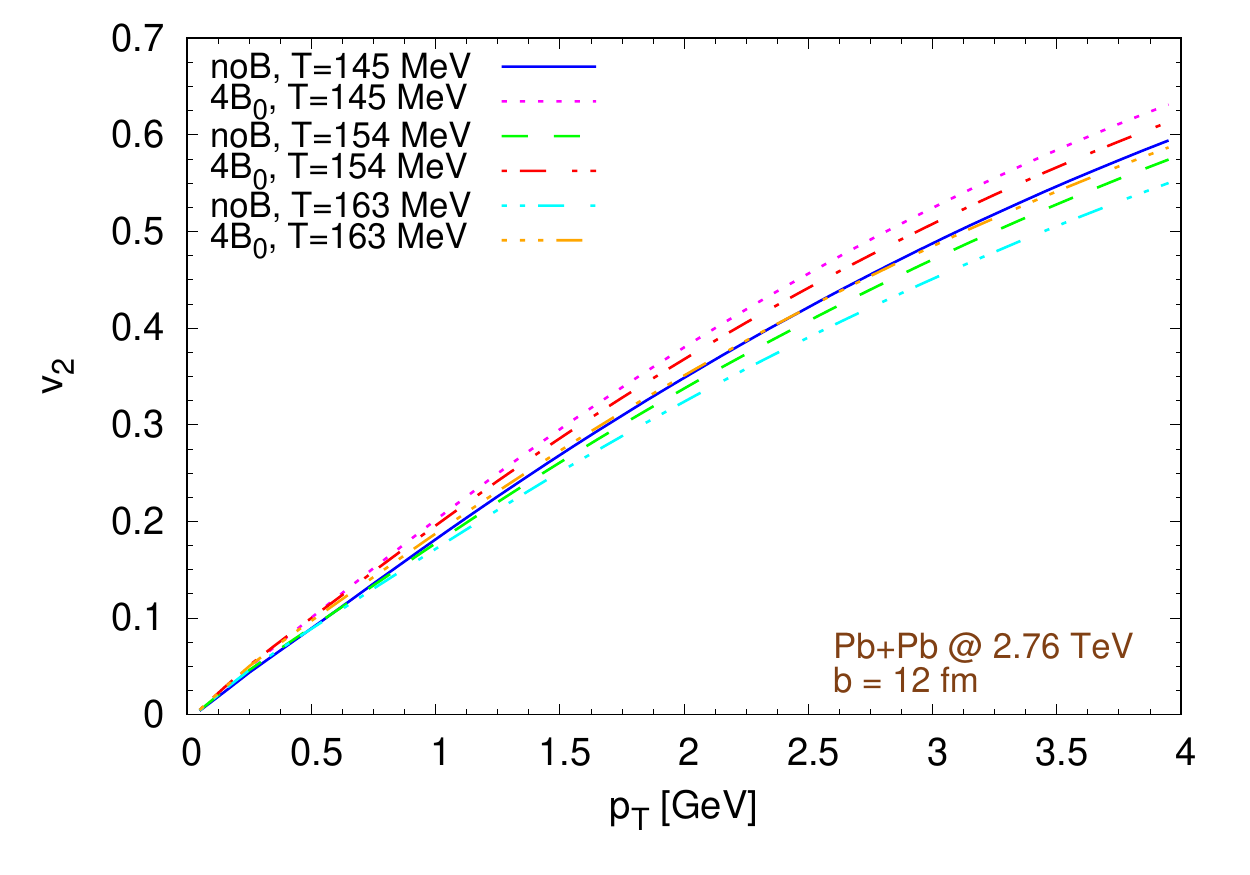}
	\end{minipage}
	\caption{Elliptic flow at mid-rapidity vs transverse momentum. The left figure refers to Au+Au collisions at $\snn=200\,\textrm{GeV}$, the right figure refers to Pb+Pb collisions at $\snn=2.76\,\textrm{TeV}$. In both cases, we run simulations with a fixed impact parameter $b=12\,\textrm{fm}$. We assume to have an initial magnetic field 4 times larger than our estimates. We explore the effect of different freeze-out temperatures. We notice how the presence of a magnetic field always enhances the elliptic flow compared to the pure hydro case.}
	\label{Tfo}
\end{figure*}

In all our simulations we choose as freeze-out temperature $154\,\textrm{MeV}$, the best value given in Ref.~\cite{Bazavov:2011nk}, with an uncertainty of $\pm9\,\textrm{MeV}$. We want to check that a different choice of the freeze-out temperature within these limits does not affect significantly the final results. To this aim, we run two additional simulations per collision energy with impact parameter $b=12\textrm{fm}$, using as values of the freeze-out temperature $145\,\textrm{MeV}$ and $163\,\textrm{MeV}$. In Fig.~(\ref{Tfo}) we compare the elliptic flow of pions computed in these cases with the standard case. The results show, as expected, that the magnitude of the elliptic flow depends inversely on the value of the freeze-out temperature (the larger the freeze-out temperature, the smaller $v_2$), however the magnetic fields always produce a similar enhancement of the elliptic flow. Temperature dependent effects, which might certainly appear in a more quantitative analysis, are out of the scope of the present work.

\section{Transformations from Minkowski to Milne/Bjorken coordinates}
\label{app:transf_law}
For the reader's convenience, we derive the transformation laws of the components of the three dimensional electromagnetic field from Minkowski coordinates $x^{\mu}=(t,x,y,z)$, with metric diag($g_{ij}$)=$(-1,1,1,1)$ to Milne coordinates $\tilde{x}^{\mu}=(\tau,x',y',\eta)$, with metric diag($g_{ij}$)=$(-1,1,1,\tau^2)$.
The Milne coordinates are related to the Minkowski coordinates by:
\begin{equation}
\left\{
\begin{array}{l}
\tau=\sqrt{t^2-z^2}\\
x'=x\\
y'=y\\
\eta=\dfrac{1}{2}\ln \dfrac{t+z}{t-z}
\end{array}\right.,
\end{equation}
while the opposite transformation is given by:
\begin{equation}
\left\{
\begin{array}{l}
t=\tau\cosh \eta\\
x=x'\\
y=y'\\
z=\tau\sinh \eta
\end{array}\right.,
\end{equation}
From now on a tilde character will be used to distinguish a tensor or a vector in Milne/Bjorken coordinates ($\tilde{F}$) from the same object in Minkowski coordinates ($F$). Keeping this convention in mind, we express the Faraday tensor in Minkowski and Milne coordinates as:
\bea
F^{\mu\nu}&=&
\begin{pmatrix}
0&E^x&E^y&E^z\\
-E^x&0&B_z&-B_y\\
-E^y&-B_z&0&B_x\\
-E^z&B_y&-B_x&0
\end{pmatrix}\,, \nonumber \\
\tilde{F}^{\mu\nu}&=&
\begin{pmatrix}
0&\tilde{E}^x&\tilde{E}^y&\tilde{E}^z\\
-\tilde{E}^x&0&\tilde{B}_z/\tau&-\tilde{B}_y/\tau\\
-\tilde{E}^y&-\tilde{B}_z/\tau&0&\tilde{B}_x/\tau\\
-\tilde{E}^z&\tilde{B}_y/\tau&-\tilde{B}_x/\tau&0
\end{pmatrix}\,.
\eea
We recall that the formula to change coordinates for a rank 2 contravariant tensor is:
\begin{equation}
\tilde{F}^{\mu\nu}=\dfrac{\partial \tilde{x}^{\mu}}{\partial x^{\rho}} \dfrac{\partial \tilde{x}^{\nu}}{\partial x^{\sigma}}F^{\rho \sigma}.
\end{equation}
The only non trivial derivatives are:
\bea
\dfrac{\partial \tau}{\partial t}&=&\dfrac{t}{\tau}=\cosh \eta\\ 
\dfrac{\partial \tau}{\partial z}&=&-\dfrac{z}{\tau}=-\sinh \eta\\ 
\dfrac{\partial \eta}{\partial t}&=&-\dfrac{z}{\tau^2}=-\dfrac{\sinh \eta}{\tau}\\ 
\dfrac{\partial \eta}{\partial z}&=&-\dfrac{t}{\tau^2}=\dfrac{\cosh \eta}{\tau} \,.
\eea
We obtain:
\bea
\tilde{E}^x&=&\cosh \eta E^x - \sinh \eta B_y \label{Ex_full} \\
\tilde{E}^y&=&\cosh \eta E^y + \sinh \eta B_x \label{Ey_full} \\
\tilde{E}^z&=&\dfrac{E^z}{\tau} ( \cosh^2 \eta - \sinh^2 \eta ) = \dfrac{E^z}{\tau} \label{Ez_full}
\eea
\bea
\tilde{B}^z &=& B_z \tau \dfrac{1}{\tau^2}=\dfrac{B_z}{\tau}=\dfrac{B^z}{\tau} \label{Bz_full} \\
\tilde{B}_y &=&\tilde{B}^y=\cosh \eta B_y - \sinh \eta E^x \label{By_full} \\
\tilde{B}_x &=& \tilde{B}^x=\cosh \eta B_x + \sinh \eta E^y \,. \label{Bx_full}
\eea
In the ideal MHD framework:
\begin{equation}
E_i=-\varepsilon_{ijk}v^jB^k,
\end{equation}
therefore:
\begin{align}
E_x&=v^zB^y-v^yB^z\\
E_y&=v^xB^z-v^zB^x\\
E_z&=v^yB^x-v^xB^y
\end{align}
Now we want to express the four velocity $u_M^{\mu}$, in Minkowski coordinates, but using Milne/Bjorken variables, of a fluid with Cartesian velocity components $v^x=v^y=0$ and $v^z=z/t$: 
\begin{equation}
v^z=v_z=\dfrac{z}{t}=\dfrac{\tau \sinh \eta}{\tau \cosh \eta}=\tanh \eta
\end{equation}
\begin{equation}
\gamma=\dfrac{1}{\sqrt{1-v_z v^z}}=\dfrac{1}{\sqrt{1-\tanh^2 \eta}}=\cosh \eta
\end{equation}
\begin{equation}
u_M^{\mu}=(\gamma, \gamma v^x, \gamma v^y, \gamma v^z)=(\cosh \eta, 0, 0, \sinh \eta)\,.
\end{equation}
When converting in Milne/Bjorken coordinates, we get:
\begin{equation}
u_B^{\mu}=\dfrac{\partial \tilde{x}^{\mu}}{\partial x^{\nu}}u_M^{\nu}
\end{equation}
and we obtain:
\begin{equation}
u_B^0=\dfrac{\partial \tau}{\partial t}u_M^{0}+\dfrac{\partial \tau}{\partial z}u_M^{3}=\cosh \eta \cosh \eta-\sinh \eta \sinh \eta=1,
\end{equation}
\begin{equation}
u_B^3=\dfrac{\partial \eta}{\partial t}u_M^{0}+\dfrac{\partial \eta}{\partial z}u_M^{3}=-\dfrac{\sinh \eta}{\tau}\cosh \eta + \dfrac{\cosh \eta}{\tau} \sinh \eta=0,
\end{equation}
so, at the end:
\begin{equation}
u_B^{\mu}=(1,0,0,0),
\end{equation}
which is the initial velocity profile used in this work.
In this case the transformation laws~(\ref{Ex_full}-\ref{Ez_full}) simplify to:
\begin{align}
E_x&=\tanh B^y \label{Ex_simpl}\\
E_y&=-\tanh B^x \label{Ey_simpl}\\
E_z&=0,\label{Ez_simpl}
\end{align}
while Eqs.~(\ref{Bx_full}-\ref{By_full}) simplify to:
\bea
\tilde{B}^x&=&\cosh \eta B^x + \sinh \eta (-\tanh \eta) B^x \nonumber \\
&=&\dfrac{\cosh^2 \eta - \sinh^2 \eta}{\cosh \eta}B^x=\dfrac{B^x}{\cosh \eta}\,,
\eea
\bea
\tilde{B}^y&=&\cosh \eta B^y - \sinh \eta (\tanh \eta) B^y\nonumber \\
&=&\dfrac{\cosh^2 \eta - \sinh^2 \eta}{\cosh \eta} B^y=\dfrac{B^y}{\cosh \eta}\,.
\eea

\section{Computation of the electric charge density in the fluid rest frame}
\label{app:chargedepcomp}
We consider two different methods to compute the electric charge density in the comoving frame of the fluid: the first one exploits directly the Maxwell equations, while the second one is based on the vorticity. 
In the first one we never make any special assumption about the electrical conductivity, therefore it is valid also in the non ideal case, without any change. In the second one the final simplified equation is valid only in the ideal MHD limit, otherwise additional terms (included in the derivation) should be considered.
\subsection{Method based on the Maxwell equations}
\label{maxwell_method}
After introducing the unit vector of the Eulerian observer as
\begin{equation}
	n_{\mu}=(-1,0,0,0),\qquad n^{\mu}=(1,0,0,0),\label{eq:unit_Eul}
\end{equation}
we can decompose the four velocity of the fluid as
\begin{equation}
	u^{\mu}=\Gamma n^{\mu}+\Gamma v^{\mu},\label{eq:vel_dec}
\end{equation}
where $v$ is the three velocity of the fluid and $\Gamma$ the corresponding Lorentz factor. From now on, we will use Latin indexes $i,j,k$ for the three vectors.

The Eulerian observer can split the four current as
\begin{equation}
	I^{\mu}=\rho_e n^{\mu}+J^{\mu}\label{eq:current_lab},
\end{equation}
in which the charge density $\rho_e$ and the current $J^{\mu}$ are measured in the lab frame.

On the other hand, as in Eq.~(\ref{eq:el_current}), we can also decompose the electric current in the proper electric charged density $\tilde{\rho_e}$ transported by the flow of the fluid and the conduction current $j^{\mu}$. The conduction current is orthogonal to the fluid four velocity $u_{\mu}$ (i.e. $j^{\mu}u_{\mu}=0$), therefore 
\begin{equation}
\tilde{\rho_e}=-I^{\mu}u_{\mu}. \label{eq:charge_com}
\end{equation}
If we compute the charge density in the comoving frame with Eq.~(\ref{eq:charge_com}) with $I^{\mu}$ and $u_{\mu}$ decomposed in the 3D+1 formalism (Eqs.~(\ref{eq:current_lab},\ref{eq:vel_dec})), we obtain
\begin{equation}
	\tilde{\rho_e}=\Gamma(\rho_e-J^i v_i).\label{eq:conv_rho_tilde}
\end{equation}
We can compute the charge density in the lab frame $\rho_e$ by using Eq.~(\ref{eq:maxwell2}), obtaining
\begin{equation}
	\rho_e=\partial_{k}(E^k)=q,\label{eq:comp_charge_dens_lab}
\end{equation}
where the $E^k$ are the components of the three dimensional electric field,
while from Eq.~(\ref{eq:maxwell1}) we get the three current $J^i$ in the lab frame as
\begin{equation}
-\tau J^i=\partial_{\tau}(\tau E^i)-\partial_j (\frac{[ijk]}{\tau} \tau B_k),\label{eq:comp_current_lab}
\end{equation}
where the $B_k$ are the components of the three dimensional magnetic field,
or, with a small rearrangement of the terms:
\begin{equation}
	J^i=\dfrac{\partial_j ([ijk]B_k)-\partial_{\tau}(\tau E^i)}{\tau}.\label{eq:J_in_lab_frame}
\end{equation}
Assembling the previous equation together, we arrive to the final expression for electric charge density in the fluid comoving frame as:
\begin{equation}
	\tilde{\rho_e}=\Gamma\left(\partial_{k}(E^k)-\dfrac{1}{\tau}\left[\partial_j ([ijk]B_k)-\partial_{\tau}(\tau E^i)\right]v_i\right).\label{eq:final}
\end{equation}
We recall that, in ideal MHD, the electric field in the lab frame is a quantity given by Eq.(\ref{eq:ohm2}).

In cartesian coordinates (but still using natural units), Eq.~(\ref{eq:final}) has the simplified expression:
\begin{equation}
	\tilde{\rho_e}=\Gamma\left[\nabla \cdot \vec{E}-\left(\nabla\times\vec{B}-\dfrac{\partial \vec{E}}{\partial t}\right)\cdot \vec{v} \right].\label{eq:final_simple}
\end{equation}
\subsection{Method based on the vorticity}
\label{vorticity_method}
Here we briefly summarize the same method described in the appendix of Ref.~\cite{Mignone:2019ebw}.  
After defining the \emph{kinematic vorticity} four-vector as \cite{2013rehy.book.....R}
\begin{equation}
\omega^\lambda = \epsilon^{\mu \nu \lambda \kappa} \nabla_\mu u_\nu \, u_\kappa =
\epsilon^{\mu \nu \lambda \kappa} \partial_\mu u_\nu \, u_\kappa,
\label{eq:kinematic_vorticity_def}
\end{equation}
we split the covariant derivative of the fluid velocity as
\begin{equation}
\nabla_\mu  u_\nu = - u_\mu a_\nu + \dfrac{1}{2} \epsilon_{\mu \nu \lambda \kappa}
\omega^\lambda u^\kappa,
\end{equation}
where $a^\mu = (u^\nu \nabla_\nu) u^\mu$ is the acceleration. We notice that $\omega^\mu u_\mu = 0$ and  $a^{\mu} u_{\mu}=0$, i.e. both the kinematic vorticity and the acceleration are orthogonal to fluid velocity.

If we use the definitions of $e^\mu$ and $b^\mu$, i.e. the electric and magnetic fields in the fluid comoving frame, given in Eqs.~(\ref{eq:e_4vec},\ref{eq:b_4vec}), we find the relation
\begin{equation}
F^{\mu \nu} \nabla_\mu  u_\nu =  e^\mu a_\mu + b^\mu \omega_\mu \label{eq:vorticity_relation}.
\end{equation}
By using Eq.~ (\ref{eq:e_4vec}) and Eq.~(\ref{eq:maxwell1}), we rewrite the divergence of the electric field in the fluid comoving frame as
\begin{equation}
\nabla_\mu e^\mu = \nabla_\mu (F^{\mu\nu}  u_\nu) =  - I^\mu u_\mu + F^{\mu\nu}
\nabla_\mu  u_\nu .
\end{equation}
We recall that, according to Eq.~(\ref{eq:charge_com}), $-I^{\mu}u_{\mu}=\tilde{\rho_e}$, where $\tilde{\rho_e}$ is the electric charge density in the fluid comoving frame, therefore:
\begin{equation}
\tilde{\rho_e}=\nabla_\mu e^\mu - F^{\mu\nu} \nabla_\mu  u_\nu.
\end{equation}
If we replace the last term of the previous equation with Eq.~(\ref{eq:vorticity_relation}), we end up with the expression
\begin{equation}\label{eq:q0}
\tilde{\rho_e}  =  \nabla_\mu e^\mu - e^\mu a_\mu  - b^\mu \omega_\mu.
\end{equation}
From Eq.~(\ref{eq:ohm}) we already know  that, in the ideal MHD approximation $e^\mu = 0$. Moreover, if we assume that Ohm's law 
\begin{equation}
e^{\mu}=\eta_r j^{\mu}
\end{equation}
holds, with the resistivity $\eta_r$ constant, then $\nabla_\mu e^\mu=\eta_r \nabla_\mu j^\mu$.
Therefore, if, as in the ideal MHD limit, we assume that $\eta_r=0$, Eq.(\ref{eq:q0}) simplifies into
\begin{equation}\label{eq:q0_final}
\tilde{\rho_e}  = - b^\mu \omega_\mu.
\end{equation}

\section{Charge dependent freeze-out}
\label{app:chargedepfo}
In the charge dependent freeze-out we modify the distribution functions of pions $f^0$ according to a factor $\lambda$, dependent on the local charge density, so that the distribution function of positive and negative pions become $f^+=(1+\lambda)f^0$ and $f^-=(1-\lambda)f^0$, obtaining at the end the same total number of pions $2f^0$, but with different abundancies of $\pi^+$ and $\pi^-$.

From a practical point of view, first we evaluate the expected average $\pi^{\pm}$ density $n_\pi$ in the fluid rest frame according to a Bose-Einstein distribution
\begin{equation}
n_\pi=\dfrac{1}{2\pi^2(\hbar c)^3}\int_0^\infty \frac{p^2}{e^{[(\sqrt{m^2+p^2}-\mu)/T]}-1}\mathrm{d}p.
	\label{Eq:BE_distr}
\end{equation}
Assuming a chemical potential $\mu\approxeq0$, given our chosen freezeout temperature $T=154\,\MeV$, for the charged pion mass $m=139.6\,\MeV$ we get $n_\pi\approxeq 0.04462$, which corresponds to a charge density $\rho^0_c=\pm  n_\pi \sqrt{4\pi\alpha \hbar c}\approxeq 0.006\, (\textrm{GeV}\textrm{fm})^{1/2}$.
If the net charge density $\rho_c$ in a freeze-out hypersurface cell is non zero, we assume that this due to a charged pion production imbalance, i.e. for one half to an enhancement of pion production with the same sign of the charge and for one half to a suppression of pion production with the opposite sign charge by a factor  $1 + s\cdot\rho_c/(2\rho^0_c)$, where $s$ is the sign of the electric charge of the pion. We do not take into account other hadron species.\\
This approach has several shortcomings: 1) as we already mentioned, it deals only with the number of produced pions, but not with their momentum distribution, assumed to be isotropic in the fluid comoving frame, thus neglecting all local electric currents 2) the computation of the electric charge density is imperfect, as we have demonstrated by the differences in the two methods employed here, which are particularly evident in the right side of Fig.~(\ref{v2_charged}). Moreover, there is no mechanism to prevent unrealistic charge densities, with enhancing/suppressing factors larger than 1, albeit typically we are far from this situation, 3) although pions are the most abundant species of hadrons produced in high energy heavy ion collisions, many other kinds are also produced (for a more complete hadronization which also respects local charge conservation, see e.g.~\cite{Oliinychenko:2019zfk}) 4) since the ions are positively charged, the QGP is not neutral (this, of course, is a general issue of the RMHD approach).
\bibliography{bibliofile}
\bibliographystyle{h-physrev.bst}

\end{document}